\documentclass[preprint,3p,fleqn,sort]{elsarticle}
\usepackage[labelfont=bf, singlelinecheck=false,
            justification=justified]{caption}
\usepackage{xcolor}
\usepackage{amsmath}
\usepackage{amssymb}
\usepackage{subcaption}
\usepackage{mathrsfs}
\usepackage{stmaryrd}
\usepackage{booktabs}
\usepackage{microtype}
\usepackage{nicefrac}
\usepackage{siunitx}
\usepackage{mathtools}
\usepackage{multirow}
\usepackage{enumitem}
\usepackage{fixltx2e}
\usepackage{setspace}
\usepackage{placeins}
\usepackage{etoolbox}
\usepackage[font=small]{caption}

\robustify\bfseries

\let\rho\varrho

\def\jump{\FloatBarrier}

\newcommand{\referee}[1]{#1}

\usepackage{lineno,hyperref}
\modulolinenumbers[5]

\renewcommand{\vec}[1]{\ensuremath{\boldsymbol #1}}
\newcommand{\mat}[1]{\ensuremath{\mathbf{#1}}}
\newcommand{\avg}[1]{\ensuremath{\left(#1\right)^\mathrm{A}}}
\newcommand{\avgln}[1]{\ensuremath{\left(#1\right)^\mathrm{ln}}}
\newcommand{\derivative}[2]{\frac{\mathrm{d} #1}{\mathrm{d} #2}}
\newcommand{\pderivative}[2]{\frac{\partial #1}{\partial #2}}

\newcommand{\ie}{\textit{i.e.}~}
\newcommand{\eg}{\textit{e.g.}~}

\begin{document}

\begin{frontmatter}
\title{A Novel High-Order, Entropy Stable, 3D AMR MHD Solver with Guaranteed Positive Pressure}
\author[physik]{Dominik Derigs\corref{mycorrespondingauthor}}
\cortext[mycorrespondingauthor]{Corresponding author}
\ead{derigs@ph1.uni-koeln.de}
\author[mathematik]{Andrew R.~Winters}
\author[mathematik]{Gregor J.~Gassner}
\author[physik]{Stefanie Walch}
\address[physik]{I.\,Physikalisches Institut, Universit\"at zu K\"oln, Z\"ulpicher Stra\ss{}e~77, 50937 K\"oln}
\address[mathematik]{Mathematisches Institut, Universit\"at zu K\"oln, Weyertal 86-90, 50931 K\"oln}

\numberwithin{equation}{section}

\begin{abstract}
We describe a high-order numerical magnetohydrodynamics (MHD) solver built upon a novel non-linear entropy stable numerical flux function that supports eight travelling wave solutions. By construction the solver conserves mass, momentum, and energy and is entropy stable. The method is designed to treat the divergence-free constraint on the magnetic field in a similar fashion to a hyperbolic divergence cleaning technique. The solver described herein is especially well-suited for flows involving strong discontinuities. Furthermore, we present a new formulation to guarantee positivity of the pressure. We present the underlying theory and implementation of the new solver into the multi-physics, multi-scale adaptive mesh refinement (AMR) simulation code \texttt{FLASH} (\url{http://flash.uchicago.edu}). The accuracy, robustness and computational efficiency is demonstrated with a number of tests, including comparisons to available MHD implementations in \texttt{FLASH}.
\end{abstract}

\begin{keyword}
magnetohydrodynamics \sep \texttt{FLASH} \sep entropy stable \sep finite volume schemes \sep pressure positivity
\end{keyword}

\end{frontmatter}
\section{Introduction}

Modelling complex non-linear astrophysical phenomena is a central task in the field of astrophysics where laboratory experiments are very difficult if not entirely impossible. Examples of interesting phenomena include the study of stellar evolution, like the star formation process and supernovae explosions, pre-stellar accretion discs and many more. Using simulations allows us to study the internals of complex systems that cannot been seen in experiments and observations.

In astrophysics, a flow involving magnetised gas is typically ionized, compressible, and often supersonic. Since the interstellar gas has essentially infinite conductivity \cite{Goldsmith1970}, we treat the flow by solving the ideal magnetohydrodynamics (MHD) equations. From the hyperbolic nature of the ideal MHD equations, it is known that discontinuous solutions may develop even from smooth initial data. Obtaining stable numerical results for the variety of physical flow regimes is extremely challenging, particularly for the natural requirement that the numerical scheme must be both accurate and robust. In this paper, we present a novel three-dimensional high-order, conservative, quasi-multifluid, entropy stable, eight wave MHD solver developed for the numerical modelling of MHD flows. It is equally well suited for one, two, or three-dimensional hydrodynamics (HD) and MHD simulations.

The core of the novel MHD solver is the use of entropy stable flux functions developed in \cite{Winters2016}. Entropy stable algorithms {\referee{have the benefit that}}, by construction, the numerical method is nearly isentropic in smooth regions and entropy is guaranteed to be increasing near discontinuities. Thus, the numerics precisely follow the physics of the second law of thermodynamics. Another advantage of entropy stable approximations is that one can limit the amount of dissipation added to the numerical approximation to guarantee entropy stability. The development and investigation of entropy stable algorithms for the ideal MHD equations has been considered by several authors \cite{Bouchut2007,Chandrashekar2015,Rossmanith2013,Winters2016}.

The entropy stable formulation also addresses the issue of divergence cleaning for approximate solutions of the ideal MHD equations. The proof of entropy stability in \cite{Winters2016} required an additional source term that acts analogously to a hyperbolic divergence cleaning technique \cite{dedner2002}. That is, errors introduced into the divergence-free condition are advected away with the fluid velocity.

The scheme handles another major robustness issue in numerical approximations of state-of-the-art high-order MHD solvers -- the possible appearance of negative pressures. Negative pressures are a numerical artifact arising due to the problem of finite numerical precision. This phenomenon has been reported frequently in the literature \cite{Christlieb2015,Huazheng2015,Xisto2014,Ersoy2013,Spicera2013,Waagan2011,Zhang2010,Wheatley2010,Li2008_2,Li2008,Toth2000,Janhunen2000,Balsara1999_2,balsara2012,Zachary1994,Ryu1993,Einfeldt1991}.
In current codes, negative pressures are avoided by adding artificially high amounts of dissipation or by introducing non-conservative low pressure limits.
Negative pressures can arise due the fact that the internal energy is obtained by subtracting the kinetic and magnetic energies from the conserved total energy. In many situations, such as high Mach number or low plasma $\beta$ flows ($\beta \propto p/\lVert\vec{B}\rVert^2$), the internal energy can be several orders of magnitude smaller then either the kinetic or magnetic energies. Thus, discretisation errors in the total energy could be significant enough to result in negative pressures. The inevitable consequence is the failure of the numerical scheme.

We describe how the novel solver uses the entropy as an auxiliary equation to eliminate this issue 
and derive a novel expression for the pressure which completely avoids the subtraction problem.
The new pressure positivity guaranteeing formulation is not tied to any specific numerical flux function. It remains general and it is straightforward to retrofit into any existing HD/MHD schemes if the underlying numerical approximation is constructed in a way that satisfies certain criteria on the entropy (see Sec. \ref{scn:PositivePressure}).

The new solver achieves high-order accuracy in space and time while remaining attractive from a computational point of view. The numerical scheme is extended to high-order in space with spatial reconstruction techniques. In particular, we use a third order spatial approximation with the newly developed reconstruction technique of Schmidtmann et al.~\cite{Schmidtmann2015}. High-order accuracy in time is obtained using the family of strong stability preserving (SSP) Runge-Kutta methods developed by Gottlieb et al.~\cite{Gottlieb2005}.

We provide here details of the novel solver as well as its implementation into the multi-scale multi-physics simulation code \texttt{FLASH} \citep{FLASH2000,FLASH2009}. \texttt{FLASH} is publicly available and has a wide international user base.
The remainder of this paper is organized as follows: Sec.~\ref{scn:Background} provides the necessary background information to discuss the novel numerical solver. In Sec.~\ref{scn:MHDSolver} we describe, in detail, the new solver. The most important aspects of which are the entropy stable numerical fluxes and the new pressure positivity guaranteeing formulation. Sec.~\ref{scn:NumResults} presents a variety of numerical results that demonstrate the utility of the new solver. We compare our results to already available MHD implementations in \texttt{FLASH} where applicable.
Sec.~\ref{scn:Conclusion}, presents our concluding remarks.

\section{Governing Equations and Discretisation}\label{scn:Background}

We first provide the necessary background to discuss the novel MHD solver. This includes a brief description of the ideal MHD equations, the concept of entropy conservation and stability, and an outline of the finite volume scheme used for the spatial discretisation.

\subsection{Ideal MHD Equations}\label{scn:idealMHD}

The ideal MHD model assumes that a fluid is a good electric conductor and neglects non-ideal effects, \eg viscosity or resistivity. It is governed by a system of conservation laws
\begin{align}\label{3DIDEALMHD}
\pderivative{}{t}\begin{bmatrix} \rho \\[0.05cm] \rho\vec{u} \\[0.05cm] E \\[0.05cm] \vec{B} \end{bmatrix} + &\nabla\cdot\begin{bmatrix} \rho\vec{u} \\[0.05cm]
\rho(\vec{u}\otimes\vec{u}) + \left(p+\frac{1}{2}\|\vec{B}\|^2\right)\mat{I}-\vec{B}\otimes\vec{B}  \\[0.05cm]
\vec{u}\left(E + p + \frac{1}{2}\|\vec{B}\|^2 \right) - \vec{B}(\vec{u}\cdot\vec{B}) \\[0.05cm]
\vec{B}\otimes\vec{u} - \vec{u}\otimes\vec{B}
\end{bmatrix} = 0, \\
&\nabla\cdot\vec{B} = 0, \label{eq:divB}
\end{align}
where $\rho$, $\rho\vec{u}$, and $E$ are the mass, momenta, and total specific energy of the plasma system, $p$ is the thermal pressure, $\mat{I}$ is the identity matrix, and $\vec{B}$ is the magnetic field, also referred to as magnetic flux density. \referee{Since our velocities are non-relativistic, Maxwell's displacement current may be ignored in the Lorentz force term. We consider the non-dimensional form of the ideal MHD equations. Details concerning physical units can be found in \ref{app:Units}.}

Numerical methods for multidimensional ideal MHD must satisfy some discrete version of the divergence-free condition \eqref{eq:divB}. There are several approaches to control the error in $\nabla\cdot\vec{B}$ and in depth review of many methods can be found in T\'{o}th \cite{Toth2000}. The thermal pressure is related to the conserved quantities through the ideal gas law for problems in which relativistic, viscous, and resistive effects can be neglected:
\begin{equation}\label{eq:pressure}
p = (\gamma-1)\left(E - \frac{\rho}{2}\|\vec{u}\|^2 -\frac{1}{2}\|\vec{B}\|^2 \right)
\end{equation}
with the ratio of specific heats $\gamma > 1$.

Note that if we take the divergence of Faraday's equation
the magnetic continuity equation
\begin{equation}\label{eq:magneticconti}
\frac{\partial}{\partial t}(\nabla\cdot\vec{B})+\nabla\cdot\big(\vec{u}(\nabla\cdot\vec{B})\big) = 0,
\end{equation}
is obtained. From \eqref{eq:magneticconti} we see that the divergence of the magnetic field may be treated as an advected scalar. \referee{Thereby, the robustness and accuracy of a numerical scheme can be significantly improved \cite{Powell1999}. This improvement is primarily because the advection of the generated errors prevents the accumulation at fixed locations.} The eigenmode which is advected with the flow in \eqref{eq:magneticconti} is referred to as the \emph{divergence wave}.

We include the Janhunen source term \cite{Janhunen2000} in the ideal MHD equations \eqref{3DIDEALMHD} which is proportional to $\nabla\cdot\vec{B}$. \referee{The use of a source term to control the error in the divergence free condition has known issues, such as errors can build up at stagnation points as well as in periodic or closed domains. The only mechanism present to remove these divergence errors is the numerical dissipation of a scheme, but true hyperbolic divergence cleaning methods can remove such limitations \cite{dedner2002}. However, the Janhunen source term} preserves the conservation of mass, momentum, total energy \referee{and allows for the construction of an entropy stable approximation \cite{Winters2016}. We explicitly ``clean'' magnetic field divergence errors in a post-processing step, as will be described later.} The governing equations in conjunction with the Janhunen source term are now a system of balance laws
\begin{equation}\label{JanhunenSource}
\pderivative{}{t}\begin{bmatrix} \rho \\[0.05cm] \rho\vec{u} \\[0.05cm] E \\[0.05cm] \vec{B} \end{bmatrix} + \nabla\cdot\begin{bmatrix} \rho\vec{u} \\[0.05cm]
\rho(\vec{u}\otimes\vec{u}) + \left(p+\frac{1}{2}\|\vec{B}\|^2\right){\color{black}{\mat{I}}}-\vec{B}\otimes\vec{B}  \\[0.05cm]
\vec{u}\left(E + p + \frac{1}{2}\|\vec{B}\|^2 \right) - \vec{B}(\vec{u}\cdot\vec{B}) \\[0.05cm]
\vec{B}\otimes\vec{u} - \vec{u}\otimes\vec{B}
\end{bmatrix} = -(\nabla\cdot\vec{B})\begin{bmatrix} 0\\\vec{0}\\0\\\vec{u}\end{bmatrix}.
\end{equation}
\referee{Note that the expression ``source term'' is common in this context, even though the term actually involves spatial derivatives.}

To simplify the discussion of the new solver we first consider the modified ideal MHD system \eqref{JanhunenSource} in one spatial dimension
\begin{equation}\label{eq:hydro:1DMHD}
\frac{\partial}{\partial t} \vec{Q} + \frac{\partial}{\partial x} \vec{F} = \vec{\Upsilon},
\end{equation}
where $\vec Q = \vec Q(x,t)$ is the vector of conservative variables, $\vec{F}(\vec{Q})$ the flux vector, and $\vec{\Upsilon}(\vec{Q})$ is the vector source term
\begin{equation}
\vec{Q} = 
\begin{bmatrix}\rho \\ \rho u \\ \rho v \\ \rho w \\ E \\ B_1 \\ B_2 \\ B_3  \end{bmatrix}
,\qquad
\vec{F} =
\begin{bmatrix}\rho \, u \\ \rho u^2 + p + \frac{1}{2} \lVert \vec{B} \rVert^2 -B_1^2 \\ \rho \, u \, v - B_1  B_2 \\ \rho \, u \, w - B_1  B_3 \\ u \left(E + p + \frac{1}{2} \lVert \vec{B} \rVert^2\right) - B_1 \left(\vec{u} \cdot \vec{B}\right) \\ 0 \\ u\, B_2 - v \, B_1 \\ u \, B_3 - w\, B_1  \end{bmatrix},
\qquad\vec{\Upsilon} = -\pderivative{B_1}{x}\begin{bmatrix}
0\\
0\\
0\\
0\\
0\\
u\\
v\\
w
\end{bmatrix}.
\end{equation}
In Sec.~\ref{scn:multidim} we provide a detailed discussion of the multi-dimensional extension of the solver.

\subsection{Entropy Conservation and Stability}\label{scn:Entropy}

This section serves as a brief introduction to entropy and numerical partial differential equations. A thorough review of this topic has been presented by Tadmor \cite{tadmor2003}. Work specifically related to entropy and the ideal MHD equations can be found in \cite{Chandrashekar2015,Winters2016}.

It is well-known that solutions of balance laws like \eqref{eq:hydro:1DMHD} may develop discontinuities in finite time,  so we consider solutions of the balance laws \eqref{eq:hydro:1DMHD} in the weak sense.
Unfortunately, the weak solution is not unique. Thus, we require an additional admissibility condition on the solution to guarantee that the numerical approximation will converge to a weak solution that is consistent with the second law of thermodynamics. In the case of ideal MHD a suitable condition can be defined in terms of the physical entropy density, as defined by Landau \cite[p.~315]{Landau1959}, divided by the constant $(\gamma -1)$ for convenience, \ie
\begin{equation}\label{eq:entropy}
S(\vec{Q}) = \frac{\rho s}{\gamma - 1} \quad \mbox{with} \quad s = \ln\big(p \rho^{-\gamma}\big),
\end{equation}
where $\gamma$ is the adiabatic index and $s$ is the entropy per particle. The approximation obeys the second law of thermodynamics and is based on an entropy condition for two regimes:
\begin{enumerate}
\item For smooth solutions, one can design numerical methods to be \textbf{\emph{entropy conservative}} if, discretely, the local changes of entropy are the same as predicted by the continuous entropy conservation law
\begin{equation}
\pderivative{}{t}S + \pderivative{}{x}\mathscr{F} = 0,
\end{equation}
where we define the corresponding entropy flux
\begin{equation}
\mathscr{F}(\vec{Q}) = uS = \frac{\rho u s}{\gamma-1}\ .
\end{equation}

\item For discontinuous solutions, the approximation is said to be \textbf{\emph{entropy stable}} if the entropy always possesses the correct sign and the numerical scheme produces more entropy than an entropy conservative scheme and satisfies the entropy inequality
\begin{equation}\label{eq:EntropyInequality}
\pderivative{}{t}S + \pderivative{}{x}\mathscr{F} \ge 0.
\end{equation}
\end{enumerate}
From the second law of thermodynamics, kinetic as well as magnetic energy can be transformed irreversibly into heat (internal energy). If additional dissipation is not included in an entropy conservative method, spurious oscillations will develop near discontinuities as energy is re-distributed at the smallest resolvable scale \citep{Mishra2011}. A numerical scheme requires a diffusion operator to match such a physical process.

For the entropy stable solver discussed in this paper we use the provably entropy stable approximate Riemann solver derived in \cite{Winters2016}.

\subsection{Finite Volume Scheme}\label{scn:FV}

The finite volume method is a discretisation technique for partial differential equations especially useful for the approximation of systems of hyperbolic conservations laws. The finite volume method is designed to approximate conservation laws in their integral form, \textit{e.g.},
\begin{equation}
\int_V\vec{Q}_t\,\mathrm{d}x + \int_{\partial V} \vec{F}\cdot\hat{\vec{n}}\,\mathrm{d}S = 0.
\end{equation}
In one spatial dimension we divide the interval, $V$, into cells
\begin{equation}
V_i = \left[x_{i-1/2},x_{i+1/2}\right],
\end{equation}
and the integral equation of a balance law with a source term becomes
\begin{equation}\label{FVSource}
\frac{\mathrm{d}}{\mathrm{d}t}\int_{x_{i-1/2}}^{x_{i+1/2}} \vec{Q}\,\mathrm{d}x + \big[\vec{F}^*\left(x_{i+1/2}\right) - \vec{F}^*\left(x_{i-1/2}\right) \big] = \int_{x_{i-1/2}}^{x_{i+{1/2}}} \vec{\Upsilon}\,\mathrm{d}x.
\end{equation}
A common approximation is to assume a constant solution within the cell \citep[p.~436]{LeVeque1998}:
\begin{equation}
\int_{x_{i-1/2}}^{x_{i+1/2}} \vec{Q}\,\mathrm{d}x \approx \int_{x_{i-1/2}}^{x_{i+{1/2}}} \vec{Q}_i\,\mathrm{d}x = \vec{Q}_i\Delta x_i.
\end{equation}
Note that the finite volume solution is typically discontinuous at the boundaries of the cells. To resolve this, we introduce the idea of a ``numerical flux'', $\vec{F}^*(\vec{Q}_R,\vec{Q}_L)$, often derived from the (approximate) solution of a Riemann problem. The function $\vec{F}^*$ takes the two states of the solution at an element interface and returns a single flux value. For consistency, we require that
\begin{equation}\label{consistency}
\vec{F}^*(\vec{Q},\vec{Q}) = \vec{F},
\end{equation}
that is, the numerical flux is equivalent to the physical flux if the states on each side of the interface are identical.

Next, we address the discretisation of the source term $\vec{\Upsilon}$ in \eqref{FVSource}. There is a significant amount of freedom in the source term discretisation. The explicit discretisation of the source term is given in Sec.~\ref{scn:flux1}. We note that the discrete source term \referee{at each left ($i-1/2$) and right ($i+1/2$) interface} will contribute in cell $i$. So, the semi-discrete finite volume method is
\begin{equation}\label{FVMethodOurs}
\left(\vec{Q}_t\right)_i +\frac{1}{\Delta x_i}\big[\vec{F}_{i+1/2}^* - \vec{F}_{i-1/2}^*\big] = \frac{1}{2}\left(\vec{\Upsilon}_{i-1/2} + \vec{\Upsilon}_{i+1/2}\right).
\end{equation}

\section{Description of the Novel Entropy Stable MHD Solver}\label{scn:MHDSolver}

Here we describe the \texttt{FLASH} implementation of the entropy stable \texttt{ES} solver in three spatial dimensions. In Sec.~\ref{scn:multidim} we discuss the extension of the solver to three dimensions using dimensional splitting. Sec.~\ref{scn:MUSCL} presents a spatial reconstruction scheme used to achieve a high-order approximation. We describe the explicit time integration technique in Sec.~\ref{scn:SSPRK}. The entropy conservative and entropy stable numerical flux functions are described in Sec.~\ref{scn:flux1} and Sec.~\ref{scn:flux2}, respectively. The new strategy to numerically guarantee the positivity of the pressure is described in Sec.~\ref{scn:PositivePressure}. The adaptive mesh refinement (AMR) functionality of \texttt{FLASH} and the new implementation is found in Sec.~\ref{scn:AMR}. Next, in Sec.~\ref{scn:Multifluid}, a brief summary of a quasi-multifluid implementation is provided. Sec.~\ref{scn:Gravity} describes how to couple gravity into the entropy stable solver. The treatment of the divergence-free condition in higher spatial dimensions is described in Sec.~\ref{scn:divB}. Finally, Sec.~\ref{scn:Procedure} summarizes the MHD update procedure in \texttt{FLASH}.

\subsection{Multi-dimensionality}\label{scn:multidim}

We extend the one-dimensional set of MHD equations \eqref{eq:hydro:1DMHD} to two or three spatial dimensions. In the case of an underlying grid structure that is logically rectangular\footnote{not strictly rectangular since cells of different spatial sizes are allowed to coexist on the same grid} (like Cartesian grid geometries) a simple and efficient way of extending the one-dimensional Riemann solver to higher spatial dimensions is to use \emph{dimensional splitting}. The method of dimensional splitting has become popular in fluid dynamics as it allows us to apply our knowledge about one-dimensional systems directly to multi-dimensional systems.
Using the dimensional splitting method, one-dimensional problems along each coordinate direction are solved in turn to determine the fluxes across the faces of a finite volume cell.
It has proven to be an inexpensive way of extending one-dimensional high-resolution methods to higher dimensions \citep[p.~103]{LeVeque1998}.

\referee{We experience that in multi-physics simulations, commonly performed using \texttt{FLASH}, the MHD solver accounts for less than 10\% of the overall CPU time (\eg \cite{Walch2014}). Thus, an MHD discretisation which allows large time steps is beneficial for the overall computational efficiency of the multi-physics framework. It is well-known that dimensionally split schemes give larger time steps than comparable unsplit schemes where the dimensionality directly enters the CFL condition. Although the technique of dimensional splitting reduces the accuracy of the solver to formally second-order, the overall increase in efficiency is often favourable for practical applications.}

If the three-dimensional semi-discrete problem can be written in the form of
\begin{align}
\left(\vec{Q}_t\right)_i + \mat{A} (\vec{Q}) + \mat{B}(\vec{Q}) + \mat{C}(\vec{Q}) &= 0, \label{eq:totalUpdate}\\
\intertext{then the total update \eqref{eq:totalUpdate} can be split up into an \emph{x-sweep}}
\left(\vec{Q}_t\right)_i + \mat{A}(\vec{Q}) &= 0 ,
\intertext{a \emph{y-sweep}}
\left(\vec{Q}_t\right)_i + \mat{B} (\vec{Q}) &= 0,
\intertext{and a \emph{z-sweep}}
\left(\vec{Q}_t\right)_i + \mat{C} (\vec{Q}) &= 0,
\end{align}
where $\mat{A}(\vec{Q})$, $\mat{B}(\vec{Q})$, and $\mat{C}(\vec{Q})$ are operators for the vector of quantities $\vec Q$ in $x$, $y$, and $z-$directions, respectively. Each of the sweep operators is a compact notation to write the numerical flux and source term contributions for a given spatial direction. For example, the operator $\mat{A}(\vec{Q})$ in three dimensions has the form
\begin{equation}
\mat{A}(\vec{Q}) = \frac{1}{\Delta x_i}\left(\vec{F}_{i+1/2,j,k}^* - \vec{F}_{i-1/2,j,k}^*\right) - \frac{1}{2}\left(\vec{\Upsilon}_{i-1/2,j,k} + \vec{\Upsilon}_{i+1/2,j,k}\right).
\end{equation}
Therefore, in each sweep direction, separate solutions of the Riemann problem and source term values are computed to update the quantities stored in $\vec{Q}^n$ according to \eqref{FVMethodOurs}.

To compute the sweeps in $y$- and $z$-directions, any direction dependent quantities, \ie velocity and magnetic field components, are rotated in order to solve them with the same algorithm that is used for the $x$-sweep.

\subsection{Spatial Reconstruction}\label{scn:MUSCL}

\begin{figure}[!ht]
	\centering
	\includegraphics[scale=1]{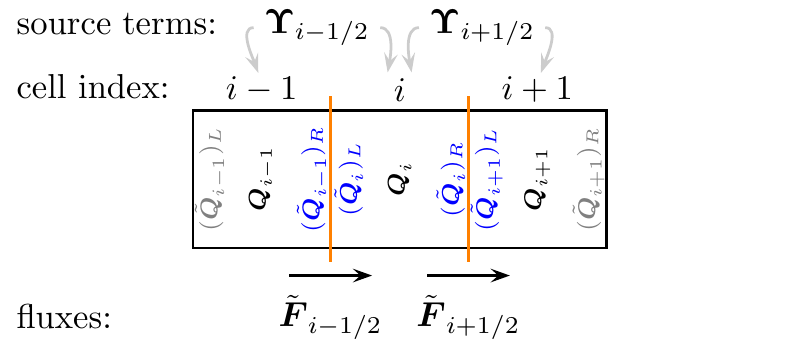}
	\caption{Graphical representation of the quantities used in \eqref{eq:Ftilde} and \eqref{eq:high_order_update}. Reconstructed quantities used for the computation of the numerical fluxes are highlighted in blue. The cell-centred quantities are printed in black.}
	\label{fig:flux}
\end{figure}

The finite volume method used by the \texttt{FLASH} framework approximates the solution with quantities which are constant within each cell. If one considers these values as point-wise approximations of the solution located at each cell centre, this method computes the numerical interface fluxes at a distance of ${\Delta x}/{2}$ from an interface.
Rather than using piecewise constant data, we use \emph{reconstructed} quantities within each cell, $(\tilde{\vec{Q}}_i)_{\rm L,R}$. Reconstruction functions, $(\vec{p}_i)_\mathrm{L,R}$, allow the computation of the approximated interface quantities
\begin{equation}
	(\tilde{\vec{Q}}_{i})_\mathrm{L} = \vec{Q}_{i} - \frac{1}{2}(\vec{p}_i)_\mathrm{L}, %
	\quad\mbox{and}\quad
	(\tilde{\vec{Q}}_{i})_\mathrm{R} = \vec{Q}_{i} + \frac{1}{2}(\vec{p}_i)_\mathrm{R}.
	\label{eq:reconstruction} %
\end{equation}
The reconstructed quantities \eqref{eq:reconstruction} are then used to compute high-order accurate numerical fluxes in the finite volume scheme \eqref{FVMethodOurs}, \ie
\begin{equation}\label{eq:Ftilde}
	\tilde{\vec{F}}_{i-1/2} = \vec{F^*}\Big( \big(\tilde{\vec{Q}}_{i-1} \big)_\mathrm{R},\, \big(\tilde{\vec{Q}}_i \big)_\mathrm{L} \Big) \quad\mbox{and}\quad
	\tilde{\vec{F}}_{i+1/2} = \vec{F^*}\Big( \big(\tilde{\vec{Q}}_{i} \big)_\mathrm{R},\, \big(\tilde{\vec{Q}}_{i+1} \big)_\mathrm{L} \Big).
\end{equation}
The resulting high-order accurate semi-discrete approximation, \referee{reorganizing \eqref{FVMethodOurs}}, is of the form
\begin{equation}\label{eq:high_order_update}
\left(\vec{Q}_t\right)_i = \frac{1}{\Delta x_i}\left(\tilde{\vec{F}}_{i-1/2} - \tilde{\vec{F}}_{i+1/2}\right) + \frac{1}{2}\left(\vec{\Upsilon}_{i-1/2} + \vec{\Upsilon}_{i+1/2}\right),
\end{equation}
as illustrated in Figure~\ref{fig:flux}.

\begin{figure}[!ht]
	\centering
	\includegraphics[scale=1.25]{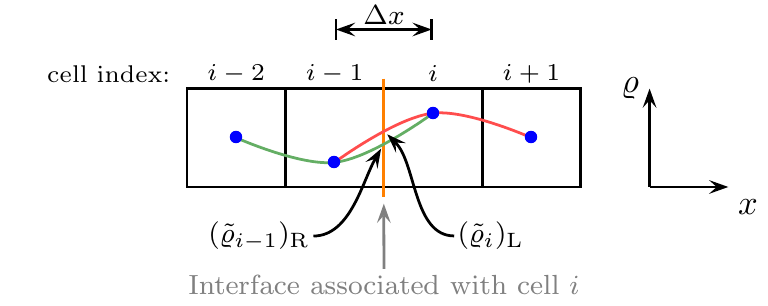}
	\caption{Principle of our spatial reconstruction. This example shows the parabolic reconstruction of a specific density pattern. The cell-centred quantities, $\rho_{i-2}$, $\rho_{i-1}$, $\rho_{i}$, and $\rho_{i+1}$, are represented by blue dots. Our scheme uses a local three-point stencil and is thereby computationally very efficient.}
	\label{fig:reconstruction}
\end{figure}

For our reconstruction we use the third order accurate shock capturing limiting procedure for numerical solutions of hyperbolic conservation laws recently described by Schmidtmann et al.~\cite{Schmidtmann2015}. Their scheme utilizes a local piecewise-parabolic reconstruction away from discontinuities (see Fig.~\ref{fig:reconstruction}) and reads
\begin{align}
	(\vec{p}_i)_\mathrm{L} = \vec{p}(\vec{Q}_{i-1}, \vec{Q}_i, \vec{Q}_{i+1}) &= +\frac{2}{3} \vec{Q}_{i-1} - \frac{1}{3}\vec{Q}_{i} - \frac{1}{6} \vec{Q}_{i+1} = \frac{2\vec{\delta_{i-\frac{1}{2}}} - \vec{\delta}_{i+\frac{1}{2}}}{3}, \\	(\vec{p}_i)_\mathrm{R} = \vec{p}(\vec{Q}_{i+1}, \vec{Q}_i, \vec{Q}_{i-1}) &= -\frac{1}{6} \vec{Q}_{i-1} - \frac{1}{3}\vec{Q}_{i} + \frac{2}{3} \vec{Q}_{i+1} = \frac{2\vec{\delta_{i+\frac{1}{2}}} - \vec{\delta}_{i-\frac{1}{2}}}{3},
\end{align}
with
\begin{equation}
\vec{\delta}_{i-\frac{1}{2}} = \vec{Q}_{i} - \vec{Q}_{i-1} \quad\mbox{and}\quad \vec{\delta}_{i+\frac{1}{2}} = \vec{Q}_{i+1} - \vec{Q}_{i} .
\end{equation}
However, such a reconstruction is known to cause oscillations in non-smooth solutions. This can be seen as a direct consequence of Godunov's Theorem \cite{Godunov1959}.
To avoid oscillations, we use the limiting procedure of Schmidtmann et al.~\cite{Schmidtmann2015} to switch to a lower-order accurate reconstruction near large gradients, shocks and discontinuities.

\subsection{Strong Stability Preserving Time Integration}\label{scn:SSPRK}

The solution of a system of hyperbolic conservation laws may not be smooth.
In such cases inaccurate time-integration schemes can suffer from poor performance such as an excessively small time step size due to the presence of spurious oscillations as well as the progressive smearing, clipping or squaring of the numerical approximation.
To alleviate such performance issues, we consider a third order accurate explicit high-order \emph{strong-stability-preserving} (SSP) low-storage Runge-Kutta time-integration scheme \cite{Gottlieb2005}. Such schemes are also referred to in the literature as \emph{total variation diminishing} (TVD) \cite{Shu1988}. However, Gottlieb et al.~\cite{Gottlieb2001} showed this moniker is misleading as their strong stability property holds in any norm and not only the TVD norm.
We complete the discretisation of the reconstructed method \eqref{eq:high_order_update} with the third order SSP Runge-Kutta scheme:
\begin{align}\label{eq:CompleteDisc}
	\vec{Q}' &= \vec{Q}^n + \Delta t \cdot \vec{Q}_t(\vec{Q}^n),\\
	\vec{Q}'' &= \frac{3}{4}\vec{Q}^n + \frac{1}{4} \Big( \vec{Q}' + \Delta t \cdot \vec{Q}_t(\vec{Q}') \Big),\\
	\vec{Q}^{n+1} &= \frac{1}{3}\vec{Q}^n + \frac{2}{3} \Big( \vec{Q}'' + \Delta t \cdot \vec{Q}_t(\vec{Q}'') \Big).
\end{align}
SSP Runge-Kutta schemes consist of convex combinations of explicit forward Euler integration. Thus, the family of methods are guaranteed to be stable under the same time step restriction \cite{Gottlieb2005}. \referee{We find that the third order SSP Runge-Kutta time integration enables us to use larger time steps, which is favorable in our multi-physics framework.}

To select a stable time step for a computational run we use the CFL condition
\begin{equation}\label{eq:CFL_timestep_3D}
\Delta t \le \mathtt{CFL} \cdot \min\bigg[\frac{\Delta x}{\lambda_{\mathrm{max}}^x}, \frac{\Delta y}{\lambda_{\mathrm{max}}^y}, \frac{\Delta z}{\lambda_{\mathrm{max}}^z}\bigg],
\end{equation}
where $\lambda^d_\mathrm{max}$ is the speed of the largest wave at time step $n$ travelling in $d = \{x,y,z\}$ direction, $\mathtt{CFL}$ is the user-definable CFL coefficient, $\mathtt{CFL} \in (0,1] $. If $\lambda_\mathrm{max}$ is known exactly, then the choice $\mathtt{CFL} = 1.0$ may be adequate \citep[p.~222]{Toro2009}. However, $\lambda_\mathrm{max}$ is usually computed in some approximate way. Thus, a more conservative choice for the CFL coefficient is typically used in practice (\eg $\mathtt{CFL} = 0.8$).

\subsection{Entropy Conserving Numerical Flux}\label{scn:flux1}

For the entropy analysis of the ideal MHD equations the divergence-free condition is incorporated into the system of conservation laws as a source term \cite{Godunov1972,Janhunen2000}. Both the Powell \cite{Powell1999} and Janhunen \cite{Janhunen2000} source terms treats the magnetic field as an advected scalar. However, the Janhunen source term remains conservative in the momentum and total energy equations and restores the positivity of the Riemann problem as well as Lorentz invariance \citep{Dellar2001}.

The discussion of the entropy conserving numerical flux function of \cite{Winters2016} requires the introduction of some notation. We introduce the jump $\llbracket \cdot \rrbracket$, the arithmetic mean $\avg{\cdot}$ as well as the logarithmic mean $\avgln{\cdot}$ of the left/right states, denoted by $(\cdot)_\mathrm{L}$ and $(\cdot)_\mathrm{R}$, respectively. These operators are defined as
\begin{equation}\label{eq:Averages}
	\llbracket \cdot \rrbracket = (\cdot)_\mathrm{R} - (\cdot)_\mathrm{L}, \qquad
	\avg{\cdot} = \frac{(\cdot)_\mathrm{L} + (\cdot)_\mathrm{R}}{2}, \qquad\mbox{and}\qquad \avgln{\cdot} = \frac{\llbracket \cdot \rrbracket}{\llbracket \ln(\cdot) \rrbracket} .
\end{equation}
A numerically stable procedure to compute the logarithmic mean is described by Ismail and Roe \cite[Appendix B]{ismail2009}. For convenience we also introduce
\begin{equation}\label{eq:Paramters}
	z_1 = \sqrt{\frac{\rho}{p}}, \qquad\mbox{and}\qquad z_5 = \sqrt{\rho p}.
\end{equation}

\subsubsection{Source Term Discretisation}

It was shown in \cite{Winters2016} that the Janhunen source term can be used to design numerical schemes that guarantee the discrete conservation of the entropy density for the ideal MHD equations.
Guaranteeing this discrete conservation of the entropy density requires a particular discretisation of the Janhunen source term:
\begin{equation}\label{eq:sourceterm}
\frac{1}{2}\left(\vec{\Upsilon}_{i-1/2} + \vec{\Upsilon}_{i+1/2}\right),
\end{equation}
with
\begin{equation}\label{eq:sourceterm2}
\begin{aligned}
\vec{\Upsilon}_{i-1/2} = -{\llbracket B_1 \rrbracket_{i-1/2}} \begin{bmatrix}0\\0\\0\\0\\0\\
\frac{(u z_1^2)^\mathrm{A} \avg{B_1}}{(\Delta x z_1^2 B_1)^\mathrm{A}}\\[0.15cm]
\frac{(v z_1^2)^\mathrm{A} \avg{B_2}}{(\Delta x z_1^2 B_2)^\mathrm{A}}\\[0.15cm]
\frac{(w z_1^2)^\mathrm{A} \avg{B_3}}{(\Delta x z_1^2 B_3)^\mathrm{A}}\end{bmatrix}_{i-1/2},
\quad\mbox{and}\quad
\vec{\Upsilon}_{i+1/2} = -{\llbracket B_1 \rrbracket_{i+1/2}} \begin{bmatrix}0\\0\\0\\0\\0\\
\frac{(u z_1^2)^\mathrm{A} \avg{B_1}}{(\Delta x z_1^2 B_1)^\mathrm{A}}\\[0.15cm]
\frac{(v z_1^2)^\mathrm{A} \avg{B_2}}{(\Delta x z_1^2 B_2)^\mathrm{A}}\\[0.15cm]
\frac{(w z_1^2)^\mathrm{A} \avg{B_3}}{(\Delta x z_1^2 B_3)^\mathrm{A}}\end{bmatrix}_{i+1/2}.
\end{aligned}
\end{equation}

\subsubsection{Entropy Conserving Flux Function}

The recently developed provably entropy conserving flux of Winters and Gassner \cite{Winters2016} reads:
\begin{equation}\label{eq:ec_flux}
\vec{F}^\mathrm{*,ec} = \begin{bmatrix}
	\hat \rho \hat u_1 \\
	\hat p_1 + \hat \rho \hat u_1^2 + \frac{1}{2} \left(\mathring B_1 + \mathring B_2 + \mathring B_3\right) - \mathring B_1 \\
	\hat \rho \hat u_1 \hat v_1 - \widehat{B_1 B_2} \\
	\hat \rho \hat u_1 \hat w_1 - \widehat{B_1 B_3} \\
	\frac{\gamma}{\gamma - 1}\, \hat u_1 \hat p_2 + \frac{1}{2}\, \hat \rho \hat u_1 \left( \hat u_1^2 + \hat v_1^2 + \hat w_1^2 \right) + \hat u_2 \left( \hat B_2^2 + \hat B_3^2 \right) - \hat B_1 \left(\hat v_2 \hat B_2 + \hat w_2 \hat B_3\right) \\
	0 \\
	\hat u_2 \hat B_2 - \hat v_2 \hat B_1 \\
	\hat u_2 \hat B_3 - \hat w_2 \hat B_1
\end{bmatrix},
\end{equation}
with the averaged quantities and products
\allowdisplaybreaks
\begin{equation}\label{eq:Averages2}
\begin{aligned}
	\hat \rho &= \avg{z_1}  \avgln{z_5},
	&\hat p_1 &= \frac{\avg{z_5}}{\avg{z_1}},
	&\hat p_2 &= \frac{\gamma + 1}{2 \gamma} \frac{\avgln{z_5}}{\avgln{z_1}} + \frac{\gamma - 1}{2 \gamma} \frac{\avg{z_5}}{\avg{z_1}}, \\
	\hat u_1 &= \frac{\avg{z_1 u}}{\avg{z_1}},
	& \hat v_1 &= \frac{\avg{z_1 v}}{\avg{z_1}},
	& \hat w_1 &= \frac{\avg{z_1 w}}{\avg{z_1}},\\
	\hat u_2 &= \frac{\avg{z_1^2 u}}{\avg{z_1^2}},
	&\hat v_2 &= \frac{\avg{z_1^2 v}}{\avg{z_1^2}},
	&\hat w_2 &= \frac{\avg{z_1^2 w}}{\avg{z_1^2}}, \\
	\hat B_1 &= \avg{B_1},
	&\hat B_2 &= \avg{B_2},
	&\hat B_3 &= \avg{B_3}, \quad\ \widehat{B_1 B_2} = \avg{B_1 B_2},\\
	\mathring B_1 &= \avg{B_1^2},
	&\mathring B_2 &= \avg{B_2^2},
	&\mathring B_3 &= \avg{B_3^2}, \quad\, \widehat{B_1 B_3} = \avg{B_1 B_3}.
\end{aligned}
\end{equation}
In the case of smooth solutions, the entropy conserving flux \eqref{eq:ec_flux} conserves the entropy density of the system up to the precision of the scheme. In order for the numerical scheme to be applicable for possibly non-smooth solutions we must extend the purely entropy \emph{conserving} flux to become an entropy \emph{stable} flux.

\subsection{Entropy Stabilization}\label{scn:flux2}

Entropy conserving approximations suffer breakdown in the presence of discontinuities, which results in large oscillations in post-shock regions. Therefore, we require dissipation to be added to the approximation in an entropy consistent manner to guarantee discrete satisfaction of the entropy inequality \eqref{eq:EntropyInequality}. The work \cite{Winters2016} derived two provably entropy stable approximate Riemann solvers for the ideal MHD equations. In this work we present a new hybrid entropy stable approximation that continuously combines these two entropy stable fluxes. This introduces explicit non-linearity to permit the calculation of sharp shock fronts and contact discontinuities.

\subsubsection{Entropy Stable Flux Functions}\label{scn:flux2a}

To build an entropy stable approximation we use the entropy conservative approximation \eqref{eq:ec_flux} as a baseline. In particular the work \cite{Winters2016} presented two possible dissipation terms that can be added to the entropy conserving scheme:
\begin{description}
	\item[\texttt{ES-Roe}:] a \emph{matrix dissipation} entropy stabilization. Similar to a Roe type method it selectively applies dissipation to each of the travelling wave solutions, particularly close to shocks.
	\item[\texttt{ES-LLF}:] a \emph{scalar dissipation} entropy stabilization. A simple, local Lax-Friedrichs type dissipation mechanism. Due to the simplicity of \texttt{ES-LLF} it cannot distinguish between the various waves present in the MHD flow and can, therefore, lead to a severe smearing of the approximation near discontinuities.
\end{description}

Here we outline the construction of the \texttt{ES-Roe} stabilization. The \texttt{ES-LLF} stabilization follows almost immediately. To build the matrix dissipation term we first select the dissipation matrix to be $|\widehat{\mat{A}}|$. That is the absolute value of the flux Jacobian for the ideal MHD 8-wave formulation:
\begin{equation}\label{fluxJacobianConservativeVars}
\widehat{\mat{A}} = \vec{F}_{\vec{Q}} + \mat{P} = \mat{A} + \mat{P},
\end{equation}
where $\mat{A}$ is the flux Jacobian for the homogeneous ideal MHD equations and $\mat{P}$ is the Powell source term \cite{Powell1999} written in matrix form, i.e.
\begin{equation}\label{PowellMatrix}
\mat{P}\pderivative{\vec{Q}}{x} = \begin{bmatrix}
0 & 0 & 0 & 0 & 0 & 0 & 0 & 0 \\
0 & 0 & 0 & 0 & 0 & B_1 & 0 & 0 \\
0 & 0 & 0 & 0 & 0 & B_2 & 0 & 0 \\
0 & 0 & 0 & 0 & 0 & B_3 & 0 & 0 \\
0 & 0 & 0 & 0 & 0 & \vec{u}\cdot\vec{B} & 0 & 0 \\
0 & 0 & 0 & 0 & 0 & u & 0 & 0 \\
0 & 0 & 0 & 0 & 0 & v & 0 & 0 \\
0 & 0 & 0 & 0 & 0 & w & 0 & 0
\end{bmatrix}
\pderivative{}{x}\begin{bmatrix}
\rho \\
\rho u\\
\rho v \\
\rho w\\
\rho e \\
B_1\\
B_2\\
B_3
\end{bmatrix} = \pderivative{B_1}{x}\begin{bmatrix}
0 \\
B_1 \\
B_2\\
B_3\\
\vec{u}\cdot\vec{B} \\
u\\
v\\
w
\end{bmatrix}.
\end{equation}
\referee{The design of the entropy stable matrix dissipation term requires the specific form of the flux Jacobian \eqref{fluxJacobianConservativeVars} because it must be possible to relate the eigenvectors of \eqref{fluxJacobianConservativeVars} to the entropy Jacobian matrix \cite{merriam1989}. This relationship is referred to as creating entropy scaled eigenvectors, \eg \cite{merriam1989,barth1999}. To ensure that this entropy scaling exists, the system of PDEs must be symmetrizable. It is known that the Powell source term restores the symmetric property to the ideal MHD system \cite{barth1999,Godunov1972}, whereas the Janhunen source term does not restore symmetry. However, both source terms allow to contract the MHD equation to the entropy evolution equation and hence both source terms can be used to construct entropy conserving (or stable) discretisations. We choose the Janhunen source term to construct the entropy conservative discretisation, as this gives us conservation of mass, momentum and energy unlike a method based on the Powell source term. Thus, the consistent symmetric part of the flux is based on the Janhunen source term. As long as the additional stabilisation term is guaranteed to dissipate entropy, the scheme is entropy stable. Hence, for the design of the stabilisation term only, we are considering the flux Jacobian that incorporates the Powell source term, as this guarantees that the entropy scaled eigenvectors exist for the ideal MHD system, which is necessary in order to get the Roe type dissipation term.}

The matrix type stabilization term requires the eigenstructure of the dissipation matrix \eqref{fluxJacobianConservativeVars}
\begin{equation}\label{eq:eigendecomp}
\widehat{\mat{A}} = \widehat{\mat{R}}\mat{D}\widehat{\mat{R}}^{-1}.
\end{equation}
The matrix $\widehat{\mat{A}}$ supports eight propagating plane-wave solutions:
\begin{itemize}
\item two fast magnetoacoustic waves ($\pm f$),
\item two slow magnetoacoustic waves ($\pm s$),
\item two Alfv\'{e}n waves ($\pm a$),
\item an entropy wave ($E$),
\item a divergence wave ($D$).
\end{itemize}
It is known that a naively scaled set of right eigenvectors will exhibit several forms of degeneracy that are carefully described by Roe and Balsara \cite{roe1996}. We follow the same rescaling procedure of Roe and Balsara to improve the numerical behaviour of the fast/slow magnetoacoustic eigenvectors. The matrix of right eigenvectors is
\begin{equation}
\widehat{\mat{R}} = \left[\,\widehat{\vec{r}}_{+\mathrm{f}} \,|\, \widehat{\vec{r}}_{+\mathrm{a}} \,|\, \widehat{\vec{r}}_{+\mathrm{s}} \,|\, \widehat{\vec{r}}_\mathrm{E} \,|\, \widehat{\vec{r}}_\mathrm{D} \,|\, \widehat{\vec{r}}_{-\mathrm{s}} \,|\, \widehat{\vec{r}}_{-\mathrm{a}} \,|\, \widehat{\vec{r}}_{-\mathrm{f}} \, \right],
\end{equation}
with the eigenvectors $\widehat{\vec{r}}$, and corresponding eigenvalues $\lambda$ \cite{barth1999,roe1996,Winters2016}
\begin{itemize}
\item[] \underline{Entropy and Divergence Waves}: $\lambda_{E,D} = u$
\begin{equation}\label{entropyAS}
\widehat{\vec{r}}_E = \begin{bmatrix} 1 \\ u \\v \\w \\ \frac{\|\vec{u}\|^2}{2} \\[0.05cm]0 \\0 \\0 \end{bmatrix},\quad\widehat{\vec{r}}_D = \begin{bmatrix} 0 \\ 0 \\0 \\0 \\B_1 \\[0.05cm]1 \\0 \\0 \end{bmatrix},
\end{equation}
\item[] \underline{Alfv\'{e}n Waves}: $\lambda_{\pm a} = u\pm b_1$
\begin{equation}\label{AlfvenAS}
\widehat{\vec{r}}_{\pm a} = \begin{bmatrix}
0 \\
0 \\
\pm \rho^{\frac{3}{2}}\,\beta_3 \\
\mp \rho^{\frac{3}{2}}\,\beta_2 \\
\mp \rho^{\frac{3}{2}}(\beta_2 w - \beta_3 v) \\
0 \\
-\rho \beta_3 \\
\rho \beta_2
\end{bmatrix},
\end{equation}
\item[] \underline{Magnetoacoustic Waves}: $\lambda_{\pm f,\pm s} = u\pm c_{f,s}$
\begin{equation}\label{MHDAS}
\widehat{\vec{r}}_{\pm f} = \begin{bmatrix}
\alpha_f\rho \\[0.1cm]
\alpha_f\rho(u \pm c_{f}) \\[0.1cm]
\rho\left(\alpha_f v \mp \alpha_s c_s \beta_2 \sigma(b_1) \right) \\[0.1cm]
\rho\left(\alpha_f w \mp \alpha_s c_s \beta_3 \sigma(b_1) \right) \\[0.1cm]
\Psi_{\pm f} \\[0.1cm]
0 \\[0.1cm]
\alpha_s a \beta_2 \sqrt{\rho} \\[0.1cm]
\alpha_s a \beta_3 \sqrt{\rho}
\end{bmatrix},
\qquad
\widehat{\vec{r}}_{\pm s} = \begin{bmatrix}
\alpha_s\rho \\[0.1cm]
\alpha_s\rho\left(u \pm c_s\right) \\[0.1cm]
\rho\left(\alpha_s v \pm \alpha_f c_f \beta_2 \sigma(b_1)\right) \\[0.1cm]
\rho\left(\alpha_s w \pm \alpha_f c_f \beta_3 \sigma(b_1)\right) \\[0.1cm]
\Psi_{\pm s} \\[0.1cm]
0 \\[0.1cm]
-\alpha_f a \beta_2 \sqrt{\rho} \\[0.1cm]
-\alpha_f a \beta_3 \sqrt{\rho}
\end{bmatrix},
\end{equation}
\end{itemize}
where we introduced several convenience variables
\begin{equation}\label{eq:alotofequations}
\begin{aligned}
\Psi_{\pm\mathrm{s}} &= \frac{\alpha_\mathrm{s} \rho \lVert\vec{u}\rVert^2}{2} - a \alpha_\mathrm{f} \rho b_\perp + \frac{\alpha_\mathrm{s} \rho a^2}{\gamma-1} \pm \alpha_\mathrm{s} c_\mathrm{s} \rho u \pm \alpha_\mathrm{f} c_\mathrm{f} \rho \sigma(b_1) (v \beta_2 + w \beta_3), \\
\Psi_{\pm\mathrm{f}} &= \frac{\alpha_\mathrm{f} \rho \lVert\vec{u}\rVert^2}{2} + a \alpha_\mathrm{s} \rho b_\perp + \frac{\alpha_\mathrm{f} \rho a^2}{\gamma-1} \pm \alpha_\mathrm{f} c_\mathrm{f} \rho u \mp \alpha_\mathrm{s} c_\mathrm{s} \rho \sigma(b_1) (v \beta_2 + w \beta_3),  \\
c_\mathrm{a}^2& = b_1^2, \quad c_\mathrm{f,s}^2 = \frac{1}{2}\left((a^2+b^2) \pm \sqrt{(a^2+b^2)^2 - 4a^2 b_1^2}\right), \quad a^2 = \gamma \, \frac{p}{\rho}, \\
b^2 &= b_1^2 + b_2^2 + b_3^2, \quad b_\perp^2 = b_2^2 + b_3^2, \quad \vec{b} = \frac{\vec{B}}{\sqrt{\rho}}, \quad \beta_{1,2,3} = \frac{b_{1,2,3}}{b_\perp},\\
\alpha_\mathrm{f}^2 &= \frac{a^2 - c_\mathrm{s}^2}{c_\mathrm{f}^2 - c_\mathrm{s}^2}, \quad \alpha_\mathrm{s}^2 = \frac{c_\mathrm{f}^2 -a^2}{c_\mathrm{f}^2 - c_\mathrm{s}^2},\quad
\sigma(\omega) = \begin{cases}
+1 &\mbox{if } \omega \ge 0, \\
-1 &\text{otherwise}
\end{cases}.
\end{aligned}
\end{equation}
In \eqref{eq:alotofequations}, for the wave speed computation $c_\mathrm{f,s}^2 $, the plus sign corresponds to the fast magnetoacoustic speed, $c^2_\mathrm{f}$, and the minus sign corresponds to the slow magnetoacoustic speed, $c^2_\mathrm{s}$.

The entropy stable dissipation term is built from three components:
\begin{itemize}
\item Entropy scaled matrix of right eigenvectors: $\mathring{\mat{R}} = \widehat{\mat{R}} \sqrt{\mat{T}}$, where $\mat{T}$ is the diagonal scaling matrix
\begin{equation}
	\mat{T} = \mathrm{diag}\left({\frac{1}{{2 \rho \gamma}}},\, {{\frac{p}{2 \rho^3}}},\, {\frac{1}{{2 \rho \gamma}}},\, {\frac{\rho (\gamma - 1)}{\gamma}},\, {\frac{p}{\rho}},\, {\frac{1}{{2 \rho \gamma}}},\, {\frac{p}{2 \rho^3}},\, {\frac{1}{{2 \rho \gamma}}}\right).
\end{equation}
For the complete motivation and details on the entropy scaling of eigenvectors see Barth \cite{barth1999}.
\item Diagonal matrix of eigenvalues: For \texttt{ES-Roe} each wave component is weighted with a different eigenvalue, whereas \texttt{ES-LLF} weights all wave components identically
\begin{subequations}
	\begin{gather}
	|\mat{D}_\mathtt{ES-Roe}| = \mathrm{diag}\big(|\lambda_{+f}|,|\lambda_{+a}|,|\lambda_{+ s}|,|\lambda_\mathrm{E}|,|\lambda_\mathrm{D}|,|\lambda_{-s}|,|\lambda_{-a}|,|\lambda_{-f}|\big),\\
	|\mat{D}_\mathtt{ES-LLF}| = \mathrm{diag}\big(\lambda_\mathrm{max},\lambda_\mathrm{max},\lambda_\mathrm{max},\lambda_\mathrm{max},\lambda_\mathrm{max},\lambda_\mathrm{max},\lambda_\mathrm{max},\lambda_\mathrm{max}\big).
	\end{gather}
\end{subequations}
	The maximum eigenvalue $\lambda_\mathrm{max}$ is given by
\begin{equation}
	\lambda_\mathrm{max} = \max\big(|\lambda_{+f}|,|\lambda_{+a}|,|\lambda_{+ s}|,|\lambda_\mathrm{E}|,|\lambda_\mathrm{D}|,|\lambda_{-s}|,|\lambda_{-a}|,|\lambda_{-f}|\big).
\end{equation}
\item Jump in the entropy vector: $\llbracket \vec{v} \rrbracket$\par
The term $\llbracket \vec{v} \rrbracket$ is the jump between left and right states of the entropy vector, which is defined as a vector field whose components are partial derivatives of the entropy density \eqref{eq:entropy} with respect to the fluid quantities $\vec{Q}$,
\begin{equation}\label{eq:EntVars}
	\vec{v} = \derivative{S}{\vec{Q}} = -\left[\frac{\gamma - s}{\gamma - 1}-\frac{\rho \lVert\vec{u}\rVert^2}{2 p},\, \frac{\rho u}{p},\, \frac{\rho v}{p},\, \frac{\rho w}{p},\, -\frac{\rho}{p},\, \frac{\rho B_1}{p} ,\, \frac{\rho B_2}{p} ,\, \frac{\rho B_3}{p} \right]^\intercal,
\end{equation}
with the physical entropy, $s$, defined in \eqref{eq:entropy}.
\end{itemize}
The general form of the \texttt{ES-Roe}, and \texttt{ES-LLF} numerical flux functions is
\begin{align}\label{eq:ESflux}
		\vec{F}_\mathtt{ES} &= \vec{F}^\mathrm{*,ec} + \frac{1}{2} \mathring{\mat{R}} |\mat{D}| \mathring{\mat{R}}^T \llbracket \vec{v} \rrbracket,
\end{align}
where $\vec{F}^\mathrm{*,ec}$ is the entropy conserving numerical flux \eqref{eq:ec_flux}. Note that the only difference between the \texttt{ES-Roe} and \texttt{ES-LLF} stabilizations is in the selection of the diagonal matrix of eigenvalues $\mat{D}$. The matrix of right eigenvectors and the eigenvalues are discretely computed from the previously defined average quantities \eqref{eq:Averages2} to ensure consistency, in the presence of vanishing magnetic fields, with the entropy stable Euler solver of Ismail and Roe \cite{ismail2009}.

\subsubsection{Hybrid Entropy Stabilization}\label{scn:flux3}

Chandrashekar \cite{Chandrashekar2012} points out that most, if not all, schemes which resolve grid aligned stationary contact discontinuities exactly suffer from the carbuncle effect as the profiles around shocks can exhibit spurious oscillations. This can also be true in our case as our flux function guarantees only the correct sign of the entropy but not necessarily the correct amount of entropy production. However, a flux function must generate \emph{enough} entropy across a shock to guarantee monotonicity \cite{IsmailCarbuncle2009}. The usual fix for this problem, \ie increasing the amount of induced dissipation, causes poor resolution of features of boundary layers or near shocks.
Another possibility is to switch the numerical scheme to a more dissipative one only near shocks and use a high resolution Riemann solver in smooth parts of the flow \cite{Quirk1994}. It is straightforward to implement such an idea in the current context because entropy stable schemes have the freedom to select the eigenvalues that essentially control the amount of dissipation.

The local Lax-Friedrichs type scalar dissipation, \texttt{ES-LLF}, effectively suppresses the carbuncle phenomenon. However, we want to use the more accurate Roe type matrix dissipation, \texttt{ES-Roe}, in regions without large pressure jumps to be able to track smooth parts of the solutions with more accuracy. To achieve this goal we construct a \emph{hybrid entropy stabilization} scheme, called \texttt{ES-Hybrid}, that blends the \texttt{ES-Roe} and the \texttt{ES-LLF} scheme continuously. In the hybrid scheme, a new diagonal matrix of eigenvalues is defined as
\begin{equation}
|\mat{D}_\mathtt{ES-Hybrid}(\Xi)| = (1 - \Xi) |\mat{D}_\mathtt{ES-Roe}| + \Xi |\mat{D}_\mathtt{ES-LLF}|,
\end{equation}
with the limits
\begin{equation}
	\begin{aligned}
	\lim\limits_{\Xi \rightarrow 0} |\mat{D}_\mathtt{ES-Hybrid}(\Xi)| &= |\mat{D}_\mathtt{ES-Roe}|,\\
	\lim\limits_{\Xi \rightarrow 1} |\mat{D}_\mathtt{ES-Hybrid}(\Xi)| &= |\mat{D}_\mathtt{ES-LLF}|.
	\end{aligned}
\end{equation}
As was done in \cite{Chandrashekar2012}, we define the parameter $\Xi \in [0,1]$ using a simple local pressure jump indicator
\begin{equation}\label{eq:Indicator}
	\Xi = \Bigg| \frac{p_\mathrm{L} - p_\mathrm{R}}{p_\mathrm{L} + p_\mathrm{R}} \Bigg|^{1/2}.
\end{equation}
From the design of the pressure indicator \eqref{eq:Indicator}, the scheme uses mainly the less dissipative \texttt{ES-Roe} scheme for smooth parts of the flow (but also near \eg contact discontinuities), while the more dissipative \texttt{ES-LLF} entropy-stabilization is used near strong shocks.

\subsection{Pressure Positivity Guaranteeing Formulation}\label{scn:PositivePressure}

We next address the issue that negative pressures may be introduced by a numerical scheme. This has been described in previous publications, \eg \cite{Ryu1993,Balsara1999_2, balsara2012}. \referee{We present here a general and physically motivated solution to the specific numerical issue of negative pressures.}

In a classical higher order Godunov method, the internal energy and thereby the thermal pressure, $p_{\rm th}$, is obtained by subtracting the kinetic and magnetic energies from the conserved total energy \eqref{eq:pressure}.
In many situations, as in high-Mach number or low plasma-beta flows ($\beta \propto p/\lVert\vec{B}\rVert^2$), the internal energy can be several orders of magnitude smaller then the kinetic or magnetic energies. Thus, discretisation errors in the total energy might be significant enough to result in negative pressures leading to a failure of the numerical scheme. This problem is often addressed by enforcing low pressure limits. However, it is questionable if the simulation can then still give a physically meaningful solution. Therefore, it is important to design a conservative pressure positivity \emph{guaranteeing} scheme that is physically convincing.

\subsubsection{Previous Investigations}

Ryu et al.~\cite{Ryu1993} state that in regions where the gas is very cold compared to the bulk kinetic energy, the flow cannot be treated using the total energy approach as the errors in calculating the total energy can be larger than the internal energy itself. In order to overcome this difficulty, they solve an entropy conservation equation and extract the pressure directly wherever the internal energy is much less than the kinetic energy, \ie$E_{\rm th}/E_{\rm kin} \ll 1$. They present several different criteria used to select whether to compute the pressure from the total energy or from their entropy formulation.

Balsara and Spicer \cite{Balsara1999_2} extend the idea of Ryu et al.~to MHD flows and present two strategies to prevent negative pressures. Their ``strategy 1'' is to use the pressure computed from the entropy density only in those cells where the thermal pressure could potentially become negative. In all other cells, they use the thermal pressure given by \eqref{eq:pressure}. Their ``strategy 2'' uses the pressure computed from the entropy everywhere except in regions near strong magnetosonic shocks or a flow configuration that may develop such shocks.
\referee{They justify the validity of their approach by noting that their work deals with magnetospheric problems. There are no shocks present in a magnetosphere, but there still remains a positivity problem.}

Li \cite{Li2008_2} extends the ideas of Balsara and Spicer to an implementation of two new equations: The entropy equation used in \cite{Balsara1999_2} and an internal energy equation. Similar to the previously mentioned works, he points out that these equations should not be used close to or within regions that contain shocks. His shock detection scheme sets a floor value for the internal energy in cells near a shock.

\referee{Balsara \citep{balsara2012} presents a general strategy to address problems where the positivity of density and pressure is uncertain. His work addresses the problem that positivity can be lost when using reconstruction schemes. He presents a self-adjusting strategy for enforcing the positivity of the pressure. However, we realize that the positivity problem of pressure can also be encountered with schemes that are first-order accurate in space and as such do not utilize a reconstruction scheme. This is due to the problem of subtracting large numbers with accordingly large discretisation errors as stated above.}

The work of Ersoy et al.~\cite{Ersoy2013} relies on improving the resolution in problematic regions. They utilize a discrete measure of the entropy in order to stabilize their computation. They use the local \emph{entropy production} as a mesh refinement criterion on their computational grid.

\subsubsection{Derivation of a New Pressure Formulation}

The current solver is built from an entropic perspective. Thus, at any time in the computation, we can compute the entropy density as well as the discrete entropy flux. With these tools we determine a value for the pressure that is guaranteed to be positive. From the computed entropy density for each cell within the computational domain, we use \eqref{eq:entropy} to derive a new expression for the pressure in the cells:
\begin{equation}\label{entropic_pressure}
p_{s} = \exp\left[\frac{\gamma-1}{\rho} S + \gamma \ln(\rho)\right].
\end{equation}
From \eqref{entropic_pressure}, it is immediately clear that this ``entropy pressure'' will always be positive as $\exp(x) > 0,\ x\in\mathbb{R}$. Hence, our solver fulfils the desired property of being pressure positivity guaranteeing under all circumstances.

We note that our scheme can be used in a similar way as described by Balsara and Spicer \cite{Balsara1999_2}. However, our scheme includes a proper treatment of the entropy at shocks. It is applicable in all regions of the flow and not only in sufficiently smooth regions.

The current scheme is: Use the ``normal'' scheme as long as the internal energy is large enough after the update with the criterion $E_{\rm int}/E_{\rm tot} > \texttt{smalleint}$ with the user-definable parameter $\texttt{smalleint}$ that defaults to $0.01$. If the internal energy is smaller than the criterion, we switch to the entropy pressure formulation without violating the conservation of total energy.

\subsubsection{Implementation of the Entropy Pressure Formulation}

It is straightforward for a given semi-discrete finite volume method to compute the entropy update of the method. This is because we know how to convert the equations into entropy space. We contract the semi-discrete equations (possibly including some reconstruction technique) \eqref{eq:high_order_update} with the entropy vector \eqref{eq:EntVars} to obtain
\begin{equation}
\vec{v}^T\left(\vec{Q}_t\right)_i = \vec{v}^T\left(\frac{1}{\Delta x_i}\left(\tilde{\vec{F}}_{i-1/2} - \tilde{\vec{F}}_{i+1/2}\right) + \frac{1}{2}\left(\vec{\Upsilon}_{i-1/2} + \vec{\Upsilon}_{i+1/2}\right)\right).
\end{equation}
From the chain rule and definition of the entropy vector \eqref{eq:EntVars} we know that
\begin{equation}
\vec{v}^T\left(\vec{Q}_t\right)_i = \left({S}_t\right)_i.
\end{equation}
So, we have an expression for the time evolution of the entropy density
\begin{equation}\label{eq:EntUpdate}
\left({S}_t\right)_i = \vec{v}^T\left(\frac{1}{\Delta x_i}\left(\tilde{\vec{F}}_{i-1/2} - \tilde{\vec{F}}_{i+1/2}\right) + \frac{1}{2}\left(\vec{\Upsilon}_{i-1/2} + \vec{\Upsilon}_{i+1/2}\right)\right),
\end{equation}
where, as was shown in \cite{Winters2016}, the entropy stable fluxes provide a discrete approximation to the spatial derivative of the entropy flux, \ie
\begin{equation}
\pderivative{}{x}(uS) \approx \vec{v}^T\left(\frac{1}{\Delta x_i}\left(\tilde{\vec{F}}_{i-1/2} - \tilde{\vec{F}}_{i+1/2}\right) + \frac{1}{2}\left(\vec{\Upsilon}_{i-1/2} + \vec{\Upsilon}_{i+1/2}\right)\right).
\end{equation}
Thus, we see that \eqref{eq:EntUpdate} is a consistent, discrete update for the entropy. This new discrete equation for the entropy density can be added to the MHD system \eqref{JanhunenSource} and evolved in time with the other fluid quantities. We reiterate that, by construction, the entropy stable approximation will guarantee that entropy is consistent with the second law of thermodynamics everywhere. Thus, the proposed positive pressure guaranteeing method is valid in any region of the flow.
The implemented procedure is:
\begin{enumerate}
	\item We update the entropy density ${S}^{n+1}$ with the same time integration scheme used to obtain $\vec{Q}^{n+1}$.
	\item If the updated energies violate the criterion $E^{n+1}_{\rm int}/E^{n+1}_{\rm tot} > \texttt{smalleint}$, we use \eqref{entropic_pressure} to get $p_s^{n+1}$.
	\item Finally, we recompute the updated internal energy from $p_s^{n+1}$ to make the scheme consistent.
\end{enumerate}

\subsection{AMR Functionality}\label{scn:AMR}

\texttt{FLASH} incorporates an adaptive mesh refinement (AMR) strategy using the \texttt{PARAMESH} library \cite{PARAMESH2006}, through which the grid is organized in a block structured, oct-tree adaptive grid.
The presented entropy stable solver is fully incorporated into \texttt{FLASH}'s AMR functionality to optimize computational costs. For completeness we briefly discuss the underlying AMR structure, and parallelization in \texttt{FLASH}. With AMR, the local spatial resolution can be dynamically controlled. This allows the maximization of the computational efficiency of the overall simulation as higher resolution is placed only where it is needed.

\begin{figure}[!htb]
	\begin{minipage}[t]{0.45\textwidth}
		\centering
		\includegraphics[height=6cm]{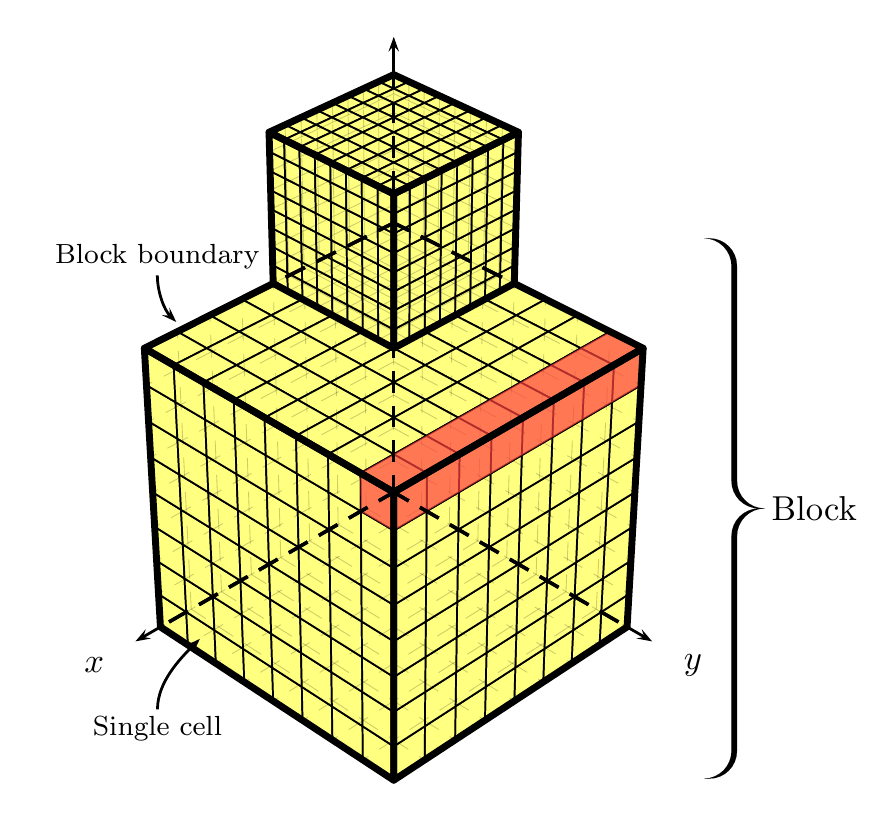}
		\caption{Two blocks with different levels of refinement (i.e.~mesh resolution). A single $x$ sweep is highlighted in red. The guard cells are not shown in this figure.}
		\label{fig:blocks3D}
	\end{minipage}\hfill
	\begin{minipage}[t]{0.45\textwidth}
		\centering
		\includegraphics[height=6cm]{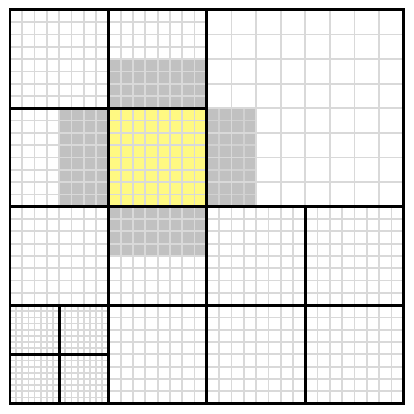}
		\caption{An adaptive grid 2D simulation with different levels of refinement. The interior cells of one of the blocks are highlighted in yellow. The according guard cells are shown in grey. Guard cells that extend into blocks having a different grid size are interpolated.}
		\label{fig:blocks2D}
	\end{minipage}
\end{figure}

Parallelization is achieved by dividing the computational domain into several \emph{blocks} (sub-domains). A block contains a number of computational cells (\texttt{NXB}, \texttt{NYB}, and \texttt{NZB} in the $x$, $y$, and $z-$direction, respectively). The default block contains \texttt{NX|Y|ZB} = 8. Each block is surrounded by a fixed number of \emph{guard cells} in each spatial direction, providing the block with information from its neighbouring blocks.
The complete computational domain consists of a number of blocks (most likely with different physical sizes). The three-dimensional structure of the blocks is sketched in Fig.~\ref{fig:blocks3D}, while a simple two-dimensional slice through an adaptive grid is shown in Fig.~\ref{fig:blocks2D}. Three rules apply in the creation of refined blocks:
\begin{enumerate}
	\item A refined block must be one-half of the size of the parent block in each spatial dimension (\eg each refinement of a block gives 8 additional blocks in three-dimensional computations).
	\item Refined blocks must fit within the parent block and are not allowed to overlap into other blocks (they have to be aligned).
	\item Blocks sharing a common border are not allowed to differ in more than one level of refinement.
\end{enumerate}
Each block contains all information about local and neighbouring cells, making the blocks with the surrounding guard cells self-contained computational domains which allows efficient parallel computation using the Message Passing Interface (MPI) framework. We configure AMR in such a way that adaptive refinement is allowed after each two consecutive time steps ($\mathtt{nrefs} = 2$).

\subsection{Quasi-Multifluid Implementation}\label{scn:Multifluid}

The ability to track the exact composition of a gas is of central importance in astrophysical simulations as they include detailed chemical networks to treat heating, cooling, as well as molecule formation and destruction to mimic the behaviour of the interstellar medium (ISM) \citep{Gatto2015,Walch2014,Glover2014}.

In order to track the different chemical species in the gas, advection equations of the form
	\begin{equation}
		\frac{\partial X_\ell \rho}{\partial t} + \nabla \cdot (X_\ell \rho \, \vec{u}) = 0,
	\end{equation}
are solved, where $X_\ell$ is the fractional abundance of the $\ell$\textsuperscript{th} species (\texttt{H}, \texttt{H\textsuperscript{+}}, \texttt{H}\textsubscript{2}, \texttt{He}, etc.) with the unity constraint $\sum_\ell X_\ell = 1$. For each species the flow of the quantity is calculated by multiplying the fractional abundances of the species in the cells with the total density fluxes. Our scheme was originally devised for a perfect gas with a constant ratio of specific heats, $\gamma$. We generalize our scheme for a multi-species fluid with variable $\gamma$ by adopting a mean value of $\gamma$ at the cell interfaces as suggested by Murawski \citep{Murawski2003}.

\referee{We implement the multi-species advection in a similar way as recommended by Plewa \& M\"uller \cite{Plewa1999} (known as Consistent Multi-fluid Advection (CMA) method). That is, we ensure that the species fluxes are consistent during the advection. Note that many existing schemes instead normalize the abundances after the advection step. However, as Glover et al.~\cite{Glover2010} pointed out, this procedure lacks any formal justification and can lead to large systematic errors in the abundance of the least abundant chemical species.}

In addition to the multifluid approach using different chemical species, we implement mass tracer fields (also called mass scalars or tracerfields). These are field variables which are advected similar to species mass fractions by an equation of the form
	\begin{equation}
		\frac{\partial \psi \rho}{\partial t} + \nabla \cdot (\psi \rho \, \vec{u}) = 0 \, ,
	\end{equation}
where $\psi$ is the mass fraction, and $\psi \rho$ is the partial density of the traced mass.

Our implementation of the mass tracer fields into the MHD solver allows the use of any number of such fields. Thus, the mass tracer fields are a flexible tool for tracing different mass quantities according to individual requirements. For example, a mass tracer field could be used to follow the distribution of metals in the interstellar gas with virtually no additional computational costs.

\subsection{Coupling to Gravity}\label{scn:Gravity}

The inclusion of gravity in the ideal MHD equations \eqref{3DIDEALMHD} introduces a force into the right-hand side of the momentum equations
\begin{equation}\label{eq:momentumeq_with_gravity}
\pderivative{}{t}\rho\vec{u} + \nabla\cdot\left[\rho(\vec{u}\otimes\vec{u}) + \left(p+\frac{1}{2}\|\vec{B}\|^2\right)\mat{I}-\vec{B}\otimes\vec{B}\right] = -\rho \nabla \phi,
\end{equation}
where the gravitational potential $\phi$ satisfies Poisson's equation
\begin{equation}\label{eq:Grav}
	\nabla^2 \phi = 2 \pi G \rho,
\end{equation}
with the universal gravitational constant $G$ that is an empirical physical constant involved in the calculations of gravitational forces between two bodies.

\texttt{FLASH} provides several algorithms for solving the Poisson equation \eqref{eq:Grav}. We tested our implementation with a Barns \& Hut tree-based algorithm implemented by R.~W\"unsch (\texttt{Poisson/BHTree}) \cite{BarnsHut} and the Fourier transform-based multigrid algorithm Poisson solver (\texttt{Poisson/Multigrid}) \cite{Ricker2008}.

\subsection{\referee{Magnetic Field Divergence Treatment}}\label{scn:divB}
Within the MHD equations \eqref{3DIDEALMHD}, the divergence free condition of the magnetic field \eqref{eq:divB} is not modelled directly. While this constraint is physically fulfilled at any time, we will see that care must be taken to fulfill this constraint numerically.

The extension to higher spatial dimensions, as described in Sec.~\ref{scn:multidim}, has been performed in a straightforward manner by relying on the Cartesian grid structure.
In one dimension the divergence-free condition implies that the longitudinal component of the magnetic field is constant over time. However, this conclusion does not generalize to two and three spatial dimensions.

Instead, due to discretisation errors, a non-zero divergence of the magnetic field occurs over time which inevitably leads to the issue that the conservation of the magnetic flux cannot be maintained. These discretisation errors effectively generate numerical \emph{magnetic monopoles} that grow exponentially during the computation and cause the magnetic field to no longer be solenoidal. From the equations of ideal MHD \eqref{3DIDEALMHD} it is clear that these monopoles cause an artificial force parallel to $\vec{B}$.

In Sec.~\ref{scn:idealMHD}, we noted that errors in the divergence-free condition are dealt with by treating the divergence of the magnetic field as an additional fluid quantity to prevent accumulation of errors when the divergence is non-zero in the computational domain. The eigenmode which is advected with the flow in \eqref{eq:magneticconti} is referred to as the \emph{divergence wave}.
This procedure might be understood as a form of \emph{divergence cleaning} for the magnetic field. However, numerical experiments show that this approach might not be sufficient to maintain adequate divergence-free magnetic fields throughout simulations.

Concerning divergence cleaning, there are different techniques available (see \eg \cite{Altmann12}). One particular example is the elliptic projection, based on the Helmholtz decomposition, originally developed by Chorin \cite{Chorin1967}. Brackbill and Barnes \cite{Brackbill1980} and Marder \cite{Marder1987} developed a \emph{projection method} in the context of the MHD equations. This method effectively suppresses the growth of unphysical magnetic monopoles locally as shown by Murawski \cite{Murawski2003} and T\'oth \cite{Toth2000}.
The projection method has successfully been applied by \eg Zachary et al.~\cite{Zachary1994}, Balsara \cite{Balsara1998}, and more recently by Crockett et al.~\cite{Crockett2005}.
We implement the projection method for divergence cleaning as a separate post-processing step and note that our original scheme remains unchanged.

The general downside of this scheme may be the high computational costs caused by the projection approach. Our implementation is based of the realization that although the divergence problem is of elliptical character, its influence is only local.
Accordingly, we design our implementation of the projection method in a way that is purely local and thereby computationally favourable.

To enforce the divergence-free constraint we subtract the portion of the magnetic field that violates $\nabla\cdot\vec{B} = 0$.
Suppose that the divergence of the magnetic field in the computation is non-zero. An easy fix to this problem is the addition of a correction field $\tilde{\vec{B}}$ such that
\begin{equation}
	\nabla \cdot \big(\vec{B} + \tilde{\vec{B}}\big) = 0. \label{eq:divSum}
\end{equation}
To guarantee physical consistency of the magnetic field correction, it is clear that $\tilde{\vec{B}}$ must not cause any additional current
\begin{equation}
	\vec{j} \propto \nabla \times \tilde{\vec{B}} = 0.
\end{equation}
Hence, we conclude that $\tilde{\vec{B}}$ must have the form
\begin{equation}
	\tilde{\vec{B}} = \nabla \phi, \label{eq:nablaPhi}
\end{equation}
where $\phi$ is a scalar potential. Combining \eqref{eq:divSum} and \eqref{eq:nablaPhi} we obtain
\begin{equation}
	\Delta \phi = \nabla^2 \phi = -\nabla \cdot \vec{B}.
\end{equation}
Note that the Laplace operator, $\Delta$, has the physical interpretation for non-equilibrium diffusion as the extent to which a point represents a source or sink of some concentration.
The resulting scalar potential can then be used to evaluate $\tilde{\vec{B}}$ according to \eqref{eq:nablaPhi} and, thus, clean the magnetic field $\vec{B}$:
\begin{equation}
	\vec{B} \Big|_{\nabla \cdot \vec{B} = 0} = \vec{B} + \tilde{\vec{B}}.
\end{equation}

By the projection of the cell-centred magnetic fields onto the space of divergence-free magnetic fields, one is left with fields at the next time step which are divergence-free to very good approximation. We note that projecting the magnetic field in the way described is consistent with the underlying cell-centred scheme.

An alternative approach to divergence cleaning that should be mentioned is the constraint transport method developed by Evans and Hawley \cite{Evans1988} or Balsara and Spicer \cite{Balsara1999} (reviewed in \cite{Toth2000}). In this approach, the divergence-free constraint is satisfied by placing the staggered magnetic field at cell faces instead of cell centres. On such a grid, the MHD equations can be approximated such that they preserve numerical solenoidality of the magnetic field by construction. Note that Balsara and Kim \cite{Balsara2004b} found advantages for the staggered-mesh in their comparison between divergence-\emph{cleaning} and divergence-\emph{free} methods for stringent test cases. However, the staggered grid approach has the downside of being much more expensive in terms of storage. In addition, it is not clear if provably stable schemes can be constructed for staggered-meshes \cite{Waagan2009}.

The precise implementation of our divergence cleaning approach is described in further detail in \ref{app:divB}.
\subsection{MHD Update Procedure}\label{scn:Procedure}

On logically Cartesian grid geometries, it is straightforward to solve multi-dimensional problems as sets of one-dimensional problems by using the \emph{dimensional split} approach. This approach is the principle of the new \texttt{ES} solver. The MHD equations are solved as one-dimensional problems along each coordinate direction in turn ($x$, $y$, and $z-$sweeps) in order to determine the fluxes through the finite volume cell surfaces.
\begin{figure}[!ht]
	\centering
	\includegraphics[scale=1]{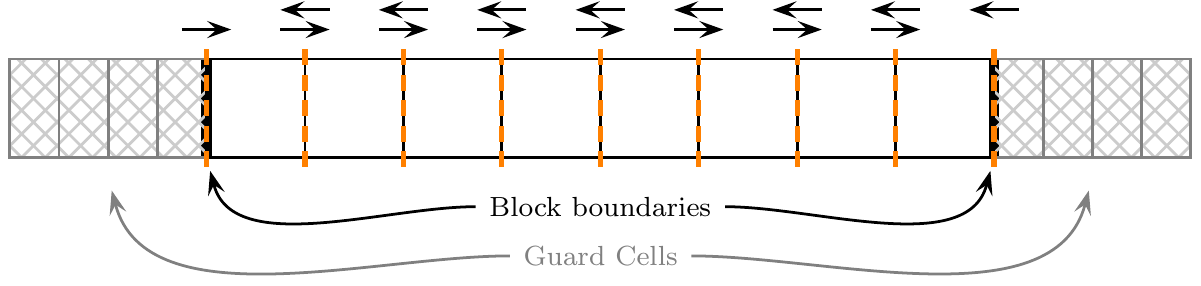}
	\caption{Principle of the one-dimensional solution update with four guard cells.}
	\label{fig:blockupdate}
\end{figure}

Each one-dimensional \emph{sweep} (like the one highlighted in Fig.~\ref{fig:blockupdate}) works as follows:
\begin{enumerate}
	\item First, the quantities are converted from primitive to conservative form (\eg velocity to momentum).
	\item For $y$ and $z-$sweeps the solution array is rotated such that we solve this sweep as if it would be an $x-$sweep. This allow us to use our one-dimensional algorithms without modification.
	\item For each cell within the array, the reconstructed quantities $(\tilde{\vec{Q}}_{i-1})_{\rm R}$ and $(\tilde{\vec{Q}}_{i})_{\rm L}$ are computed using the spatial reconstruction scheme (see Sec.~\ref{scn:MUSCL}) at the current time.
	\item Then, the entropy stable numerical fluxes as well as the source terms are computed using the algorithms described in Secs.~\ref{scn:flux1} and \ref{scn:flux2}.
\end{enumerate}
The default behaviour is to use the \texttt{ES-Hybrid} flux due to its flexibility. However, the user can easily change which flux function the computation uses with a single switch. Depending on the settings, either the entropy conserving fluxes, $\vec F^{*,\mathtt{ec}}$, the matrix dissipation entropy stable fluxes, $\vec F_\mathtt{ES-Roe}$, the scalar dissipation entropy stable fluxes, $\vec F_\mathtt{ES-LLF}$, or the hybrid entropy stable fluxes, $\vec F_\mathtt{ES-Hybrid}$ are used.
\begin{enumerate}\setcounter{enumi}{4}
	\item After this preparation, the solution array is updated using the time integration scheme described above.

	\item The updated internal energies are then derived from the updated total energy.
\end{enumerate}
We update the total energy as it is a conserved quantity. From the updated total energy, we derive the internal energy by subtracting the magnetic and kinetic energies as suggested in \cite{Colella1984}:
\begin{align}
E_\mathrm{int}^{n+1} = E_\mathrm{tot}^{n+1} - \Big( E_\mathrm{mag}^{n+1} + E_\mathrm{kin}^{n+1} \Big) \quad\mbox{with}\quad E_\mathrm{mag}^{n} = \frac{1}{2}\, \big\lVert \vec B^{n} \big\rVert^2, \quad \mbox{and}\quad E_\mathrm{kin}^{n} = \frac{1}{2}\, \rho^{n} \big\lVert \vec u^{n} \big\rVert^2
\end{align}
If the computed internal energy fails the criterion $E_{\rm int}/E_{\rm tot} > \mathtt{smalleint}$, then the total energy update is done with the pressure computed from the entropy density described in Sec. \ref{scn:PositivePressure}.

\begin{enumerate}\setcounter{enumi}{6}
	\item Finally, the variables are converted to primitive form as other \texttt{FLASH} modules expect primitive variables.
	\item In higher dimensions, the divergence cleaning procedure, described in Sec.~\ref{scn:divB}, is used to diffuse away errors in the divergence-free condition as a post-processing step.
\end{enumerate}

\section{Numerical Results}\label{scn:NumResults}

We demonstrate the utility, robustness\referee{, and accuracy} of the new solver by computing the solution to several well-known HD and MHD test problems.
The version of \texttt{FLASH} on hand is 4.3 as of 18\textsuperscript{th} July, 2015.
We consider six numerical test cases to test the performance of our new solver and compare to results obtained using already available MHD solver implementations for \texttt{FLASH}.
\referee{A test that extends the well-known Shu-Osher test to MHD is presented in Sec.~\ref{scn:ShuOsherMHD} which is used to test the \texttt{ES} scheme's artificial dissipation in 1D.}
The propagation of smooth Alfv\'{e}n waves is studied in Sec.~\ref{scn:Alfven}.
We forgo the presentation of further one-dimensional results as we felt multi-dimensional results were more valuable to the present discussion. The application of the entropy stable MHD solver to the shock tube problems of Brio and Wu \cite{Brio1988}, Ryu and Jones \cite{ryu1994}, and Torrilhon \cite{torrilhon2003} can be found in Winters and Gassner \cite{Winters2016}.
\referee{In Sec.~\ref{scn:OrszagTang} we further explore the accuracy of the method in multiple spatial dimensions by considering the Orszag-Tang vortex problem.}
The MHD rotor problem originally proposed by Balsara \citep{Balsara1999} is investigated in Sec.~\ref{scn:Rotor}. The MHD Rotor is also used in Sec. \ref{scn:Efficiency} to compare CPU timing and memory consumption of the new \texttt{ES} solver and the other schemes. Sec.~\ref{scn:Jeans} provides an example of using gravity with the new solver by considering the Jeans instability. We note that the Jeans instability is a pure HD configuration and demonstrates that the new MHD solver remains applicable to flows with vanishing magnetic fields. Finally, we test the conservation of the available MHD schemes using the involving MHD blast wave test discussed in Sec.~\ref{scn:MHDBlast}. All tests, except the Jeans instability test, are performed using dimensionless units. \referee{Each test is run with $\mathtt{CFL} = 0.8$ unless specified otherwise.}

\subsection{MHD version of Shu-Osher test (1D)}\label{scn:ShuOsherMHD}
The test proposed by Shu and Osher \cite{Shu1989} is commonly used to test a scheme's ability to resolve small-scale fluid features in the presence of a supersonic shock. A sinusoidal density/entropy perturbation is added downstream of a Mach 3 shock wave. The interaction of the shock wave with the perturbations gives rise to complex fluid features as the shock amplifies the initial oscillations. This test is an excellent testbed to measure the numerical (artificial) viscosity of a scheme. Additionally, the presence of a supersonic shock is used to demonstrate the robustness and stability of a scheme \cite{FLASHug}.
We consider a complex MHD version of the Shu-Osher problem recently developed by Susanto \cite{Susanto14}. We present the initial conditions for this test in Table~\ref{tab:ShuOsherMHD}. The left and right boundaries are taken sufficiently far from the initial discontinuity such that they do not influence the solution. This test has no analytic solution, so we compute a reference solution on a highly refined grid using the \texttt{MHD\_8Wave} solver for comparison.

\begin{table}[h]
	\centering
	\begin{minipage}[t]{0.35\textwidth}
		\begin{tabular}[t]{l|cc}
					&	{$x \le x_0$}		& {$x > x_0$}\\
			\midrule
			$\rho$	& $3.5$		& $1 + 0.2\sin(5x)$ \\
			$u$		& $5.8846$	& $0$ \\
			$v$		& $1.1198$	& $0$ \\
			$w$		& $0$		& $0$ \\
			$p$		& $42.0267$	& $1$ \\
			$B_1$	& $1$		& $1$ \\
			$B_2$	& $3.6359$	& $1$ \\
			$B_3$	& $0$		& $0$	\\
		\end{tabular}\\[-1em]
	\end{minipage}
	\begin{minipage}[t]{0.49\textwidth}
		\setlength\extrarowheight{3pt}
		\begin{tabular}[t]{|l|l|}\hline
			Domain size &$\{x_\mathrm{min},x_\mathrm{max}\} = \{-5,5\}$ \\\hline
			Initial shock location & $x_0 = -4$ \\\hline
			Boundary conditions & zero-gradient (``outflow'') \\\hline
			Simulation end time & $t_\mathrm{max} = 0.7$ \\\hline
			Adiabatic index & $\gamma = 5/3$ \\\hline
		\end{tabular}
	\end{minipage}
	\caption{Initial conditions and runtime parameters: MHD Shu-Osher test (1D)}
	\label{tab:ShuOsherMHD}
\end{table}

\begin{figure}[h]
	\centering
	\includegraphics[scale=1]{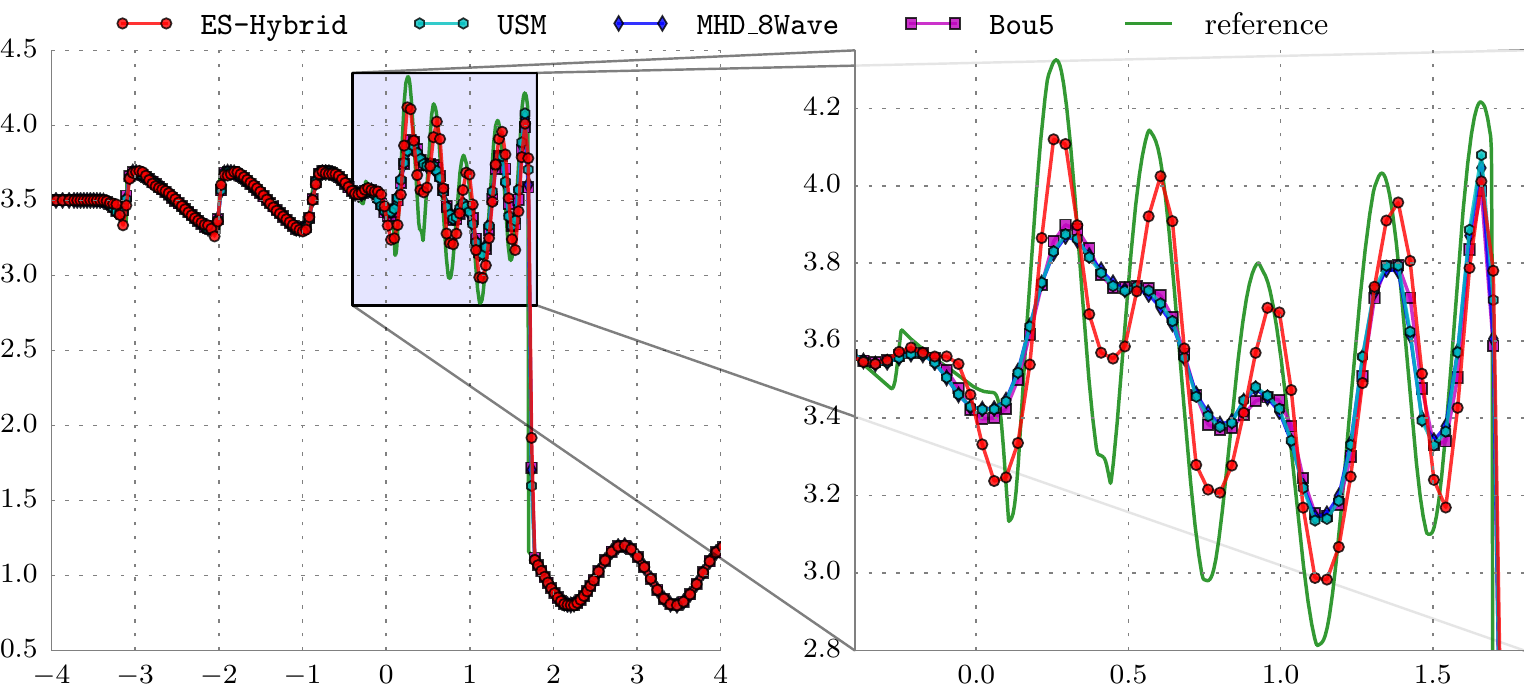}%
	\caption{Density of the MHD Shu-Osher problem at $t=0.7$. These plots be compared to Fig.~1 of \cite{Balasubramanian14} or Fig.~3.9 of \cite{Susanto14}. We used an adaptive grid resolution of up to $256$ cells. The reference solution is computed on a uniform grid of $\num{4096}$ cells.}
	\label{fig:ShuOsher}
\end{figure}

Fig.~\ref{fig:ShuOsher} shows the density at $t=0.7$ for all solvers. Using the same number of cells, we see that the \texttt{ES-Hybrid} solver captures the small-scale flow features much better than the other schemes available in \texttt{FLASH}. Also, no stability or overshoot problems are visible in the solutions.

According to previous investigations, \eg \cite{Chakravarthy2015}, a scheme is considered ``acceptable'' for capturing supersonic turbulence if the dynamics can be captured well with 400 cells. However, the entropy stable scheme is also able to resolve the dynamics of the flow with a much lower spatial resolution (the result used in Fig.~\ref{fig:ShuOsher} is 208 cells in total).
\jump

\subsection{Smooth Alfv\'en Wave (1D, 2D)}\label{scn:Alfven}

The smooth Alfv\'en wave test \citep{Toth2000} is used to compare the accuracy of MHD schemes for smooth flows. \referee{The initial circularly polarized Alfv\'en wave propagates across a periodic domain. For the 2D test, we incline the smooth Alfv\'en wave by an angle of $\alpha = \SI{45}{\degree}$ relative to the $x$-axis.}
The Alfv\'en wave speed is $|v_A|=B_\parallel/\sqrt{\rho} = 1$ and thus, the wave is expected to return to its initial state at each time $t \in \mathbb{N}$. This test is run to a final time $t_{\rm max} = 5.0$ with a CFL number of $0.6$. We introduce additional notation for a perpendicular coordinate $x_\parallel =  x\cdot\cos(\alpha) + y\cdot \sin(\alpha)$ as well as the parallel, $B_\parallel = 1.0 $, and perpendicular, $B_\perp = 0.1 \sin(2\pi x_\parallel)$, magnetic fields. The field in $z$-direction is given by $B_z = 0.1 \cos(2\pi x_\parallel)$. The initial conditions listed in Table~\ref{tab:smoothAlfven} ensure that the magnetic pressure is constant.

\referee{The ability to propagate Alfv\'en waves over long times and distances is crucial for \eg MHD turbulence simulations. If the Alfv\'en waves are damped strongly because of inherent numerical dissipation in a scheme, the code will fail to capture the resulting turbulence behaviour correctly as MHD turbulence is mainly sustained by Alfv\'en waves \citep{Balsara2014}.}

\begin{table}[h]
	\setlength\extrarowheight{3pt}
	\centering
	\begin{minipage}[t]{0.43\textwidth}
		\begin{tabular}[t]{|l|l|}\hline
			Density	$\rho$	&	1.0	\\ \hline
			Pressure $p$		&	0.1\\\hline
			Velocity $\vec{u}$	&	{\parbox{42mm}{$B_\perp \cdot \big(-\sin(\alpha),\cos(\alpha),0 \big)^\intercal$ $+B_z \cdot \big(0,0,1 \big)^\intercal$}}\\\hline
			Mag.~field $\vec{B}$& $B_\parallel \cdot (\cos(\alpha),\sin(\alpha),0)^\intercal + \vec{u}$\\ \hline
		\end{tabular}
	\end{minipage}
	\begin{minipage}[t]{0.56\textwidth}
		\begin{tabular}[t]{|l|l|}\hline
			Domain size &$\{x,y\}_\mathrm{min} = \{0.0,0.0\}$ \\
						&$\{x,y\}_\mathrm{max} = \{1/\cos(\alpha), 1/\sin(\alpha)\}$ \\ \hline
			Boundary conditions & periodic\\\hline
			Simulation end time & $t_\mathrm{max} = 5.0$ \\\hline
			Adiabatic index & $\gamma = 5/3$ \\\hline
		\end{tabular}
	\end{minipage}
		\caption{Initial conditions and runtime parameters: Smooth Alfv\'en wave test (1D, 2D)}
		\label{tab:smoothAlfven}
\end{table}

\newcommand{\vcenteredincludegraphics}[1]{\begingroup
\setbox0=\hbox{\includegraphics{#1}}%
\parbox{\wd0}{\box0}\endgroup}

\subsubsection{One dimensional test}
\referee{In the one dimensional smooth Alfv\'en wave test we check the spatial high resolution properties of our scheme. For sufficiently smooth fields, \ie in cases where discontinuous features are absent, the used reconstruction technique is designed to achieve third order accuracy (see Sec.~\ref{scn:MUSCL}).}

\begin{figure}[h]
	\centering
	\includegraphics[scale=1]{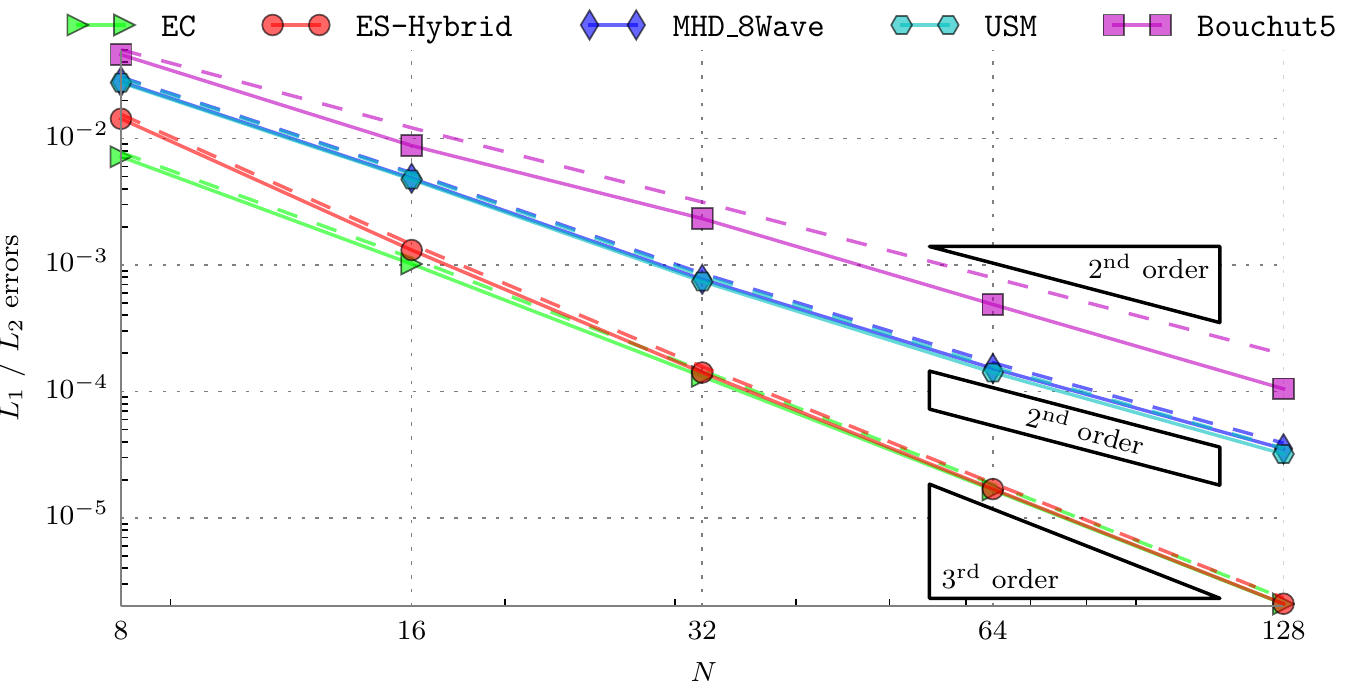}%
	\caption{$L_1$ (solid lines) and $L_2$ (dashed lines) errors measured with the smooth Alfv\'en wave test in 1D. We omit the lines for the \texttt{ES-Roe} and \texttt{ES-LLF} schemes as they are visually indistinguishable from \texttt{ES-Hybrid} (cf.~Table \ref{tab:Alfven1D}).}
	\label{fig:Alfven1DErrors}
\end{figure}

\referee{To test the accuracy of our scheme, we run several simulations with varying resolutions and compute the $L_1$ and $L_2$ errors for the quantity $B_\perp = B_y \cos(\alpha) - B_x \sin(\alpha)$ as described in \ref{scn:errors_and_eoc}. The obtained errors are plotted as a function of the number of grid points in logarithmic scale in Fig.~\ref{fig:Alfven1DErrors} and are listed in Table~\ref{tab:Alfven1D}. As can be seen, third order accuracy is achieved, already at very low resolutions.}

\begin{figure}[h]
	\centering
	\includegraphics[scale=1]{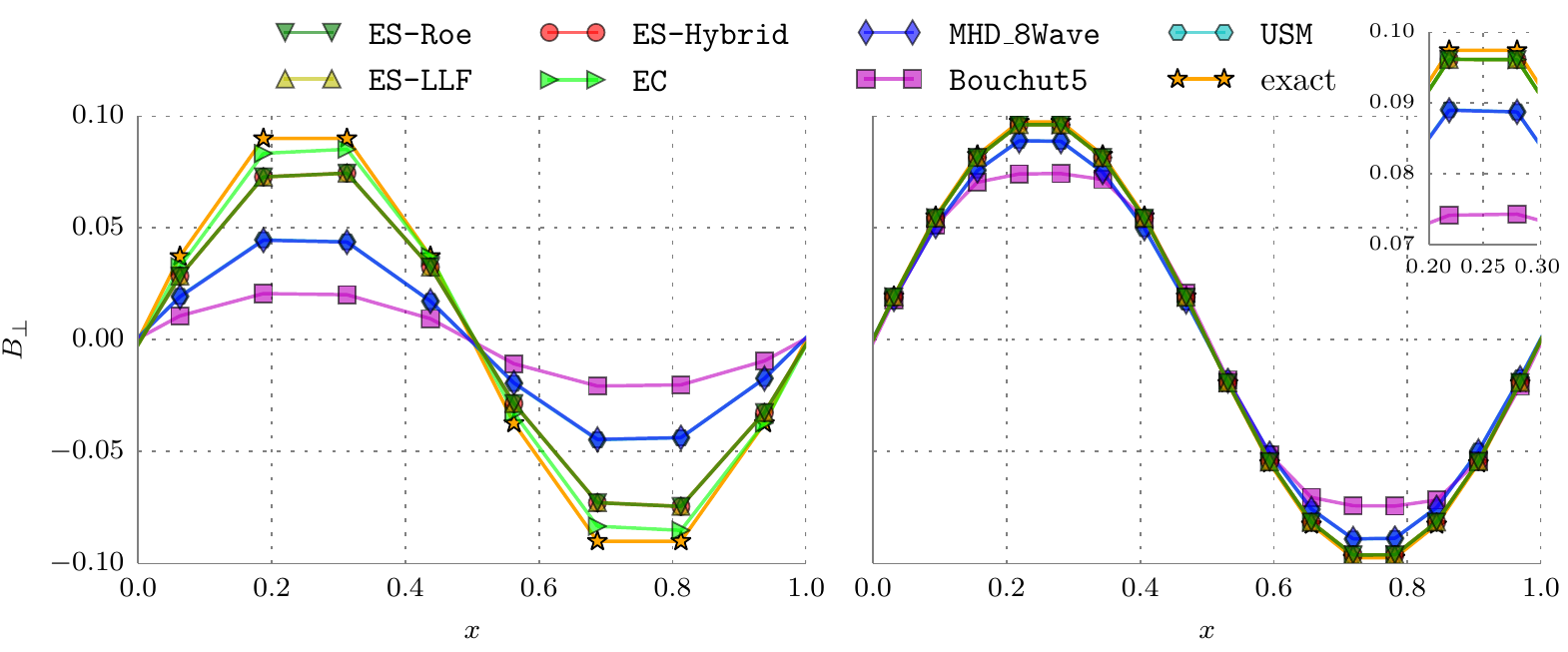}%
	\caption{Smooth Alfv\'en wave test after five crossing times. For the left plot, we used a fixed grid of $8$ cells. For the right plot we use a grid of $16$ cells. The \emph{exact} solution shows the initial configuration at the given resolution.}
	\label{fig:Alfven1D}
\end{figure}

Fig.~\ref{fig:Alfven1D} shows $B_\perp$ vs.~$x_\perp$ at time $t=5$ for the one dimensional Alfv\'en wave test.
As we know that the solution is smooth, we disable the entropy stabilisation term described in Sec.~\ref{scn:Entropy} and obtain an entropy conserving \texttt{EC} scheme (\includegraphics{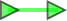}). The \texttt{EC} solution shows very little dissipation. Note that the \texttt{EC} scheme is only applicable to smooth solutions and should not be used for arbitrary flows.
We observe that the different \texttt{ES} schemes (\includegraphics{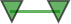}, \includegraphics{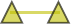}, and \includegraphics{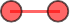}) resolve the Alfv\'en wave with the least dissipation of all tested MHD solvers (except the entropy conserving scheme) while their results are virtually identical.
The \texttt{MHD\_8Wave} implementation (\vcenteredincludegraphics{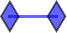}) \cite{Powell1999,FLASHug} as well as the unsplit \texttt{USM} implementation (\includegraphics{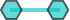}) \cite{Lee2013,Lee2009,FLASHug} are considerably more diffusive. They show second order convergence.
Finally, we note that the \texttt{Bouchut5} implementation (\includegraphics{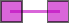}) \cite{Waagan2011} has the highest measured amount of dissipation for this one dimensional test case.

\subsubsection{Two dimensional test}
\begin{figure}[h]
	\centering
	\includegraphics[scale=1]{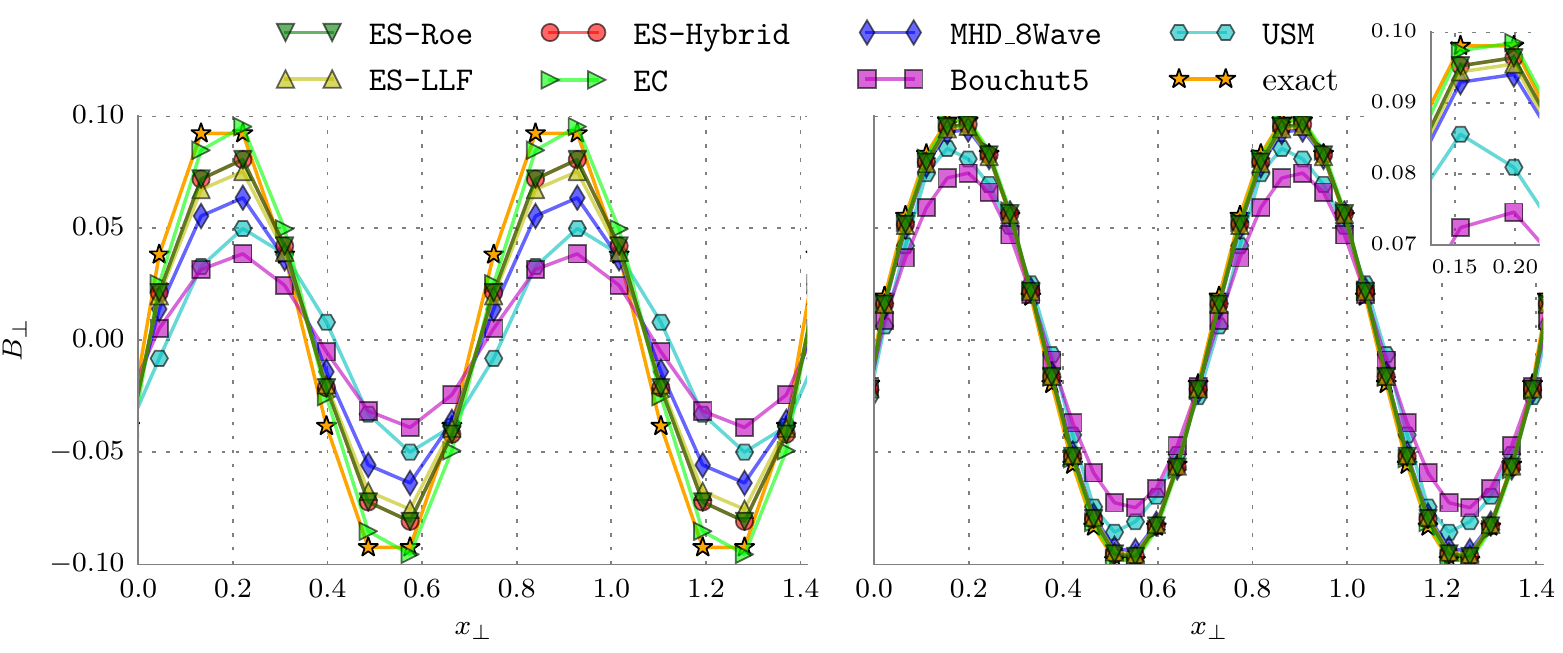}%
	\caption{Smooth Alfv\'en wave test after five crossing times. These plots be compared to Fig.~8 of \cite{Toth2000} or Fig.~4 of \cite{Chandrashekar2015}. For the left plot, we used a fixed grid of $16\times16$ cells. For the right plot we use a grid of $32\times32$ cells. The \emph{exact} solution shows the initial configuration at the given resolution. We use $\mathtt{CFL} = 0.6$ to remove artificial wave steeping effects in the \texttt{USM} solver solution.}
	\label{fig:Alfven}
\end{figure}

Fig.~\ref{fig:Alfven} shows $B_\perp$ vs.~$x_\perp$ at time $t=5$ for the two dimensional Alfv\'en wave test.
The \texttt{EC} solution (\includegraphics{python/alfven_legend_5.pdf}) shows again very little dissipation.
As before, the \texttt{ES} schemes resolve the Alfv\'en wave with the least dissipation of all tested MHD solvers while the \texttt{ES-Roe} scheme (\includegraphics{python/alfven_legend_7.pdf}) is least dissipative and the \texttt{ES-LLF} scheme (\includegraphics{python/alfven_legend_6.pdf}) is slightly more diffusive.
As expected for smooth problems, the \texttt{ES-Hybrid} scheme (\includegraphics{python/alfven_legend_4.pdf}) gives results that are identical to those computed using the \texttt{ES-Roe} scheme.
The \texttt{MHD\_8Wave} implementation (\vcenteredincludegraphics{python/alfven_legend_3.pdf}) gives similar results compared to the \texttt{ES} solver but is slightly more diffusive.
The unsplit \texttt{USM} implementation (\includegraphics{python/alfven_legend_1.pdf}) shows a higher dissipation compared to the \texttt{ES} or \texttt{MHD\_8Wave} implementations and its zero-crossing points are clearly shifted at the lower resolution run.
We find that the \texttt{Bouchut5} implementation (\includegraphics{python/alfven_legend_2.pdf}) has the highest amount of dissipation for this smooth test case.
\referee{Note that the dissipation of the Alfv\'en waves is significantly reduced in higher dimensions if multidimensional Riemann solvers with sub-structure are used, as was shown by Balsara \cite{Balsara2014}.}

\begin{figure}[h]
	\centering
	\includegraphics[scale=1]{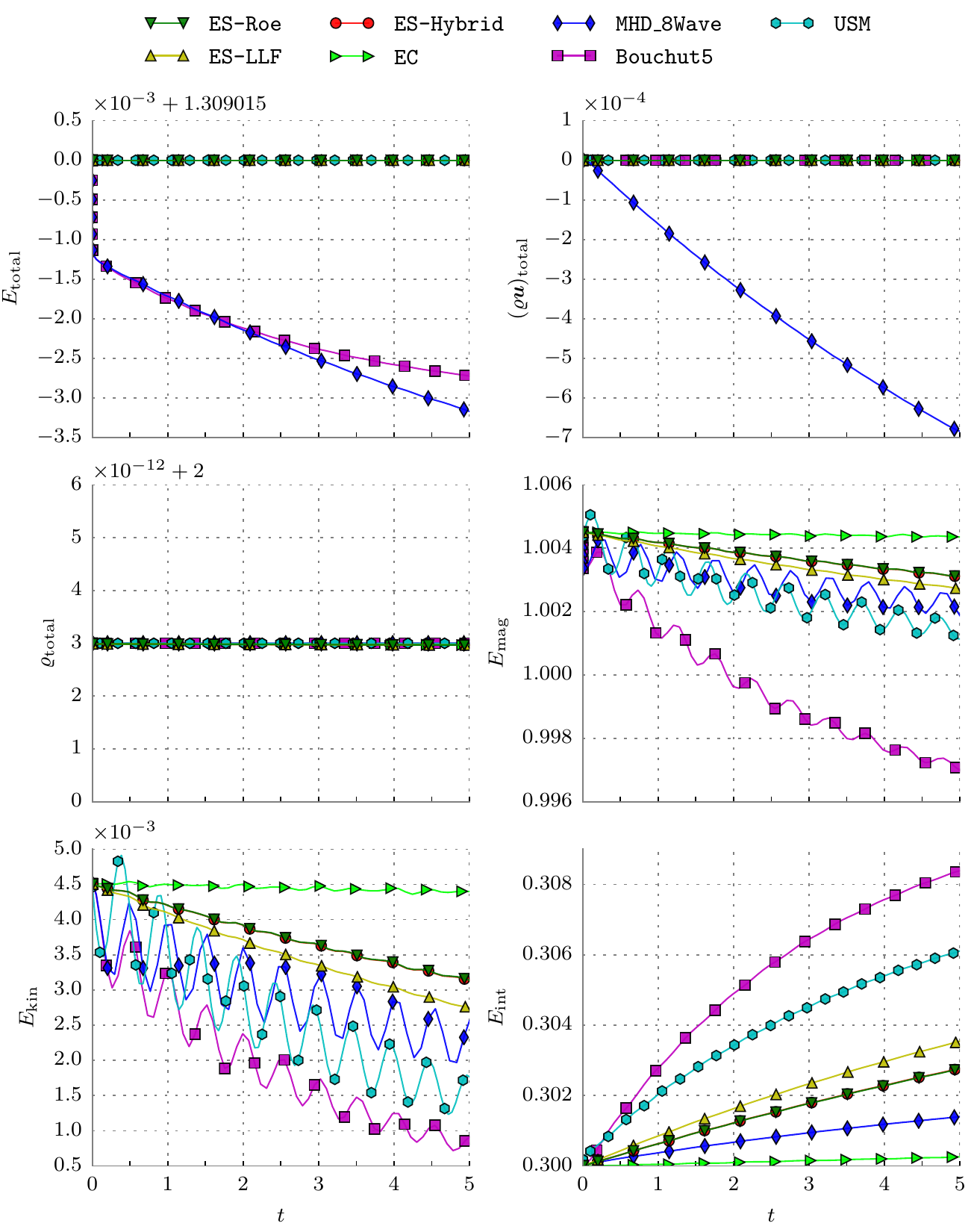}%
	\caption{Evolution of the conserved quantities as well as the individual energies in the smooth Alfv\'en wave test over five crossing times ($16\times16$ cells).}
	\label{fig:Alfven_Energy}
\end{figure}

In Fig.~\ref{fig:Alfven_Energy} we plot the evolution of the conserved quantities as well as the individual energies. Looking, for example, at the magnetic energy, $E_{\rm mag}$, it can be seen that the \texttt{EC} scheme introduces the least amount of dissipation. It is followed by our entropy stable schemes \texttt{ES-Roe}/\texttt{ES-Hybrid} and \texttt{ES-LLF}. The \texttt{MHD\_8Wave} and the \texttt{USM} implementations show higher dissipation while the \texttt{Bouchut5} implementation shows the highest amount of dissipation. If one would only look at the internal energy, one might conclude that the \texttt{MHD\_8Wave} solver introduces even less dissipation that the \texttt{ES} schemes. However, one has to be cautious with such a conclusion because both \texttt{MHD\_8Wave} as well as \texttt{Bouchut5} fail to preserve total energy conservation.

We list the computed $L_1$ and $L_2$ errors for the quantity $B_\perp$ in Table~\ref{tab:Alfven}. They support our conclusions given above, \eg \texttt{ES-Roe} and \texttt{ES-Hybrid} give identical solutions for smooth problems. \referee{Due to dimensional splitting, the obtained results are only second-order accurate in space.}

\begin{table}[h]
	\small
	\centering
	\sisetup{table-format=1.2e1,table-column-width=0mm}
	\setlength{\tabcolsep}{0.575em}
		\begin{tabular}[t]{lr|Sp{0mm}Sp{0mm}Sp{0mm}Sp{0mm}Sp{0mm}Sp{0mm}S}
\toprule
&& {\texttt{EC}}	&& {\texttt{ES-Roe}}	&& {\texttt{ES-Hybrid}}	&& {\texttt{ES-LLF}}	&& {\texttt{USM}}	&& {\texttt{MHD\_8Wave}}	&& {\texttt{Bouchut5}}	\\
\midrule
\vspace*{-1.0em}
\parbox[t]{0mm}{\multirow{2}{*}{\rotatebox[origin=c]{90}{\scriptsize $N=8$\ }}}&&&&&&&&&&&&&&\\
&{${L}_1$ error}	& 7.17e-03 	&$<$& 1.43e-02 	&$=$& 1.43e-02 	&$=$& 1.43e-02 	&$<$& 2.76e-02 	&$\approx$& 2.81e-02 	&$<$& 4.60e-02 	\\
&{${L}_2$ error}	& 7.76e-03 	&$<$& 1.55e-02 	&$=$& 1.55e-02 	&$=$& 1.55e-02 	&$<$& 3.00e-02 	&$\approx$& 3.06e-02 	&$<$& 5.03e-02 	\\[.2em]
\midrule
\vspace*{-1.2em}
\parbox[t]{0mm}{\multirow{4}{*}{\rotatebox[origin=c]{90}{\scriptsize $N=16$}}}&&&&&&&&&&&&&&\\
&{${L}_1$ error}	& 1.02e-03 	&$<$& 1.31e-03 	&$=$& 1.31e-03 	&$=$& 1.31e-03 	&$<$& 4.74e-03 	&$<$& 4.84e-03 	&$<$& 8.78e-03 	\\
&{${L}_2$ error}	& 1.13e-03 	&$<$& 1.46e-03 	&$=$& 1.46e-03 	&$=$& 1.46e-03 	&$<$& 5.24e-03 	&$<$& 5.35e-03 	&$<$& 1.21e-02 	\\
&{EOC (${L}_1$)}	& 2.82	& & 3.45	& & 3.45	& & 3.45	& & 2.54	& & 2.54	& & 2.39	\\
&{EOC (${L}_2$)}	& 2.78	& & 3.41	& & 3.41	& & 3.41	& & 2.52	& & 2.52	& & 2.05	\\
\midrule
\vspace*{-1.2em}
\parbox[t]{0mm}{\multirow{4}{*}{\rotatebox[origin=c]{90}{\scriptsize $N=32$}}}&&&&&&&&&&&&&&\\
&{${L}_1$ error}	& 1.31e-04 	&$<$& 1.41e-04 	&$=$& 1.41e-04 	&$=$& 1.41e-04 	&$<$& 7.41e-04 	&$<$& 7.69e-04 	&$<$& 2.33e-03 	\\
&{${L}_2$ error}	& 1.46e-04 	&$<$& 1.57e-04 	&$=$& 1.57e-04 	&$=$& 1.57e-04 	&$<$& 8.24e-04 	&$<$& 8.56e-04 	&$<$& 3.14e-03 	\\
&{EOC (${L}_1$)}	& 2.95	& & 3.22	& & 3.22	& & 3.22	& & 2.68	& & 2.65	& & 1.92	\\
&{EOC (${L}_2$)}	& 2.95	& & 3.21	& & 3.21	& & 3.21	& & 2.67	& & 2.64	& & 1.95	\\
\midrule
\vspace*{-1.2em}
\parbox[t]{0mm}{\multirow{4}{*}{\rotatebox[origin=c]{90}{\scriptsize $N=64$}}}&&&&&&&&&&&&&&\\
&{${L}_1$ error}	& 1.66e-05 	&$\approx$& 1.69e-05 	&$=$& 1.69e-05 	&$=$& 1.69e-05 	&$\ll$& 1.41e-04 	&$<$& 1.52e-04 	&$<$& 4.87e-04 	\\
&{${L}_2$ error}	& 1.85e-05 	&$\approx$& 1.88e-05 	&$=$& 1.88e-05 	&$=$& 1.88e-05 	&$\ll$& 1.57e-04 	&$<$& 1.69e-04 	&$<$& 7.92e-04 	\\
&{EOC (${L}_1$)}	& 2.98	& & 3.06	& & 3.06	& & 3.06	& & 2.39	& & 2.34	& & 2.25	\\
&{EOC (${L}_2$)}	& 2.99	& & 3.06	& & 3.06	& & 3.06	& & 2.39	& & 2.34	& & 1.99	\\
\midrule
\vspace*{-1.2em}
\parbox[t]{0mm}{\multirow{4}{*}{\rotatebox[origin=c]{90}{\scriptsize $N=128$}}}&&&&&&&&&&&&&&\\
&{${L}_1$ error}	& 2.09e-06 	&$\approx$& 2.10e-06 	&$=$& 2.10e-06 	&$=$& 2.10e-06 	&$\ll$& 3.20e-05 	&$<$& 3.52e-05 	&$<$& 1.05e-04 	\\
&{${L}_2$ error}	& 2.32e-06 	&$\approx$& 2.33e-06 	&$=$& 2.33e-06 	&$=$& 2.33e-06 	&$\ll$& 3.55e-05 	&$<$& 3.91e-05 	&$<$& 1.97e-04 	\\
&{EOC (${L}_1$)}	& 2.99	& & 3.01	& & 3.01	& & 3.01	& & 2.14	& & 2.11	& & 2.22	\\
&{EOC (${L}_2$)}	& 2.99	& & 3.01	& & 3.01	& & 3.01	& & 2.14	& & 2.11	& & 2.01	\\
\bottomrule
		\end{tabular}
	\caption{Computed errors and experimental order of convergence (EOC) for $B_2$ after five oscillation of the Alfv\'en wave in one dimension ($t=5.0$). Sorted by increasing errors from left to right.}
	\label{tab:Alfven1D}
\end{table}
\begin{table}[h]
	\small
	\centering
	\sisetup{table-format=1.2e1,table-column-width=0mm}
	\setlength{\tabcolsep}{0.575em}
		\begin{tabular}[t]{lr|Sp{0mm}Sp{0mm}Sp{0mm}Sp{0mm}Sp{0mm}Sp{0mm}S}
			\toprule
			&& {\texttt{EC}}	&& {\texttt{ES-Roe}}  && \texttt{ES-Hybrid} &&  {\texttt{ES-LLF}}	&&	{\texttt{MHD\_8Wave}}	&& {\texttt{USM}}	&&	{\texttt{Bouchut5}}\\
			\midrule
	\vspace*{-1em}
	\parbox[t]{0mm}{\multirow{2}{*}{\rotatebox[origin=c]{90}{\scriptsize $N=16$}}}
	&&&&&&&&&&&&&&\\
	&{${L}_1$ error}& 8.34e-03 &$<$& 1.34e-02 &$=$& 1.34e-02 &$<$& 1.62e-02 &$<$& 2.43e-02 &$<$& 3.96e-02 &$<$& 4.15e-02 \\
	&{${L}_2$ error}& 9.26e-03 &$<$& 1.47e-02 &$=$& 1.48e-02 &$<$& 1.80e-02 &$<$& 2.69e-02 &$<$& 4.39e-02 &$<$& 4.54e-02 \\[.2em]
	\midrule
	\vspace*{-1.2em}
	\parbox[t]{0mm}{\multirow{4}{*}{\rotatebox[origin=c]{90}{\scriptsize $N=32$}}}
	&&&&&&&&&&&&&&\\
	&{${L}_1$ error}& 1.83e-03 &$<$& 2.36e-03 &$=$& 2.36e-03 &$<$& 2.76e-03 &$<$& 3.61e-03 &$<$& 1.11e-02 &$<$& 1.67e-02 \\
	&{${L}_2$ error}& 2.05e-03 &$<$& 2.64e-03 &$=$& 2.64e-03 &$<$& 3.06e-03 &$<$& 3.94e-03 &$<$& 1.19e-02 &$<$& 1.84e-02 \\
	&{EOC (${L}_1$)}& 2.19e+00 && 2.50e+00 && 2.51e+00 && 2.55e+00 && 2.75e+00 && 1.83e+00 && 1.31e+00 \\
	&{EOC (${L}_2$)}& 2.17e+00 && 2.48e+00 && 2.49e+00 && 2.56e+00 && 2.77e+00 && 1.88e+00 && 1.30e+00 \\ 
	\midrule
	\vspace*{-1.2em}
	\parbox[t]{0mm}{\multirow{4}{*}{\rotatebox[origin=c]{90}{\scriptsize $N=64$}}} %
	&&&&&&&&&&&&&&\\
	&{${L}_1$ error}& 4.36e-04 &$<$& 4.73e-04 &$=$& 4.73e-04 &$<$& 5.10e-04 &$<$& 6.78e-04 &$<$& 4.03e-03 &$<$& 7.79e-03 \\
	&{${L}_2$ error}& 4.93e-04 &$<$& 5.37e-04 &$=$& 5.37e-04 &$<$& 5.73e-04 &$<$& 7.38e-04 &$<$& 4.85e-03 &$<$& 8.65e-03 \\
	&{EOC (${L}_1$)}& 2.06e+00 && 2.32e+00 && 2.32e+00 && 2.44e+00 && 2.41e+00 && 1.46e+00 && 1.10e+00 \\
	&{EOC (${L}_2$)}& 2.06e+00 && 2.30e+00 && 2.30e+00 && 2.41e+00 && 2.42e+00 && 1.30e+00 && 1.09e+00 \\ 
	\midrule
	\vspace*{-1.2em}
	\parbox[t]{0mm}{\multirow{4}{*}{\rotatebox[origin=c]{90}{\scriptsize $N=128$}}} %
	&&&&&&&&&&&&&&\\
	&{${L}_1$ error}& 1.08e-04 &$\approx$& 1.10e-04 &$=$& 1.10e-04 &$\approx$& 1.12e-04 &$<$& 1.56e-04 &$<$& 1.57e-03 &$<$& 3.87e-03 \\
	&{${L}_2$ error}& 1.22e-04 &$\approx$& 1.25e-04 &$=$& 1.25e-04 &$\approx$& 1.28e-04 &$<$& 1.73e-04 &$<$& 1.89e-03 &$<$& 4.30e-03 \\
	&{EOC (${L}_1$)}& 2.01e+00 && 2.10e+00 && 2.10e+00 && 2.18e+00 && 2.12e+00 && 1.36e+00 && 1.01e+00 \\
	&{EOC (${L}_2$)}& 2.01e+00 && 2.10e+00 && 2.10e+00 && 2.16e+00 && 2.09e+00 && 1.36e+00 && 1.01e+00 \\ 
			\bottomrule
		\end{tabular}
	\caption{Computed errors and experimental order of convergence (EOC) for $B_\perp$ after five oscillation of the Alfv\'en wave \referee{in two dimensions} ($t=5.0$). Sorted by increasing errors from left to right.}
	\label{tab:Alfven}
\end{table}
\jump
\clearpage

\subsection{Orszag-Tang MHD Vortex (2D)}\label{scn:OrszagTang}

The Orszag-Tang vortex problem \citep{Orszag1979} is a two-dimensional, spatially periodic problem well suited for studies of MHD turbulence. Thus, it has become a classical test for numerical MHD schemes. It includes dissipation of kinetic and magnetic energy, magnetic reconnection, formation of high-density jets, dynamic alignment and the emergence and manifestation of small-scale structures.
The Orszag-Tang MHD vortex problem starts from non-random, smooth initial data. As the flow evolves it gradually becomes increasingly complex, forming intermediate shocks. Thus, this problem demonstrates the transition from initially smooth data to compressible, supersonic MHD turbulence. The initial data is chosen such that the root mean square values of the velocity and the magnetic fields as well as the initial Mach number are all one. The average plasma beta is $\beta=\frac{10}{3}$.

\begin{table}[h]
	\setlength\extrarowheight{3pt}
	\centering
	\begin{minipage}[t]{0.45\textwidth}
		\begin{tabular}[t]{|l|l|}
			\hline
			Density $\rho$ & $1.0$ \\
			\hline
			Pressure $p$ & $1.0/\gamma$ \\
			\hline
			Velocity $\vec{u}$ & $(-\sin(2\pi y),\,\sin(2\pi x),\,0.0)^\intercal$ \\
			\hline
			Mag.~field $\vec{B}$ & $\frac{1}{\gamma}(-\sin(2\pi y),\,\sin(4\pi x),\,0.0)^\intercal$\\
			\hline
		\end{tabular}
	\end{minipage}
	\begin{minipage}[t]{0.48\textwidth}
		\begin{tabular}[t]{|l|l|}
			\hline
			Domain size &$\{x,y\}_\mathrm{min} = \{0.0,0.0\}$ \\
						&$\{x,y\}_\mathrm{max} = \{1.0,1.0\}$ \\
			\hline
			Boundary conditions & all: periodic\\
			\hline
			Adaptive refinement on	& density, magnetic field \\
			\hline
			Simulation end time & $t_\mathrm{max} = 0.5$ \\
			\hline
			Adiabatic index & $\gamma = 5/3$\\
			\hline
		\end{tabular}
	\end{minipage}
	\vspace*{-3mm}
	\caption{Initial conditions and runtime parameters: Orszag-Tang MHD vortex test}
	\label{tab:OrszagTang}
\end{table}

\referee{Additionally, we compute the experimental convergence order for the available MHD schemes after \SI{10}{\percent} of the total runtime. At this time, the solution is already very complex but still smooth. As there is no analytic solution available, we compare to a high resolution simulation obtained using our entropy-conserving (\texttt{EC}) scheme on an uniform grid of $\num{2048}\times\num{2048}$ cells. We compute the $L_1$ and $L_2$ errors as well as the experimental order of convergence and list them in Table~\ref{tab:OrszagTang:Errors}. They coincide with our results presented in the preceding section. As in Sec.~\ref{scn:Alfven}, we find the results to be at most second-order accurate in space due to dimensional splitting.}
\begin{table}[h]
	\small
	\centering
	\sisetup{table-format=1.2e1,table-column-width=0mm}
	\setlength{\tabcolsep}{0.575em}
		\begin{tabular}[t]{lr|Sp{0mm}Sp{0mm}Sp{0mm}Sp{0mm}S}
			\toprule
			&& {\texttt{EC}} && \texttt{ES-Hybrid} && {\texttt{Bouchut5}} && {\texttt{USM}} && {\texttt{MHD\_8Wave}}\\
			\midrule
			\vspace*{-1em}
			\parbox[t]{0mm}{\multirow{2}{*}{\rotatebox[origin=c]{90}{\scriptsize $N=16$}}}
			&&&&&&&&&&\\
			&{${L}_1$ error}& 6.59e-03 &$<$& 7.00e-03 &$<$& 7.25e-03 &$<$& 8.08e-3 &$<$& 8.18e-3  \\
			&{${L}_2$ error}& 8.51e-03 &$<$& 8.71e-03 &$<$& 9.32e-03 &$<$& 1.02e-2 &$\approx$& 1.03e-2  \\[.2em]
			\midrule
			\vspace*{-1.2em}
			\parbox[t]{0mm}{\multirow{4}{*}{\rotatebox[origin=c]{90}{\scriptsize $N=32$}}}
			&&&&&&&&&&\\
			&{${L}_1$ error}& 1.60e-03 &$\approx$& 1.63e-03 &$<$& 2.08e-03 &$\approx$& 2.09e-3 &$\approx$& 2.12e-3  \\
			&{${L}_2$ error}& 2.02e-03 &$\approx$& 2.06e-03 &$<$& 2.65e-03 &$\approx$& 2.69e-3 &$\approx$& 2.63e-3  \\
			&{EOC (${L}_1$)}& 2.04 && 2.10 && 1.80 && 1.95 && 1.95  \\
			&{EOC (${L}_2$)}& 2.07 && 2.08 && 1.81 && 1.93 && 1.96  \\
			\midrule
			\vspace*{-1.2em}
			\parbox[t]{0mm}{\multirow{4}{*}{\rotatebox[origin=c]{90}{\scriptsize $N=64$}}} %
			&&&&&&&&&&\\
			&{${L}_1$ error}& 4.01e-04 &$\approx$& 3.99e-04 &$<$& 5.13e-04 &$<$& 5.55e-4 &$\approx$& 5.55e-4  \\
			&{${L}_2$ error}& 5.17e-04 &$\approx$& 5.14e-04 &$<$& 6.60e-04 &$<$& 7.25e-4 &$>$& 7.06e-4  \\
			&{EOC (${L}_1$)}& 2.00  && 2.03 && 2.02 && 1.91 && 1.93  \\
			&{EOC (${L}_2$)}& 1.97  && 2.00 && 2.01 && 1.81 && 1.90   \\
			\bottomrule
		\end{tabular}
	\caption{Computed errors and experimental order of convergence (EOC) for $p_{\rm mag} = \frac{1}{2} \lVert \vec{B} \rVert^2$ before the onset of discontinuities in the Orszag-Tang MHD vortex test ($t=0.05$).}
	\label{tab:OrszagTang:Errors}
\end{table}

\begin{figure}[!ht]
	\centering
	\includegraphics[scale=1]{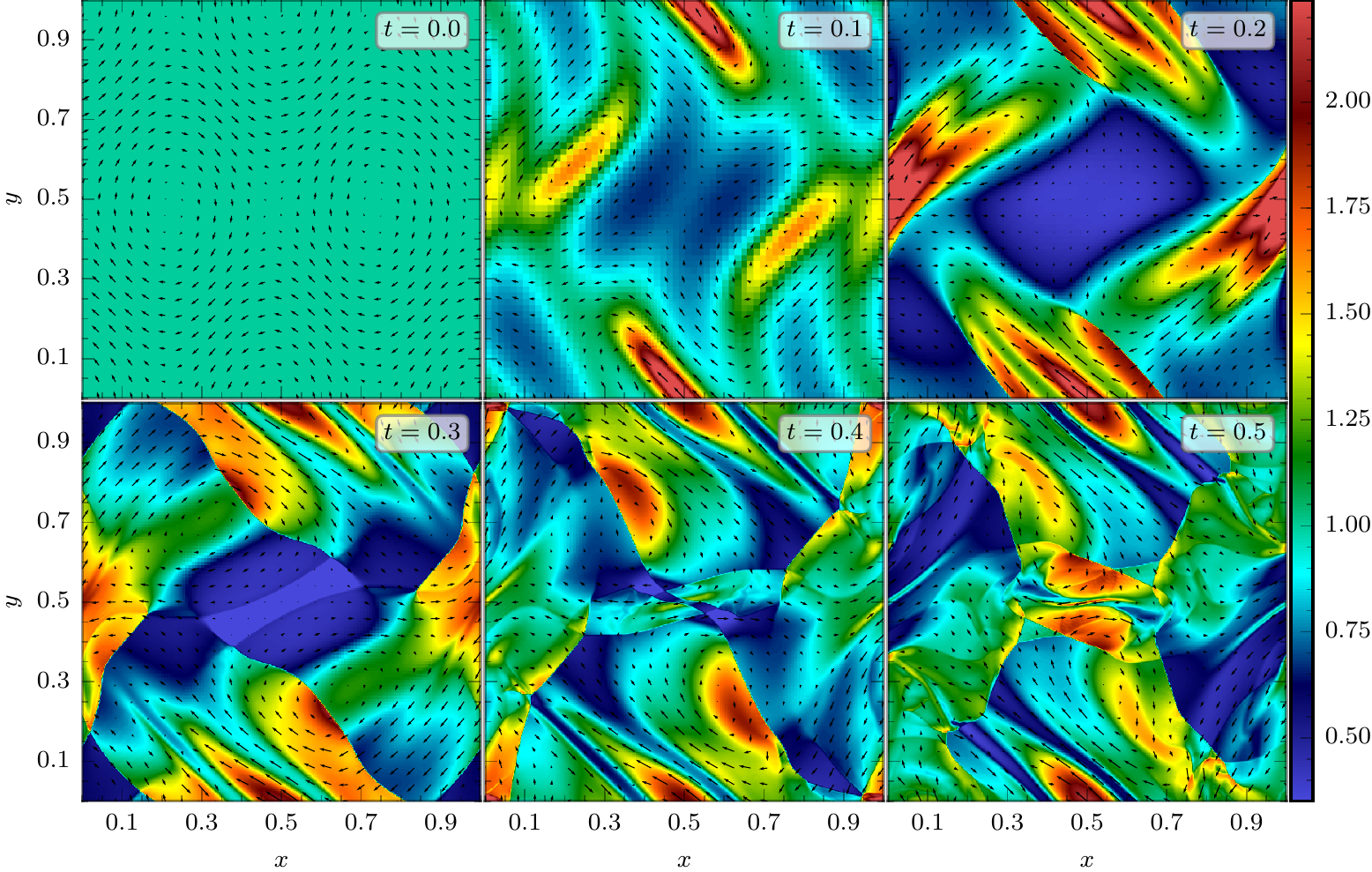}
	\caption{Orszag-Tang MHD vortex test: Density plots with superimposed magnetic field directions. \texttt{ES-Hybrid} scheme with an adaptive grid resolution up to $512\times512$. The time step is shown in the upper right corner of each plot. These plots can be compared to \eg Fig.~1 of \cite{Balbas2005}. The lower right plot ($t=0.5$) can be compared to \eg Fig.~10 of \cite{Londrillo2000} and Fig.~14 of \cite{Dai1998}. 
	}
	\label{fig:OrszagTang:time}
\end{figure}

\begin{figure}[!ht]
	\centering
	\includegraphics[scale=1]{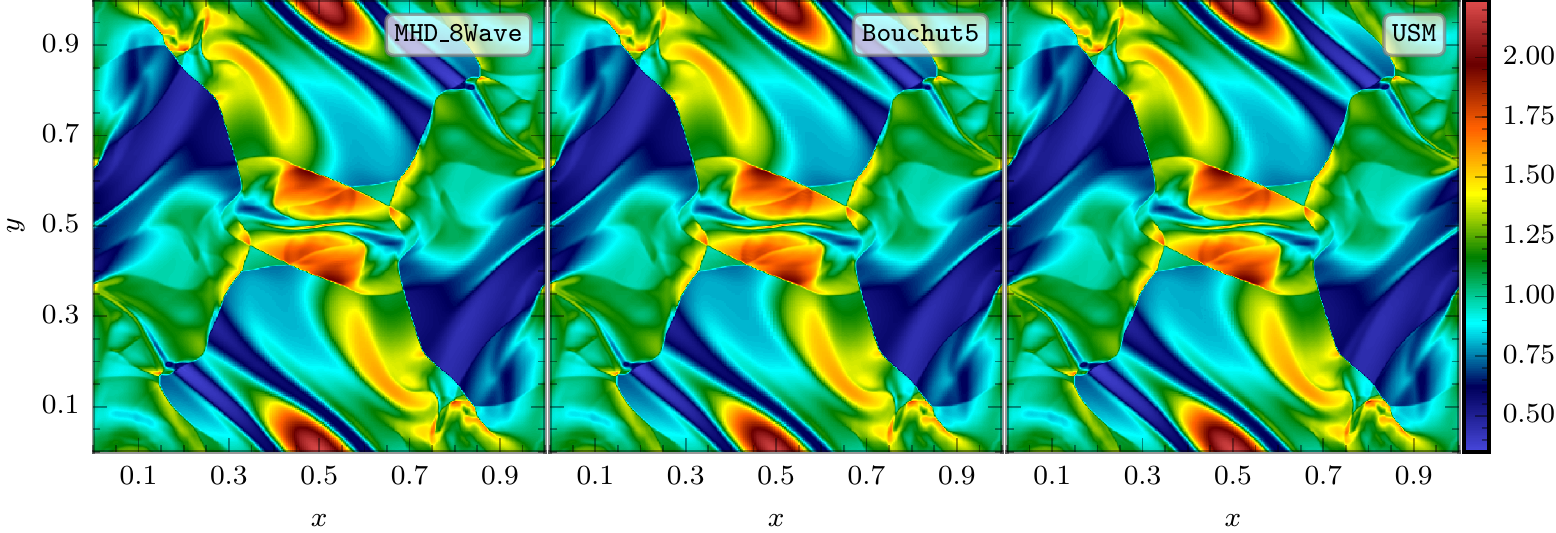}
	\caption{Orszag-Tang MHD vortex test: Density plots with an adaptive grid resolution up to $512\times512$ at $t=0.5$. The MHD solver used is given in the upper right corner of each plot.}
	\label{fig:OrszagTang:othersolvers}
\end{figure}

Fig.~\ref{fig:OrszagTang:time} displays the evolution of the density of a plasma given the initial conditions listed in Table~\ref{tab:OrszagTang}.
As the solution evolves in time, the initial vortex splits into two vortices. Sharp gradients accumulate and the vortex pattern becomes increasingly complex due to highly non-linear interactions between multiple intermediate shock waves travelling at different speeds. The result compares very well with results given in the literature, \eg \cite{Balbas2005,Londrillo2000,Dai1998}, as well as with the solution of the Orszag-Tang MHD vortex obtained using the \texttt{MHD\_8Wave}, the \texttt{Bouchut5}, and the unsplit \texttt{USM} implementations (shown in Fig.~\ref{fig:OrszagTang:othersolvers}).


\jump
\clearpage
\subsection{MHD Rotor (2D)}\label{scn:Rotor}
The MHD rotor problem \citep{Balsara1999} describes a rapidly spinning dense cylinder embedded in a magnetized, homogeneous medium at rest. Due to centrifugal forces, the dense cylinder is in non-equilibrium. As the rotor spins with the given initial rotating velocity, the initially uniform magnetic field will wind up the rotor. The wrapping of the rotor by the magnetic field leads to strong torsional Alfv\'en waves launched into the ambient fluid.
Due to the onset and propagation of strong Alfv\'en waves, this test is relevant for the understanding of star formation. The initial conditions are listed in Table~\ref{tab:Rotor}.

\begin{table}[h]
	\centering
	\begin{minipage}[t]{0.41\textwidth}
		\begin{tabular}[t]{l|ccc}
							&	{$r \le r_0$}				& {$r \in (r_0,r_1)$}		& {$r \ge r_1$}\\
			\midrule
			$\rho$		&	$10.0$ 						& $1.0 + 9.0 f(r)$			& $1.0$	\\
			$p$		&	$1.0$						& $1.0$  					& $1.0$\\
			$B_1$		&	$5/\sqrt{4\pi}$				& $5/\sqrt{4\pi}$			& $5/\sqrt{4\pi}$\\
			$B_2$		&	$0.0$						& $0.0$						& $0.0$\\
			$B_3$		&	$0.0$						& $0.0$						& $0.0$\\
			$u$		&	$-20.0 \Delta y$		& $-20.0 f(r) \Delta y$		& $0.0$\\
			$v$		&	$20.0 \Delta x$		& $20.0 f(r) \Delta x$		& $0.0$\\
			$w$		&	$0.0$		& $0.0$		& $0.0$\\
		\end{tabular}\\[.6em]
		with $r=\sqrt{(x-x_\mathrm{center})^2+(y-y_\mathrm{center})^2}$,\par
		$\Delta x = (x-x_\mathrm{center})$, $\Delta y = (y-y_\mathrm{center})$,\par
		and $f(r) = \frac{r_1-r}{r_1-r_0}$
	\end{minipage}
	\begin{minipage}[t]{0.545\textwidth}
		\setlength\extrarowheight{3pt}
		\begin{tabular}[t]{|l|l|}
			\hline
			Domain size &$\{x,y\}_\mathrm{min} = \{0.0,0.0\}$ \\
			&$\{x,y\}_\mathrm{max} = \{1.0,1.0\}$ \\
			\hline
			Inner radius	&  $r_0 = 0.1$ \\
			\hline
			Outer radius	&  $r_1 = 0.115$ \\
			\hline
			$x$-center		& $x_\mathrm{center} = 0.5$ \\
			\hline
			$y$-center		& $y_\mathrm{center} = 0.5$ \\
			\hline
			Boundary conditions & all: zero-gradient (``outflow'') \\ 
			\hline
			Adaptive refinement on	& density, magnetic field \\
			\hline
			%
			Simulation end time & $t_\mathrm{max} = 0.15$ \\
			\hline
			Adiabatic index & $\gamma = 1.4$ \\ 
			\hline
		\end{tabular}
	\end{minipage}
	\caption{Initial conditions and runtime parameters: MHD rotor test}
	\label{tab:Rotor}
\end{table}

\begin{figure}[!ht]
	\centering
	\includegraphics[scale=1]{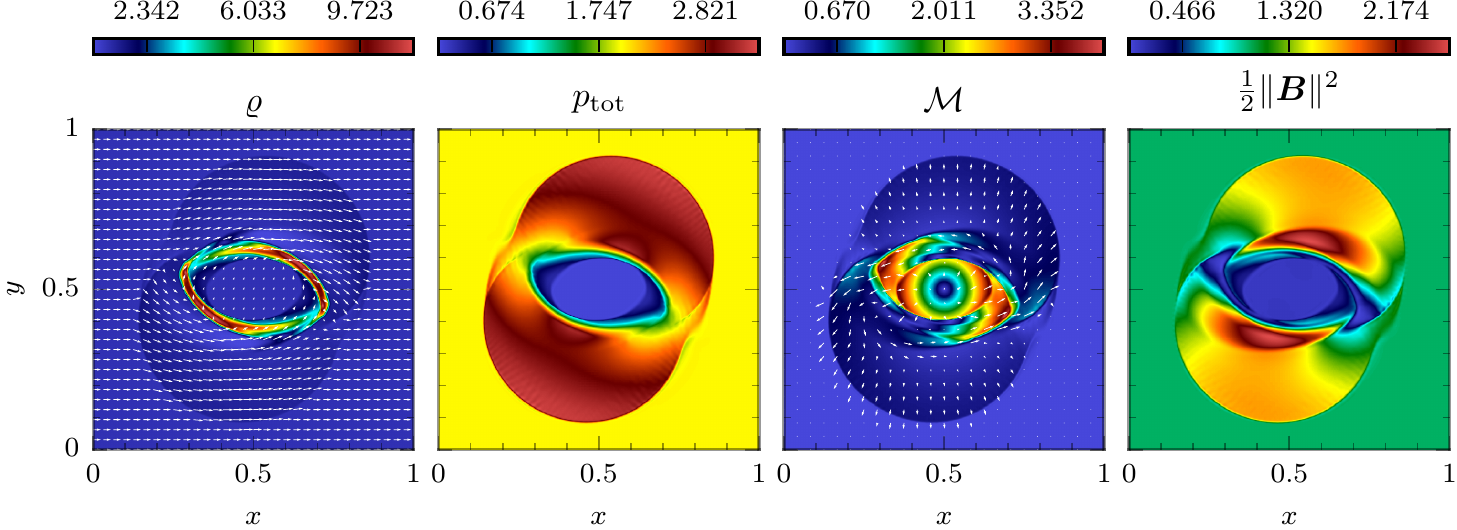}
	\caption{MHD rotor test: Adaptive grid resolution up to $512\times512$. From left to right: density $\rho$ with overlayed magnetic field, total pressure $p_{\rm tot} = p + p_{\rm mag}$, Mach number $\mathcal{M}$ with overlayed velocity vectors, and magnetic pressure $p_{\rm mag} = \frac{1}{2}\lVert\boldsymbol{B}\rVert^2$. This plot can directly be compared to Fig.~7 of \cite{Winters2016}, Fig.~14 of \cite{Londrillo2000}, and Fig.~2 of \cite{Balsara1999}.}
	\label{fig:MHDRotor:2D}
\end{figure}

\begin{figure}[!ht]
	\centering
	\includegraphics[width=\textwidth]{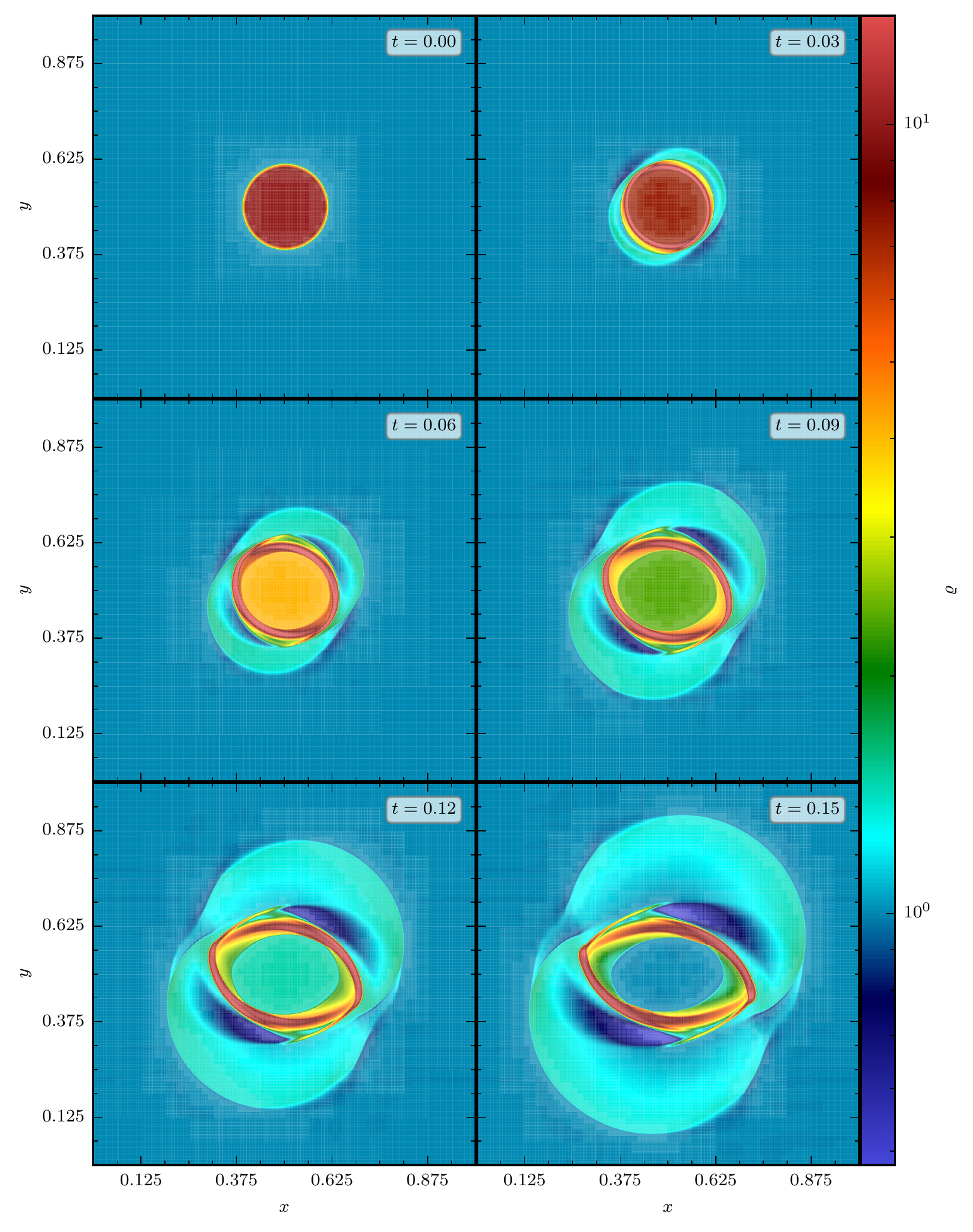}
	\caption{MHD rotor test: Density evolution on a logarithmic scale with superimposed AMR grid. Adaptive grid resolution up to level 8 (up to $1024\times1024$ cells).}
	\label{fig:MHDRotor}
\end{figure}

This test demonstrates that our solver is able to resolve torsional Alfv\'en waves, which is particularly visible in the plot of the magnetic pressure (right plot in Fig.~\ref{fig:MHDRotor:2D}).
The Mach number $M$ (see Fig.~\ref{fig:MHDRotor:2D}) shows that the rotor is, up to a certain radial distance, still in uniform rotation. Beyond this radius, the rotor has exchanged momentum with its environment and decelerated. In the left plot of Fig.~\ref{fig:MHDRotor:2D} we present the magnetic field superimposed on the fluid density, $\rho$. It is clearly seen that the magnetic field basically maintains its initial shape outside of the region of influence of the Alfv\'en waves. Inside, the magnetic field is refracted by the MHD discontinuities.
In Fig.~\ref{fig:MHDRotor} we present six snapshots of the evolution of the fluid density (as well as the AMR grid) up to the final time $t=t_\mathrm{max}$. This is shortly before the torsional Alfv\'en waves leave the computational domain and after the cylinder has rotated almost $180^\circ$.

\jump

\subsection{Comparison of Computational Efficiency (2D)}\label{scn:Efficiency}
We perform a memory and CPU time comparison on a uniform grid.
We compare the new \texttt{ES} solver implementation against the Bouchut 5 wave (\texttt{Bouchut5}) \cite{Waagan2011}, Powell's 8 wave (\texttt{MHD\_8Wave}) \cite{Powell1999}, and the unsplit staggered mesh (\texttt{USM}) \cite{Lee2013,Lee2009} solver implementations applied to the MHD Rotor problem, described in the preceding section. For this test we use identical runtime parameters. The AMR grid is fixed to level 5.

We present the results of the study in Table \ref{tab:Costcomparison}. \referee{Note that the effective computational costs are very implementation specific.}
We see that the \texttt{ES} solver uses slightly less memory than the other schemes. The \texttt{ES} solver needs more computational time per time step since the Runge-Kutta time integration scheme involves the full flux computation and spatial reconstruction procedure in each of the intermediate stages. The higher computational costs per time step can be -- at least partially -- compensated by choosing a larger \texttt{CFL} coefficient. We neglect this benefit here and run all simulations with a fixed \texttt{CFL} number to give a fair comparison.
For \texttt{USM} we use a second order accurate Roe-type Riemann solver, previously used for the numerical tests in \cite{Lee2009}.

\sisetup{detect-weight=true,detect-inline-weight=math}
\begin{table}[!ht]
	\centering
	\begin{tabular}{l|S[table-format=3.1]|S[table-format=2.2]}
	\toprule
	{Scheme} & {Memory consumption (MB)} & {CPU time (s)} \\
	\midrule
	\bfseries \texttt{ES}			& \bfseries 88.5	& \bfseries \referee{11.18}	\\
	\texttt{Bouchut5}				& 90.7				& 7.43	\\
	\texttt{MHD\_8Wave}				& 95.4				& 7.80	\\
	\texttt{USM}					& 143.5				& 12.16	\\
	\bottomrule
	\end{tabular}
	\vspace*{-2mm}
	\caption{Comparison of computational efficiency. The memory consumption is measured for the whole \texttt{flash4} process, while the computational time corresponds only to the time used by the MHD solver as given by \texttt{FLASH}'s code performance summary.}
	\label{tab:Costcomparison}
\end{table}
\jump

\subsection{Gravitational Instability}\label{scn:Jeans}
A particularly simple example of gravitational instability was discovered by Jeans \citep{Jeans1902}. This phenomenon is of great astrophysical interest in the context of star formation and cosmic structure growth. The configuration is a useful test for the coupling of multi-dimensional gravity to hydrodynamics in a computational code. The Jeans instability allows one to study the pressure dominated and gravity dominated limits as well as the numerical method's behaviour between the two limits. We start from an infinite homogeneous medium at rest and consider a small perturbation in density. We shall suppose that the initial fluctuations in density and pressure take place adiabatically, so that ${p_0} = \gamma {\rho_0}$.
The initial conditions for this test are summarized in Table~\ref{tab:JeansTest}. We use the direct multigrid fast Fourier transform Poisson solver (\texttt{Grid/GridSolvers/Multigrid/fft}) for the computation of the gravitational source term.
\begin{table}[h]
	\centering
	\setlength\extrarowheight{3pt}
	\begin{minipage}[t]{0.40\textwidth}
		\begin{tabular}[t]{|l|l|}
			\hline
			Density	$\rho$ [\si{g.cm^{-3}}]		&	$\rho_0 \cdot \left[1+\delta(\vec{x})\right]$	\\
			\hline
			Pressure $p$ [\si{dyn.cm^{-2}}]		&	$p_0 \cdot \left[1+\gamma \delta(\vec{x})\right]$\\
			\hline
			Perturbation $\delta(\vec{x})$ & $\delta_0\cdot\cos\left(\vec{k}\cdot\vec{x}\right)$ \\
			\hline
			Velocity $\vec{u}$ [\si{cm.s^{-1}}]	&	\vec{0}\\
			\hline
			Magnetic field $\vec{B}$ [\si{G}]		&	\vec{0} \\
			\hline
			\multicolumn{2}{c}{with $\rho_0 = \SI{1.5e7}{g.cm^{-3}}$, $\delta_0 = \num{1e-3}$,} \\
			\multicolumn{2}{c}{and $p_0 = \SI{1.5e7}{dyn.cm^{-2}}$}\vspace*{-3mm}
		\end{tabular}
	\end{minipage}
	\hspace{1mm}
	\begin{minipage}[t]{0.53\textwidth}
		\begin{tabular}[t]{|l|l|}
			\hline
			Domain size [\si{cm}] &$\{x,y\}_\mathrm{min} = \{0.0,0.0\}$ \\
			& $\{x,y\}_\mathrm{max} = \{1.0,1.0\}$ \\
			\hline
			Boundary conditions & all: periodic\\ 
			\hline
			Simulation end time [\si{s}] & $t_\mathrm{max} = 5.0$ \\
			\hline
			Adiabatic index & $\gamma = 5/3$ \\
			\hline
			Wave vector $\vec{k}$ [\si{cm^{-1}}] & $2\pi/\vec{\lambda}$ with $\vec{\lambda} = (0.5,0,0)^\intercal$ \\
			\hline
		\end{tabular}
	\end{minipage}
	\vspace*{-2mm}
	\caption{Runtime parameters and initial conditions: Jeans Instability test (2D)}
	\label{tab:JeansTest}
\end{table}

We obtain the dispersion relation of a self-gravitating fluid by solving the perturbed wave equations by planar wave solutions in Fourier space. 
From the relation,
\begin{equation}
\omega^2 = {a_0^2k^2-4\pi G\rho_0},
\end{equation}
we define the \emph{Jeans wavenumber}
\begin{equation}
k_J = \frac{\sqrt{4\pi G \rho_0}}{a_0} \approx 2.75
\end{equation}
with the gravitational constant, $G=\SI{6.674e-8}{cm^3.g^{-1}.s^{-2}}$,
and the adiabatic sound speed $a_0 = \sqrt{\gamma\, p_0/\rho_0} \approx \SI{1.29}{cm.s^{-1}}$ where the given numbers correspond to the initial conditions used.
The Jeans wavenumber number is very important as it defines a scale on which gravitational effect become dominant in astrophysical systems.
As long as $k > k_J$, the perturbation is stable and oscillates with a real frequency of $\omega$. This is the case with our chosen initial conditions, as $k \approx 11 > 2.75 \approx k_J$. However, if $k < k_J$, the perturbation grows exponentially in time as $\omega$ is purely imaginary. An overdense region would become denser and denser, leading to gravitational collapse \cite{Chandrasekhar1961}.

We compute the oscillation frequency, $\omega$, by measuring the time interval required for the energy to undergo exactly ten oscillations. The analytical expression for the kinetic, internal, and potential energies are provided in \ref{app:Jeans}.

\begin{figure}[h]
	\centering
	\includegraphics[scale=1]{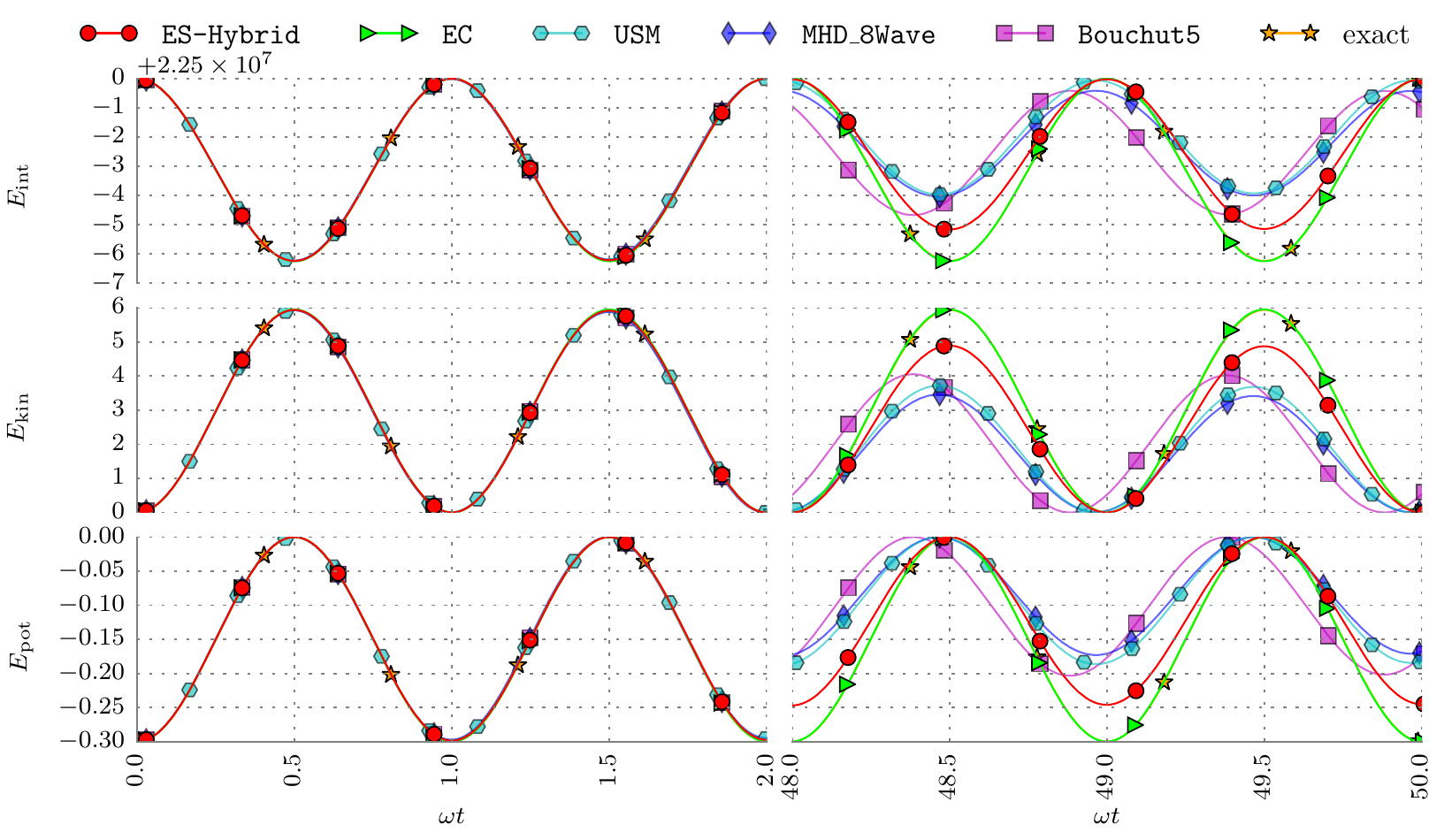}\\
	\vspace*{-3mm}
	\caption{Jeans Gravitational Instability test: Plot of internal, kinetic, and gravitational energies. (left) The energy changes for all solvers agree well early in the computation. (right) As time progresses we see the dissipation of energy by each solver differs. The \texttt{ES} solver exhibits the least dissipation of the solvers tested and shows the best agreement with the analytic solution. We use a fixed resolution of $64\times64$ cells. We give the \emph{exact} solution in \ref{app:Jeans}.}
	\label{fig:Jeans}
\end{figure}

The resulting kinetic and internal (thermal) energies as functions of $\omega t$
are shown in Fig.~\ref{fig:Jeans}. We performed the simulations using a uniform resolution of $64\times64$ grid cells and chose a small CFL coefficient of $\mathtt{CFL}=0.1$ to limit the maximum time step for all solvers. The small CFL coefficient is not chosen for stability reasons, but to ensure enough data is generated to create smooth plots. As can be seen in Fig.~\ref{fig:Jeans}, all solvers agree well in the beginning.
Again, the \texttt{EC} scheme can be used as the solution is smooth. We note that the \texttt{EC} scheme shows essentially no dissipation even at the final time $t=5.0$.
All remaining solvers dissipate energy in some capacity. We see from Fig.~\ref{fig:Jeans} that the \texttt{ES} solver is considerably less dissipative than the other tested solvers. Furthermore, the \texttt{ES} scheme agrees well with the analytic solution, while the other schemes fail to maintain the exact oscillation period at later times and dissipate much of the energy of the dynamics.
\jump

\subsection{MHD Blast Wave (2D, 3D)}\label{scn:MHDBlast}

The two-dimensional version of the MHD blast wave problem was studied by \cite{Balsara1999}. We use an extended three-dimensional version to demonstrate the robustness of our scheme in simulations involving regimes with low thermal pressures and high kinetic as well as magnetic energies in three dimensions.
This test problem leads to the onset of strong MHD discontinuities, relevant to astrophysical phenomena where magnetic fields can have strong dynamical effects.
It describes an initially circular pressure pulse. We choose here a relative magnitude of $10^4$ for comparison with \cite{Balsara1999}.
The initial conditions used are listed in Table~\ref{tab:MHDBlast}.

\begin{table}[h]
	\centering
	\begin{minipage}[t]{0.41\textwidth}
		\begin{tabular}[t]{l|ccc}
			&	{$r \le r_0$}				& {$r \in (r_0,r_1)$}		& {$r \ge r_1$}\\
			\midrule
			$\rho$	&	$1.0$ 						& $1.0$						& $1.0$	\\
			$p$		&	$1000.0$					&{$0.1+999.9 f(r)$}  			& $0.1$\\
			$B_1$	&	$\frac{100}{\sqrt{4\pi}}$	& $\frac{100}{\sqrt{4\pi}}$	& $\frac{100}{\sqrt{4\pi}}$\\
			$B_2$	&	$0.0$						& $0.0$						& $0.0$\\
			$B_3$	&	$0.0$						& $0.0$						& $0.0$\\
			$u$		&	$0.0$						& $0.0$						& $0.0$\\
			$v$		&	$0.0$						& $0.0$						& $0.0$\\
			$w$		&	$0.0$						& $0.0$						& $0.0$\\

		\end{tabular}\\[.4em]
		with $r=\sqrt{(x-x_\mathrm{center})^2+(y-y_\mathrm{center})^2}$, \par and $f(r) = \frac{r_1-r}{r_1-r_0}$
	\end{minipage}
	\begin{minipage}[t]{0.575\textwidth}
		\setlength\extrarowheight{3pt}
		\vspace*{-10pt}
		\begin{tabular}[t]{|l|l|}
			\hline
			Domain size &$\{x,y,z\}_\mathrm{min} = \{-0.5,-0.5,-0.5\}$ \\
			&$\{x,y,z\}_\mathrm{max} = \{0.5,0.5,0.5\}$ \\
			\hline
			Inner radius	&  $r_0 = 0.09$ \\
			\hline
			Outer radius	&  $r_1 = 0.1$ \\
			\hline
			Explosion center		& $\vec{x}_\mathrm{center} = (0.0,0.0,0.0)$\\
			\hline
			Boundary conditions & all: periodic \\ 
			\hline
			Adaptive refinement on	& density, pressure \\
			\hline
			Simulation end time & $t_\mathrm{max} = 0.01$ \\
			\hline
			Adiabatic index & $\gamma = 1.4$ \\ 
			\hline
		\end{tabular}
	\end{minipage}
	\caption{Initial conditions and runtime parameters: MHD blast wave test}
	\label{tab:MHDBlast}
\end{table}

\begin{figure}[!h]
	\centering
	\includegraphics[scale=1]{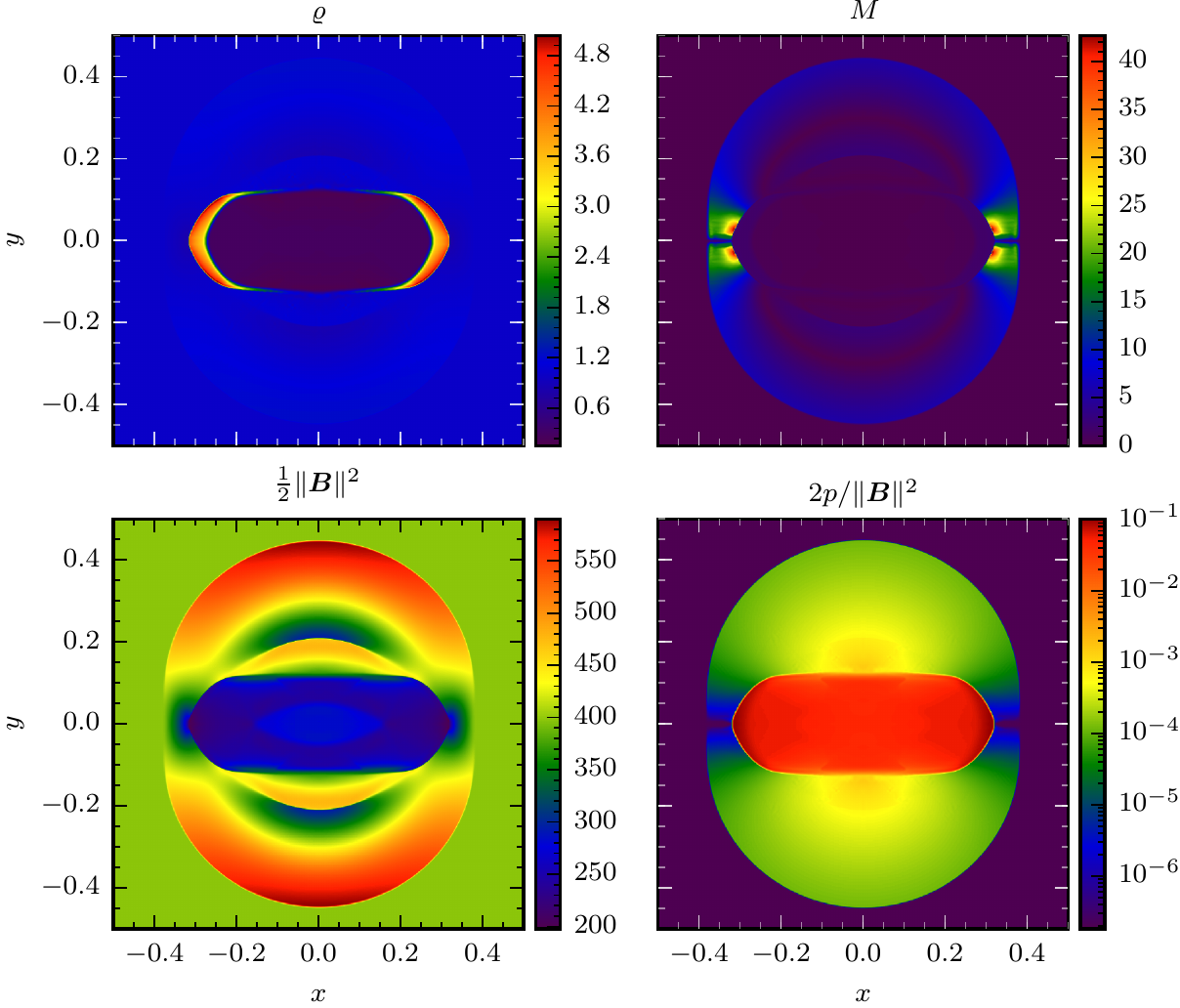}
	\caption{MHD blast wave test: Adaptive grid resolution up to $512\times512$. Top left: density $\rho$, Top right: Mach number $M$, Lower left: magnetic pressure $p_{\rm mag}=\frac{1}{2}\lVert\vec{B}\rVert^2$, Lower right: Plasma-$\beta=p/p_{\rm mag}$. The density and magnetic pressure plots can be compared to Fig.~13 of \cite{Londrillo2000}. The plots of density and magnetic pressure can directly be compared to Fig.~4 of \cite{Balsara1999}.}
	\label{fig:MHDBlast2D}
\end{figure}

The chosen initial conditions result in a very low plasma-$\beta$ parameter, $\beta = 2p/\vec{B}^2 \approx \num{2.5e-4}$. The MHD explosion emits fast magnetosonic shock waves propagating with high velocities. The explosion is highly anisotropic and the displacement of the gas in the transverse $y$ and $z-$direction is inhibited. This leads to the phenomenon that the explosion bubble is strongly distorted according to the initial magnetic field (in the $x-$direction). Furthermore, we see that the Mach number spans a broad range from 0 up to 42. As can be seen from the results in Fig.~\ref{fig:MHDBlast2D} showing the density, Mach number, magnetic pressure and the plasma-$\beta$ at $t=0.01$, the out-going blast wave shows no grid alignment effect.
We also tested the new \texttt{ES} solver with relative pressure magnitudes of $10^5$ and $10^6$ without finding any numerical defects.

\begin{figure}[!h]
	\centering
	\includegraphics[scale=1]{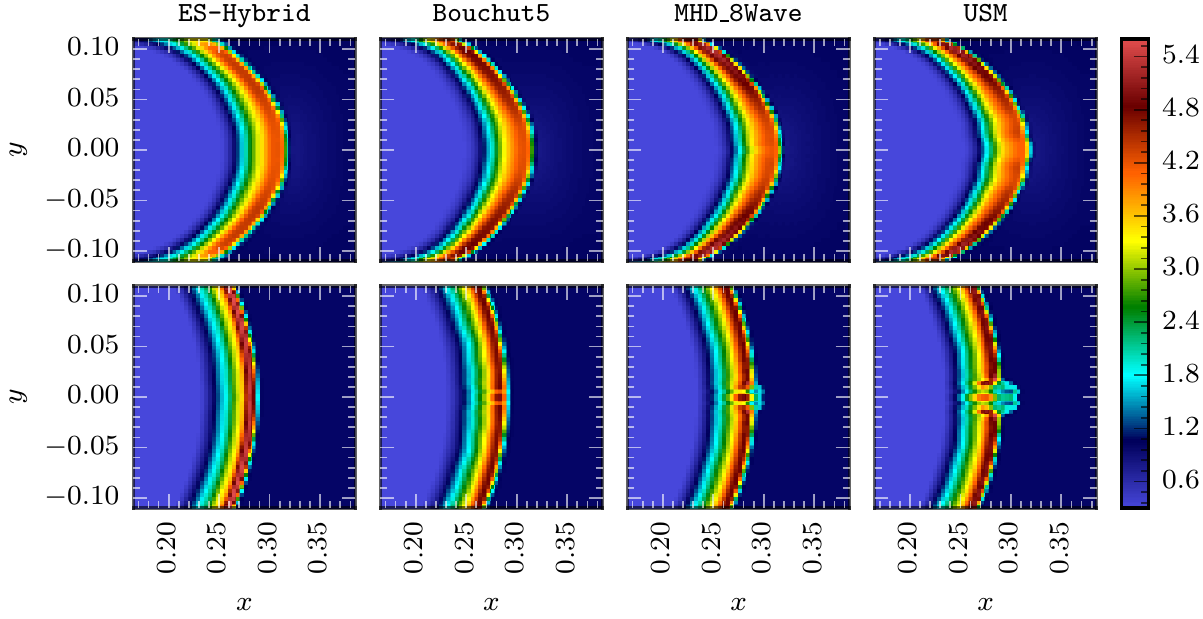}
	\caption{MHD blast wave test: Linear density plots. Top:~$B_1=100/\sqrt{4\pi}$, Bottom:~$B_1=0$}
	\label{fig:MHDBlast_carbuncle}
\end{figure}

In Fig.~\ref{fig:MHDBlast_carbuncle}, we show zoomed linear density plots of the fast expanding shock front computed with our \texttt{ES-Hybrid}, the \texttt{Bouchut5} \cite{Waagan2011}, the unsplit \texttt{USM} \cite{Lee2009,Lee2013}, and the \texttt{MHD\_8Wave} \cite{Powell1999} solvers. As can be seen, the blast wave front has a elliptical shape as is expected due to the strong influence of the magnetic field. If we run the same simulation without magnetic fields, \ie the hydrodynamic limit $\vec{B} = \vec{0}$, we observe numerical defects close to the Cartesian grid axes in all simulation runs except the one using the \texttt{ES} solver. Ismail et al.\ \cite{IsmailCarbuncle2009} showed that even schemes which have increased dissipation and do not show 1D shock instabilities can still suffer from the carbuncle phenomenon in multiple dimensions. We emphasize that there are no numerical artifacts in the hydrodynamic limit when using the \texttt{ES-Hybrid} solver.

\begin{figure}[!h]
	\centering
	\includegraphics[scale=1]{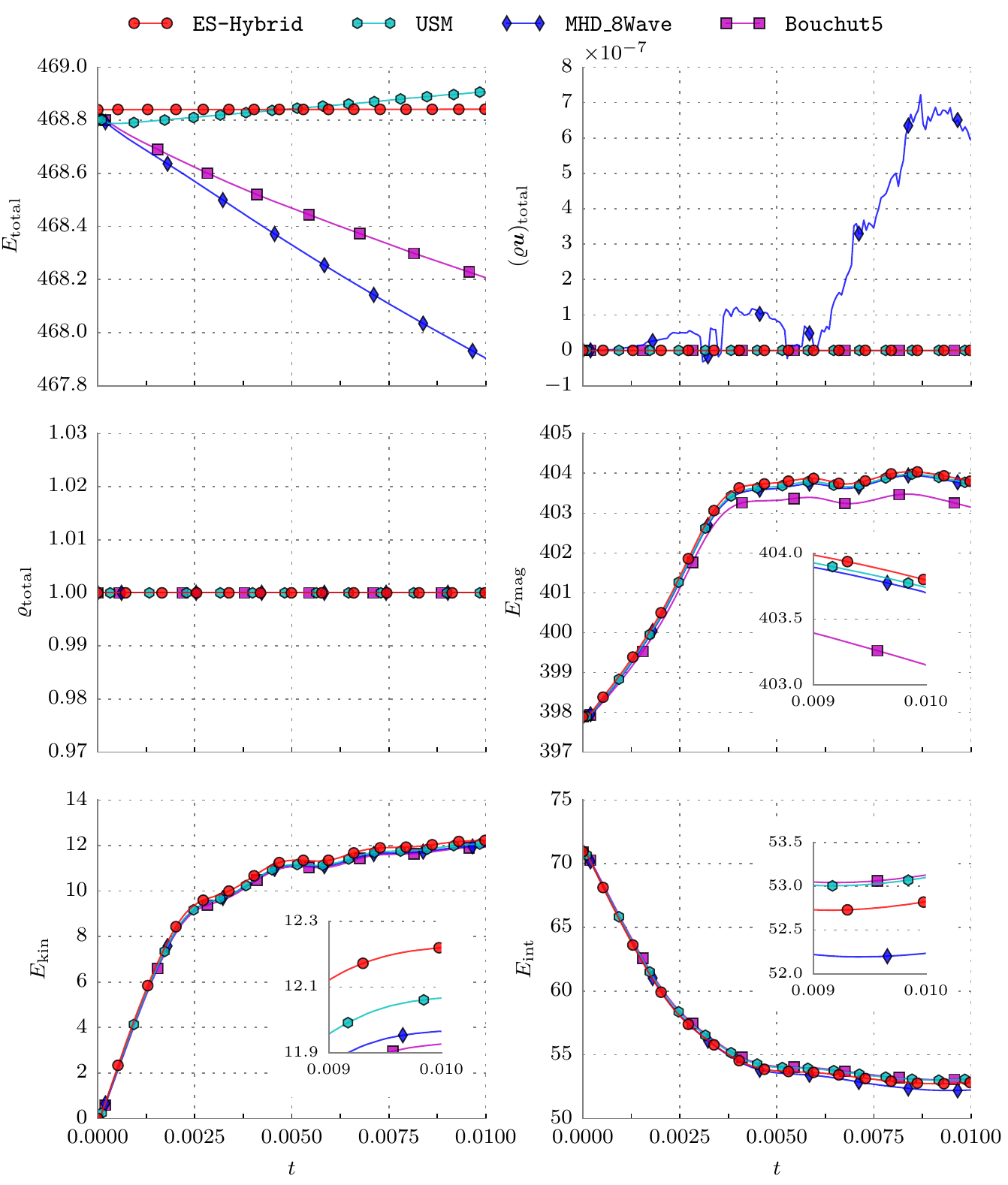}
	\caption{MHD blast wave test: Evolution of the total conserved quantities as well as the individual energies in the $B_x=100/\sqrt{4\pi}$ simulation run. It can easily be seen that the \texttt{ES} solver preserves the total energy well while the other schemes fail to preserve the total energy.}
	\label{fig:MHDBlast_energy}
\end{figure}

\begin{figure}[!h]
	\centering
	\includegraphics[width=0.55\textwidth,trim=250 150 250 150,clip]{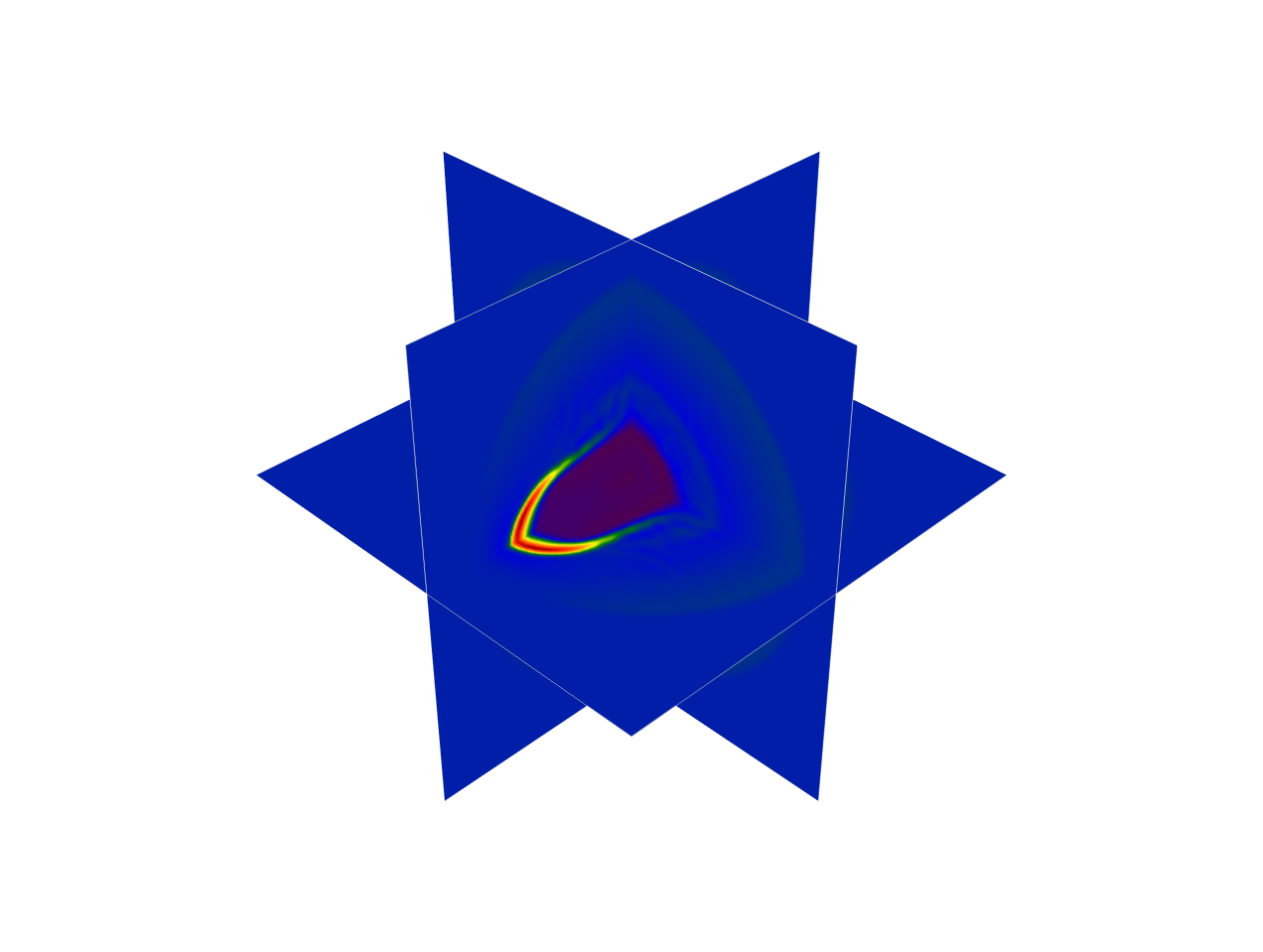} %
	\caption{MHD blast wave test: Adaptive grid resolution up to $128\times128\times128$ cells. Three-slice plot. Linear plot of density, $\rho$, using the same colour range as used in Fig.~\ref{fig:MHDBlast2D}.}
	\label{fig:MHDBlast3D}
\end{figure}

Fig.~\ref{fig:MHDBlast_energy} shows the evolution of the conserved quantities as well as the individual energies. We observe that the \texttt{ES} scheme is mass, momentum, and total energy conservative also in this extreme test. The other MHD solvers fail to maintain the the conservation of total energy by a small amount. Also, the \texttt{MHD\_8Wave} solver fails to preserve the conservation of momentum as expected with the addition of the Powell source term. All solvers conserve mass up to machine precision. Fig.~\ref{fig:MHDBlast3D} shows a 3D variant of this test \cite{FLASHug}.


\jump

\section{Conclusion}\label{scn:Conclusion}

In this work we describe and test an implementation of a high-order, entropy stable numerical method for MHD problems.
Entropy stable numerical fluxes serve as the core of the new entropy stable MHD solver. The implementation is implemented as a new module for the multi-physics \texttt{FLASH} framework \cite{FLASH2000} offering adaptive mesh refinement and large-scale, multi-processor simulations.

The new scheme is implemented with three entropy stable variants: Roe type dissipation (\texttt{ES-Roe}), local Lax-Friedrichs type dissipation (\texttt{ES-LLF}), and a hybrid dissipation term that uses a pressure switch to use  \texttt{ES-Roe} in smooth regions and the more dissipative \texttt{ES-LLF} near pressure discontinuities (\texttt{ES-Hybrid}). The integration in time uses a strong stability preserving (SSP) Runge-Kutta method.

The numerical approximation is built from an entropic point of view. Thus, at a given time, it is possible to compute the current entropy density for all cells in the computational domain. We exploit this additional knowledge of the entropy and reformulate the computation to guarantee positivity of the pressure while maintaining the conservation of the total energy of the system. This reformulation prevents non-physical negative pressures which can occur numerically if the internal energy is a small fraction of the total energy.

We presented a variety of numerical results for the new entropy stable solver implementation for HD and MHD flows in multiple spatial dimensions. These numerical tests served to demonstrate the flexibility and robustness of the new solver. The testing included a recently developed shock-tube experiment, smooth Alfv\'{e}n wave propagation, the Orszag-Tang MHD vortex, the MHD Rotor, and the strong MHD explosion test. The coupling of gravity to our new solver has been tested using the Jeans instability test. We compared the physical aspects of the numerical results, CPU timing and memory consumption of the new entropy stable scheme against the 8-wave, Bouchut 5-wave and unsplit MHD solver implementations already available for \texttt{FLASH}.

We found in these comparisons that the newly implemented entropy stable approximation was the most accurate in smooth regions of a flow. Also, it was shown that the entropy stable scheme was the only one that maintains the conservation of total energy during the computation.

The new \texttt{FLASH} MHD module is freely available upon contact with the corresponding author.

\section*{Acknowledgement}
We thank the anonymous referee for their useful comments, which helped to improve this article.
Dominik Derigs and Stefanie Walch acknowledge the support of the Bonn-Cologne Graduate School for Physics and Astronomy (BCGS), which is funded through the Excellence Initiative, as well as the Sonderforschungsbereich (SFB) 956 on the ``Conditions and impact of star formation''. Stefanie Walch thanks the Deutsche Forschungsgemeinschaft (DFG) for funding through the SPP 1573 ``The physics of the interstellar medium''.
This work has been partially performed using the Cologne High Efficiency Operating Platform for Sciences (\texttt{CHEOPS}) HPC cluster at the Regionales Rechenzentrum K\"{o}ln (RRZK), University of Cologne, Germany.
The software used in this work was developed in part by the DOE NNSA ASC- and DOE Office of Science ASCR-supported FLASH Center for Computational Science at the University of Chicago.
To create some of the figures, we have used the free visualization software \texttt{YT} \cite{Turk11}.

\bibliography{mybibfile}

\appendix
\setcounter{figure}{0}
\setcounter{table}{0}

\section{Analytic Solution of the Gravitational Instability Test}\label{app:Jeans}

We list here the analytic solution for the gravitational instability test as given in \cite{Chandrasekhar1961,FLASHug}.
\begin{align}
\intertext{Kinetic energy:}
E_{\rm kin}(t) &= \frac{1}{8} \frac{\rho_0 \delta_0^2 |\omega|^2 L^2}{k^2} \bigg( 1 - \cos(2 \omega t) \bigg)
\intertext{Internal energy:}
E_{\rm int}(t) &- E_{\rm int}(0) = - \frac{1}{8} \rho_0 a_0^2 \delta^2 L^2 \bigg( 1 - \cos(2 \omega t) \bigg)
\intertext{Potential energy:}
E_{\rm pot}(t) &= - \frac{1}{2} \frac{\pi G \rho_0^2 \delta_0^2 L^2}{k^2} \bigg( 1 + \cos(2 \omega t) \bigg)
\end{align}
where $L$ is the length of the computational domain, and $k$ is the magnitude of the wave vector $\vec{k}$.

\section{Diffusive Magnetic Field Correction}\label{app:divB}
\begin{figure}[h]
	\centering
	\includegraphics[scale=1]{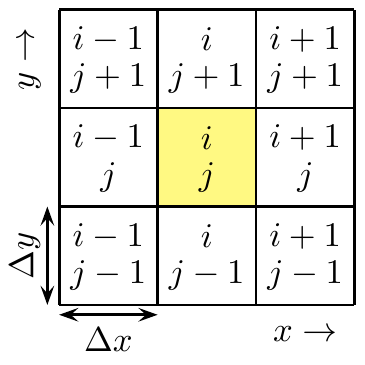}
	\caption{Illustration of index locations in 2D}
	\label{fig:divB_indices}
\end{figure}

We present here the equations used for the implementation of the diffusive divergence error method described in Sec.~\ref{scn:divB}. First and second derivatives are approximated by central second-order finite differences. Fig.~\ref{fig:divB_indices} illustrates the location of the different indices in the two dimensional case.
\vspace*{-3mm}

\definecolor{3D}{rgb}{0.8,0,0}
\begin{align}
\tilde{B_1'} = \partial_x^2 B_1 + \partial_x(\partial_y B_2) {\color{3D}\, +\, \partial_x(\partial_z B_3)} &= \frac{B_{1,i+1,j,k} - 2 B_{1,i,j,k} + B_{1,i-1,j,k}}{\Delta x^2} \notag\\
&+ \frac{\frac{B_{2,i+1,j+1,k} - B_{2,i+1,j-1,k}}{2 \Delta y} - \frac{B_{2,i-1,j+1,k} - B_{2,i-1,j-1,k}}{2 \Delta y}}{2 \Delta x} \notag\\
&{\color{3D}\: +\: \frac{\frac{B_{3,i+1,j,k+1} - B_{3,i+1,j,k-1}}{2 \Delta z} - \frac{B_{3,i-1,j,k+1} - B_{3,i-1,j,k-1}}{2 \Delta z}}{2 \Delta x}}\\[1em]
\tilde{B_2'} = \partial_y^2 B_2 + \partial_y(\partial_x B_1) {\color{3D}\, +\, \partial_y(\partial_z B_3)} &= \frac{B_{2,i,j+1,k} - 2 B_{2,i,j,k} + B_{2,i,j-1,k}}{\Delta y^2} \notag\\
&+ \frac{\frac{B_{1,i+1,j+1,k} - B_{1,i-1,j+1,k}}{2 \Delta x} - \frac{B_{1,i+1,j-1,k} - B_{1,i-1,j-1,k}}{2 \Delta x}}{2 \Delta y} \notag\\
&{\color{3D}\: +\:  \frac{\frac{B_{3,i,j+1,k+1} - B_{3,i,j+1,k-1}}{2 \Delta z} - \frac{B_{3,i,j-1,k+1} - B_{3,i,j-1,k-1}}{2 \Delta z}}{2 \Delta y}}\\[1em]
{\color{3D} \tilde{B_3'} = \partial_z^2 B_3 + \partial_z(\partial_x B_1) \, +\, \partial_z(\partial_y B_2)} &{\color{3D} \:= \frac{B_{3,i,j,k+1} - 2 B_{3,i,j,k} + B_{3,i,j,k-1}}{\Delta z^2}} \notag\\
&{\color{3D} \:+\: \frac{\frac{B_{1,i+1,j,k+1} - B_{1,i-1,j,k+1}}{2 \Delta x} - \frac{B_{1,i+1,j,k-1} - B_{1,i-1,j,k-1}}{2 \Delta x}}{2 \Delta z}} \notag\\
&{\color{3D} \:+\: \frac{\frac{B_{2,i,j+1,k+1} - B_{2,i,j-1,k+1}}{2 \Delta y} - \frac{B_{2,i,j+1,k-1} - B_{2,i,j-1,k-1}}{2 \Delta y}}{2 \Delta z}}
\end{align}
\begin{equation}
\tilde{\vec{B}} = \frac{\Delta x^2 \Delta y^2 \Delta z^2}{\Delta x^2 \Delta y^2 + \Delta x^2 \Delta z^2 + \Delta y^2 \Delta z^2}\bigg(\tilde{B_1'},\tilde{B_2'},\tilde{B_3'}\bigg)^\intercal
\end{equation}

In 2D, we instead have:
\begin{equation}
\tilde{\vec{B}} = \frac{\Delta x^2 \Delta y^2}{\Delta x^2 + \Delta y^2 }\bigg(\tilde{B_1'},\tilde{B_2'},0\bigg)^\intercal
\end{equation}

Note that in 2D computations the {\color{3D} dark red} highlighted parts are zero and can be neglected.

\jump

\section{Flowchart}\label{sec:Flowcharts}

\begin{figure}[h]
	\centering
	\begin{minipage}{0.48\textwidth}
		\centering
		\includegraphics[scale=0.8]{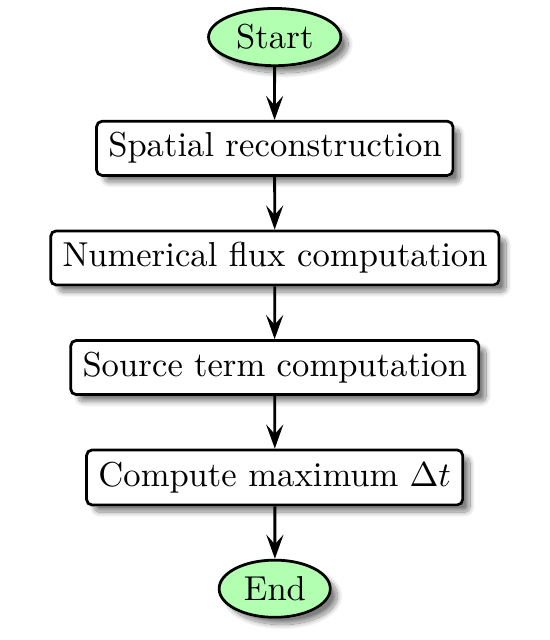}
	\end{minipage}\hfill
	\begin{minipage}{0.48\textwidth}
		\centering
		\includegraphics[scale=0.8]{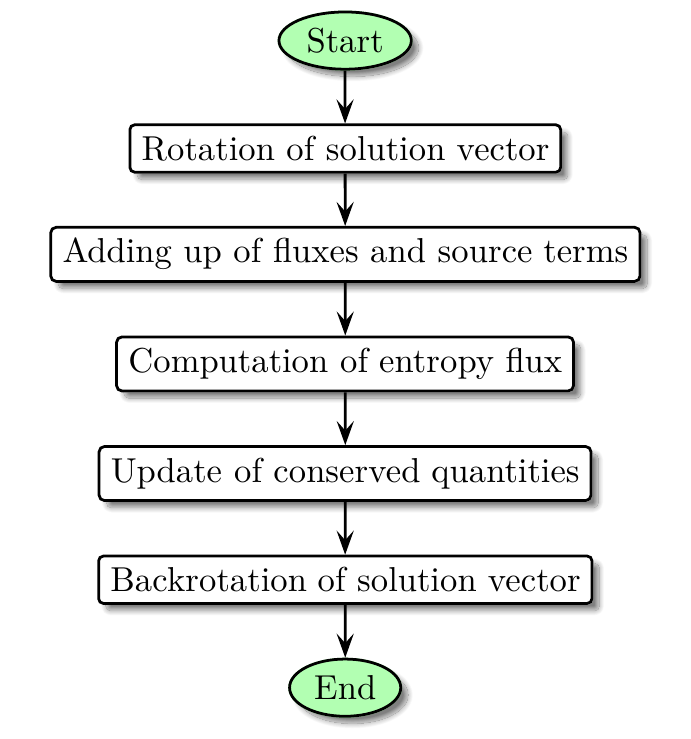}
	\end{minipage}\\[-.5em]
	\begin{minipage}{0.48\textwidth}
		\caption{Flowchart showing the flux computation procedure of the \texttt{ES} solver. For a full description see Fig.~\ref{fig:flowchart:global}.}
		\label{fig:flowchart:flux}
	\end{minipage}\hfill
	\begin{minipage}{0.48\textwidth}
		\caption{Flowchart showing the solution update procedure of the \texttt{ES} solver. For a full description see Fig.~\ref{fig:flowchart:global}.}
		\label{fig:flowchart:update}
	\end{minipage}
\end{figure}

\begin{figure}[h]
	\centering
	\includegraphics[scale=0.8]{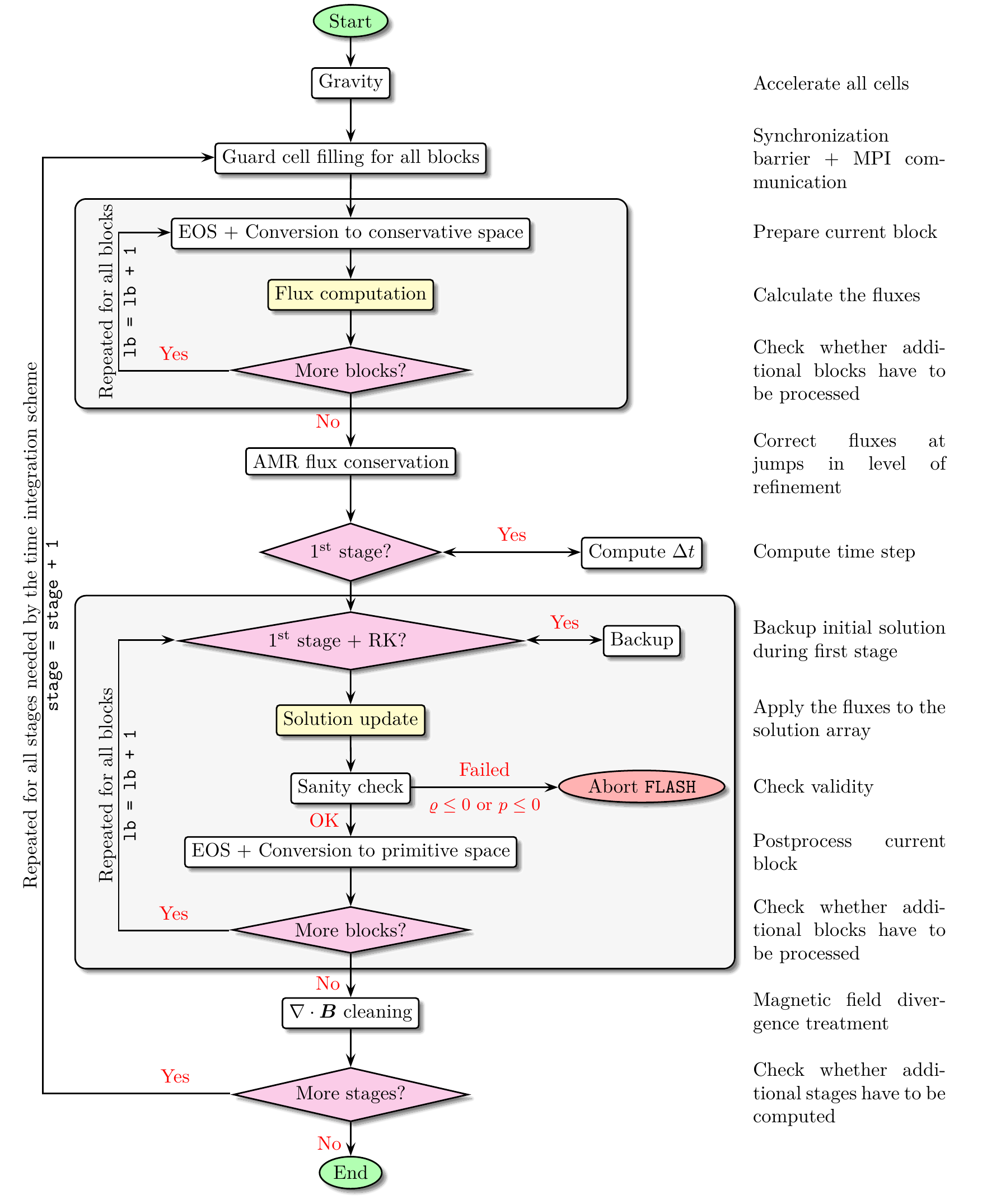}
	\caption{Flowchart showing the principle of operation of the \texttt{ES} solver. The steps ``Flux computation'' and ``Solution update'' is depicted in separate flowcharts, shown in Fig.~\ref{fig:flowchart:flux} and \ref{fig:flowchart:update}, respectively.}
	\label{fig:flowchart:global}
\end{figure}

In this section we provide flowcharts to detail and outline the workflow of the new \texttt{ES} MHD solver. We divide the flowchart description of the algorithm into three parts. The first, shown in Fig.~\ref{fig:flowchart:global}, depicts the global solver procedure for a single stage (of the possibly multi-stage) Runge-Kutta method. The second flowchart in Fig.~\ref{fig:flowchart:flux} displays the workflow for the computation of the numerical fluxes. The third flowchart in Fig.~\ref{fig:flowchart:update} displays the workflow for the update of the solution array.

\jump

\section{Dimensional MHD equations}\label{app:Units}
The dimensional MHD equations are given by
\begin{align}\label{eq:dim3DMHD}
\pderivative{}{t}
\begin{bmatrix} \rho \\[0.05cm]
\rho\vec{u} \\[0.05cm]
E \\[0.05cm] \vec{B}
\end{bmatrix}
+ &\nabla\cdot
\begin{bmatrix} \rho\vec{u} \\[0.05cm]
\rho(\vec{u}\otimes\vec{u}) + \left(p+\frac{\|\vec{B}\|^2}{2 \mu_0}\right)\mat{I}-\frac{\vec{B}\otimes\vec{B}}{\mu_0} \\[0.05cm]
\vec{u}\left(E + p + \frac{\|\vec{B}\|^2}{2 \mu_0} \right) - \frac{\vec{B}(\vec{u}\cdot\vec{B})}{\mu_0} \\[0.05cm]
\vec{B}\otimes\vec{u} - \vec{u}\otimes\vec{B}
\end{bmatrix} = 0,
\end{align}
where the thermal pressure is related to the conserved quantities through the ideal gas law:
\begin{equation}\label{eq:dimpressure}
p = (\gamma-1)\left(E - \frac{\rho}{2}\|\vec{u}\|^2 -\frac{\|\vec{B}\|^2}{2 \mu_0} \right).
\end{equation}
The unit system is determined at compilation time through the user-definable parameter \texttt{HY\_UNIT} in the \texttt{ES.h} parameter file.
In non-dimensional units, \eqref{eq:dim3DMHD} and \eqref{eq:dimpressure} are identical to \eqref{3DIDEALMHD} and \eqref{eq:pressure}.

The chosen unit system leads to different internal conversion factors within our implementation. The resulting units of the simulation quantities are listed in Table~\ref{tab:units}. Depending on the setting, the according vacuum permeability constant, $\mu_0$, is chosen. Note that both the specific energy, $E$, and the pressure, $p$, are of the same units. 

\begin{table}[h]
	\centering
	\begin{tabular}{lrccc}
	\toprule
	\multicolumn{2}{l}{Unit system} & non-dimensional & SI & CGS \\
	\multicolumn{2}{l}{\texttt{\#define HY\_UNIT}} & \texttt{0} & \texttt{1} & \texttt{2} \\
	\midrule
	Length &$\ell$ 			& 1 & \si{m} & \si{cm}\\
	Time &$t$ 				& 1 & \si{s} & \si{s}\\
	Density &$\varrho$		& 1 & \si{kg.m^{-3}} & \si{g.cm^{-3}}\\
	Velocities &$\vec{u}$ 	& 1 & \si{m.s^{-1}} & \si{cm.s^{-1}}\\
	Specific energy &$E$ 	& 1 & \si{J.m^{-3}} & \si{erg.cm^{-3}}\\
	Pressure &$p$			& 1 & \si{N.m^{-2}} & \si{dyn.cm^{-2}}\\
	Magnetic field &$B$ 	& 1 & \si{T} & \si{G}\\
	\multicolumn{2}{c}{with $\mu_0 = $}	& 1 & \SI{4\pi e-7}{T^2.m^3.J^{-1}} & \SI{4\pi}{G^2.cm^3.erg^{-1}}\\
	\bottomrule
	\end{tabular}
	\caption{Simulation units with different settings of the compilation-time parameter \texttt{HY\_UNIT}}
	\label{tab:units}
\end{table}

\section{Error norms and the experimental order of convergence}\label{scn:errors_and_eoc}
The discrete ${L}_1$- and ${L}_2$-errors are defined by
\begin{equation*}
	\lVert \Delta u(t) \rVert_{L_1} = \frac{1}{N^d} \sum_{i,j=1}^N \bigg| u_{i,j}^{\rm exact} - u_{i,j}^{\rm solution} \bigg|, 
	\quad\mbox{and}\quad
	\lVert \Delta u(t) \rVert_{L_2} = \sqrt{ \frac{1}{N^d} \sum_{i,j=1}^N \bigg( u_{i,j}^{\rm exact} - u_{i,j}^{\rm solution} \bigg)^2 }.
\label{eq:errornorm}
\end{equation*}
where $N$ is the number or grid points in each direction, and $d$ is the number of spatial dimensions. After computing the norms of the errors, we obtain the experimental order of convergence (EOC) using
\begin{equation}
{\rm EOC}(i \rightarrow j) = \frac{\log\big(\lVert \Delta u_i(t) \rVert\big) - \log\big(\lVert \Delta u_j(t) \rVert\big)}{\log\big(N_i / N_j\big)} .
\end{equation}
\end{document}